%% file: acl_latex.tex
\pdfoutput=1

\documentclass[11pt]{article}

\usepackage[preprint]{acl}

\usepackage{times}
\usepackage{latexsym}
\usepackage{tabularx}

\usepackage{booktabs}
\usepackage{array}
\usepackage{multirow}
\usepackage{graphicx}

\usepackage{makecell}
\usepackage{rotating} 

\usepackage{amsmath} 
\usepackage{multicol} 
\usepackage{enumitem} 
\usepackage{appendix} 
\usepackage{adjustbox} 
\usepackage{xspace}

\usepackage[T1]{fontenc}

\usepackage[utf8]{inputenc}

\usepackage{microtype}

\usepackage{inconsolata}

\usepackage{graphicx}
\usepackage{tcolorbox}

\newcommand{\ruhumaneval}{ruHumanEval\xspace}
\newcommand{\rucodeeval}{ruCodeEval\xspace}
\newcommand{\javatestgen}{JavaTestGen\xspace}
\newcommand{\realcode}{RealCode\xspace}
\newcommand{\realcodejava}{RealCodeJava\xspace}
\newcommand{\rucodereviewer}{RuCodeReviewer\xspace}
\newcommand{\codelintereval}{CodeLinterEval\xspace}
\newcommand{\unittests}{UnitTests\xspace}
\newcommand{\strucom}{StRuCom\xspace}
\newcommand{\codecorrectness}{CodeCorrectness\xspace}
\newcommand{\yabloco}{YABLoCo\xspace}

\def\rot{\rotatebox}

\title{MERA Code: A Unified Framework for Evaluating Code Generation Across Tasks}

\author{
 \textbf{Artem Chervyakov\textsuperscript{1}},
 \textbf{Alexander Kharitonov\textsuperscript{1}},
 \textbf{Pavel Zadorozhny\textsuperscript{1}},
 \textbf{Pavel Adamenko\textsuperscript{1}}, \\
 \textbf{Rodion Levichev\textsuperscript{1}}, 
 \textbf{Dmitrii Vorobev\textsuperscript{1}},
 \textbf{Dmitrii Salikhov\textsuperscript{1}},
 \textbf{Aidar Valeev\textsuperscript{1,8}},
 \textbf{Alena Pestova\textsuperscript{3}}, \\
 \textbf{Maria Dziuba\textsuperscript{2,3}},
 \textbf{Ilseyar Alimova\textsuperscript{7}},
 \textbf{Artem Zavgorodnev\textsuperscript{4}}, 
 \textbf{Aleksandr Medvedev\textsuperscript{4}}, 
 \textbf{Stanislav Moiseev\textsuperscript{4}},\\
 \textbf{Elena Bruches\textsuperscript{6}}, 
 \textbf{Daniil Grebenkin\textsuperscript{6}}, 
 \textbf{Roman Derunets\textsuperscript{6}}, 
 \textbf{Vikulov Vladimir\textsuperscript{5}},  
 \textbf{Anton Emelyanov\textsuperscript{1}}, \\
 \textbf{Vladimir V. Ivanov\textsuperscript{8}},
 \textbf{Dmitry Babayev\textsuperscript{1}},
 \textbf{Valentin Malykh\textsuperscript{3}}
 \textbf{Alena Fenogenova\textsuperscript{1}}
\\
\\
 \textsuperscript{1} SberAI,
 \textsuperscript{2} ITMO University,
 \textsuperscript{3} MWS AI,
 \textsuperscript{4} T-Technologies,
 \textsuperscript{5} Rostelecom, \\
 \textsuperscript{6} Siberian Neuronets,
 \textsuperscript{7} Skoltech,
 \textsuperscript{8} Innopolis University
 \small{
   \textbf{Correspondence:} \href{mailto:mera@a-ai.ru}{mera@a-ai.ru}
 }
}

\begin{document}
\maketitle
\begin{abstract}
Advancements in LLMs have enhanced task automation in software engineering; however, current evaluations primarily focus on natural language tasks, overlooking code quality. 
Most benchmarks prioritize high-level reasoning over executable code and real-world performance, leaving gaps in understanding true capabilities and risks associated with these models in production. To address this issue, we propose \textbf{MERA Code}, a new addition to the MERA benchmark family, specifically focused on evaluating code for the latest code generation LLMs. This benchmark includes 11 evaluation tasks that span 8 programming languages. 
Our proposed evaluation methodology features a taxonomy that outlines the practical coding skills necessary for models to complete these tasks. The benchmark comprises an open-source codebase for users to conduct MERA assessments, a scoring system compatible with various programming environments, and a platform featuring a leaderboard and submission system. 
We evaluate open LLMs and frontier API models, analyzing their limitations regarding practical coding tasks in non-English languages. 
We are publicly releasing MERA to guide future research, anticipate groundbreaking features in model development, and standardize evaluation procedures.
\end{abstract}

\section{Introduction}

Recent advances in large language models (LLMs) have demonstrated significant potential for automating software engineering tasks, such as code generation and documentation. However, current evaluation methods predominantly emphasize natural language understanding, often neglecting critical factors such as code quality, real-world applicability, and multilingual support. Benchmarks such as HumanEval-X~\cite{zheng2023codegeex}, MultiPL-E~\cite{cassano2022multipl}, and mxEval~\cite{athiwaratkun2023multi} address multilingualism through translated datasets but mainly focus on programming languages. They do not fully capture the interaction between natural language (such as requirements and comments) and code, which is vital for real-world development. This limitation restricts our understanding of LLMs' practical capabilities, particularly in non-English contexts. Tasks such as writing localized documentation or interpreting vague requirements necessitate proficiency in both natural and programming languages, yet no comprehensive benchmark currently evaluates these cross-disciplinary skills. To fill this gap, we introduce \textbf{MERA Code} — a comprehensive evaluation system explicitly designed for the Russian language. MERA Code provides tools to assess LLMs in realistic, practical, multilingual software development scenarios.

Our key contributions are:
\begin{itemize}[noitemsep, topsep=1pt, partopsep=1pt, parsep=3pt]
\item A reproducible evaluation methodology for LLMs;
\item A suite of 11 instruction-formatted tasks (code2text, text2code, code2code) across 8 programming languages (Python, Java, C\#, JavaScript, Go, C, C++, and Scala);
\item An open evaluation platform 
with a scoring system, framework~\footnote{The evaluation code  \url{https://github.com/MERA-Evaluation/MERA_CODE} is released under the MIT License}, and public leaderboard;
\item A comprehensive evaluation of models performance, ranging from open-source general models to proprietary coding APIs.
\end{itemize}

MERA Code serves as a foundational resource for the research and industrial community, promoting collaboration to enhance task coverage and adapt to evolving LLM capabilities. By combining natural and programming language evaluation, we support more relevant assessments of LLMs in software engineering.

\section{Related Work}
\label{sec:related}

Early benchmarks for evaluating LLM coding skills focused on short problems scored by automated test‐execution systems, such as HumanEval~\cite{chen2021evaluating} and MBPP~\cite{austin2021program}, and tasks from online platforms like APPS~\cite{hendrycksapps2021} and CodeContests~\cite{li2022competition}. As context windows grew, evaluations expanded from class-level (ClassEval~\cite{du2023classeval}) to repository-scale tasks (RepoBench~\cite{liurepobench}, SWE-Bench~\cite{jimenez2024swe}), with dynamic suites (LiveCodeBench~\cite{jainlivecodebench}, CodeElo~\cite{quan2025codeelo}) emerging to mitigate data leakage, though they still lack industrial complexity. More fundamentally, the field predominantly privileges code generation, overlooking critical tasks such as code repair, execution, and test output prediction that are partially addressed by broader initiatives (CodeXGLUE~\cite{lu2021codexglue}, LongCodeArena~\cite{bogomolov2024long}). Furthermore, while multilingual programming benchmarks (MultiPL-E~\cite{cassano2022multipl}, HumanEval-X~\cite{zheng2023codegeex}) expand code language coverage, they omit the essential multilingual natural-language aspects of software engineering—such as requirements, documentation, and review. Consequently, no existing benchmark comprehensively evaluates LLMs across the full software development lifecycle in both natural and programming languages as encountered in practice.

\section{MERA Code}
\subsection{Overview}
\label{subsec:system_demo:mera_code_overview}
MERA Code introduces a practical evaluation framework that expands the Family of MERA benchmarks\footnote{\url{https://mera.a-ai.ru/en/code}}~\cite{fenogenova-etal-2024-mera} for assessing code-oriented and general-purpose state-of-the-art (SOTA) models. This framework serves as a foundational resource for both the research and industrial communities, fostering collaboration to improve task coverage and adapt to the evolving capabilities of LLMs. 
The release includes the proposed taxonomy of the model's skills needed to solve practical coding tasks, along with a set of public and private test tasks in an instruction-based format, an open evaluation codebase, and a web platform featuring an open automatic scoring system and leaderboard.

\subsection{Methodology}
\label{subsec:system_demo:methodology}
\subsubsection{Taxonomy}
\label{subsubsec:taxonomy}
Several works proposed various taxonomies to systematize the field of software development and engineering. In \cite{zhang2023unifying}, the taxonomy aligns with the software development pipeline and considers tasks that must be solved to deliver the product. The work \cite{ruf2015classification} analyzes the field of software engineering from a perspective of programming skills required to solve specific tasks.
Our taxonomy is based on the latter approach since it allows us to decompose arbitrary tasks into a limited number of skills. Hence, the resulting taxonomy could be exhaustive yet straightforward.

\begin{figure*}[htbp]
    \centering
    \includegraphics[width=0.95\textwidth]{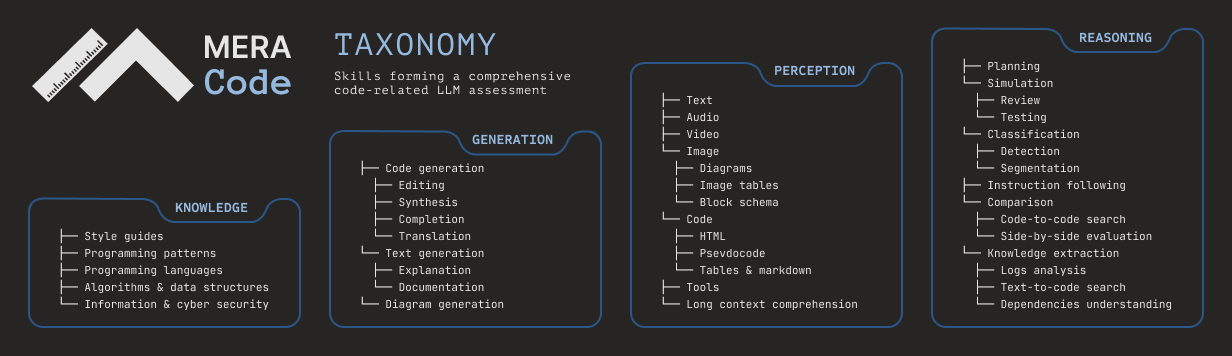}
    \caption{Taxonomy of MERA Code encompassing four foundational skills \textit{Perception / Knowledge / Reasoning / Generation} utilized in the model to address certain tasks. Detailed explanation of each skill could be found in Appendix \ref{sec:appendx_taxonomy}.}
    \label{fig:taxonomy}
\end{figure*}

A language model can be considered as a system consisting of inputs, internal state, and outputs. According to these parts, we derive four fundamental skills as a ground of our taxonomy: \textit{Perception} (input), \textit{Reasoning} and \textit{Knowledge} (internal state), and \textit{Generation} (output). All other skills form a tree, becoming increasingly niche with each new level. The taxonomy~\footnote{We recognize that the current tasks in MERA do not encompass the complete taxonomy of skills. This serves as a guide for the community to understand what is covered and what needs to be added to future tasks in a systematic manner.} presented in Figure~\ref{fig:taxonomy} is based on \cite{ruf2015classification}, \cite{zhang2023unifying}, and the authors' domain expertise.

\subsubsection{Prompts selection}
LLMs' performance might significantly depend on a given prompt \cite{shin-etal-2020-autoprompt}, \cite{gao-etal-2021-making}, \cite{fenogenova-etal-2024-mera}. 
To ensure an unbiased and robust assessment, each of the 11 tasks is paired with a set of distinct prompts. These prompts differ in syntactic framing, level of detail in the task description, and specification of the required output format. Prompts are assigned uniformly at random across all samples for a given task, with exactly one prompt applied per sample and the assignment is fixed. This strategy mitigates any inadvertent advantage or disadvantage conferred by a particular prompt style or model family affinity. Every prompt comprises a clear template for the expected output, thereby aligning model responses with the automated scoring system.
Example prompts for each task are provided in Appendix \ref{sec:appendix_ds_exp}. The benchmark is designed to evaluate performance on tasks relevant to Russian-speaking users, and as such, the prompt examples reflect real-world use cases. An ablation study~\ref{sec:prompts_en} provides evidence of the language-robust benchmark methodology based on the comparison of the Russian and English instructions.

\subsubsection{The impact of the prompts language}

Table \ref{tab:en_prompts_tasks} reports the score differences between Russian- and English-prompted runs\footnote{The \rucodereviewer, \rucodeeval, \ruhumaneval, and \strucom are excluded from the comparison, as the tasks are designed to have natural language as a part of the input data or targets (not only the instruction formulations as in the other benchmark tasks).} as median [Q1, Q3] across tasks (scores are bounded between 0 and 1). Overall, the medians are very close to zero, and the interquartile ranges are typically well below 0.1 in absolute magnitude, indicating that prompt language has little practical effect on performance for the benchmark. For most tasks, the IQR straddles or stays near zero (e.g., \codecorrectness, \unittests, \realcode), suggesting no consistent advantage of either language, with only a small positive shift on \yabloco and \codelintereval, and a mild negative tendency on \realcodejava and \javatestgen. Taken together, these results support the view that the benchmark is largely language-robust (at least between English and Russian); since the tasks are code-centric and much of the input is programming language rather than natural language, models appear to rely primarily on the code and formal constraints, making the evaluation methodology relatively insensitive to the human-language surface form of the instruction (see the detailed comparison in \ref{sec:prompts_en}).

\input{table/short_en_prompts_per_task} 

\subsubsection{Evaluation procedure}
All MERA Code tasks employ purely generative and instructive assessment where models continuously emit tokens until meeting predefined stopping conditions. Each raw output undergoes task-specific post-processing to conform with metric requirements, including mandatory extraction of markdown-fenced code blocks from responses (with fallback to full outputs when parsing fails) before being scored with metrics. All metrics utilized in MERA Code are listed in Table~\ref{tab:tasks_info} and described in Appendix\ref{sec:metric_desc}.

\textbf{Total Score} and \textbf{Private Score} are calculated as the mean value across all tasks and private tasks, respectively. For tasks that have multiple metrics, these metrics are also averaged.

\subsection{Tasks}
\label{subsec:system_demo:tasks}

The MERA Code currently encompasses 11 code tasks. The statistics are presented in Table~\ref{tab:tasks_info}. We will briefly discuss the tasks listed below. Additional details and examples can be found in Appendix~\ref{sec:appendix_ds_exp}.

\paragraph{CodeLinterEval} is a benchmark dataset designed to evaluate Python code generation and correction abilities based on linter errors, assessing models' understanding of error messages, adherence to PEP 8 style, and preservation of code logic. It contains 110 samples with source code, linter feedback, instructions, and canonical solutions, using \texttt{pass@k} as an evaluation metric.

\paragraph{CodeCorrectness} is a benchmark for evaluating LLMs’ ability to predict whether unit tests compile and execute successfully or fail. The dataset consists of 1,361 samples written in Java, Python, and Go, which were collected from both human and LLM sources and subsequently selected based on automated filtering criteria. Evaluation is performed via \texttt{EM} against execution-verified ground truth.

\paragraph{RealCode and RealCodeJava} are benchmarks for evaluating the ability of code generation models to synthesize function bodies within real-world Python and Java projects, respectively. RealCode includes 802 tasks from 95 Python repositories created in 2024, while RealCodeJava comprises 298 tasks from 27 Java repositories created between 2024 and 2025. In both cases, tasks are constructed by identifying functions covered by tests, replacing their bodies with mocks, and retaining only those where tests fail on the mock but pass on the original. This ensures that the task requires generating semantically correct and test-sensitive logic. For each task, the model must generate the full function body based on the surrounding code context. Generated completions are inserted back into the project and evaluated automatically using the \texttt{repotest}\footnote{\url{https://github.com/MERA-Evaluation/repotest}} library and the project's original test suite. The primary metric is \texttt{pass@1}.

\paragraph{JavaTestGen} is a benchmark for evaluating code generation models on the task of producing \emph{executable} JUnit~5 unit tests for real-world Java classes. The benchmark consists of 227 tasks mined from open-source GitHub repositories. Each task provides the source code of a focal Java class, its name, and the expected test class name; the model must generate a complete and executable test file. 
Evaluation is fully automated: \texttt{compile@1} measures the proportion of generations that compile successfully, while \texttt{pass@1} quantifies those that pass all tests when executed with \texttt{mvn test} inside a provided Docker environment. We also release and use \texttt{repotest}\footnotemark[\value{footnote}] to run the tests and measure correctness.

\paragraph{StRuCom} \cite{dziuba2025strucom} is the first benchmark for evaluating LLMs' ability to generate structured Russian-language code documentation. It comprises 153K examples across Python, Java, JavaScript, C\#, and Go, sourced from human-authored GitHub repositories and synthetically generated examples. The $500$ samples were carefully selected from these examples as a test set for MERA Code. Evaluation employs the \texttt{chrF} metric due to its sensitivity to Russian morphological complexity.

\paragraph{YABLoCo} \cite{valeev2025yabloco} is a benchmark that evaluates the performance of LLMs on large repositories. It focuses on C and C++ languages since other benchmarks do not fully cover them. YABLoCo consists of $208$ carefully selected functions from four large open-source repositories (bullet3, Redis, OpenSSL, LLVM). Each function is assigned a context category determined by the function's dependencies. The primary purpose of the benchmark is to evaluate the quality of LLMs' generations when various repository contexts are given alongside the prompt. The quality of the generated code is measured by the \texttt{pass@1} and \texttt{EM} metrics.

\paragraph{RuCodeReviewer}
The first benchmark for automated code review comment generation in Russian is designed to address the limitations of English-centric and often noisy datasets. The benchmark is built from a high-quality dataset of 689 merge-request diff-comment pairs sourced from an industry-oriented educational program, where professional developers review student code (Java, Python, Go, Scala). A two-stage curation process is employed, involving an LLM-assisted filter followed by verification from two independent human experts, to ensure that every comment is self-contained, actionable, and reproducible from the code diff alone. As evaluation metrics, we use \texttt{Judge@k}, \texttt{chrF}, \texttt{BLEU} in a reference-based setup. The LLM judge is validated with the help of human experts and achieves a \textbf{Spearman correlation of 0.831} with human annotators, confirming its reliability for the assessment of semantic equivalence. Details about LLM-as-a-Judge validation could be found in Appendix \ref{sec:rucodereview}.

\paragraph{UnitTests} is a multilingual dataset for evaluating model abilities to generate unit tests for functions and methods in five programming languages: Python, Java, Go, JavaScript, and C\#. It is built from GitHub repositories with open licenses. Each observation consists of a focal function/method, a unit test for it (test function/method), as well as the focal and test contexts collected from the repository for the focal and test functions, respectively. The goal of the model is to generate a unit test based on the provided focal function/method and its context, as well as the test function/method context. The evaluation metric is \texttt{CBLEU}. 

\paragraph{ruHumanEval and ruCodeEval} are presented in the MERA benchmark\footnote{\url{https://mera.a-ai.ru/ru/text/tasks}} as public and private tests, respectively. In this version, we refine the prompt's base, making it more precise for code completion tasks and eliminating ambiguous formulations that cause issues for models generating from scratch. We also change the generation processing algorithm to ensure better code blocks parsing and increase the maximum amount of new tokens to accommodate for models with higher reasoning abilities. The evaluation is conducted via \texttt{pass@k} metric.

\input{table/datasets}

\subsection{System Demo}
\label{subsec:system_demo}

The automatic scoring system is accessible on the benchmark platform~\footnote{\url{https://mera.a-ai.ru/en/code}}. The submission process involves the following steps. 

First, users must clone the MERA Code benchmark repository~\footnote{The \href{https://github.com/MERA-Evaluation/MERA\_CODE}{MERA Code} is based on the fork of the LM-evaluation harness~\cite{eval-harness}.} and create submission files using a shell script and the provided evaluation code. Two repository branches are available: one enables local scoring of public tasks with reference outputs, while the other excludes complex dependencies to facilitate submission to the automated scoring system. 

Second, they must register on the benchmark platform and upload their submission files through the interface in their account for automatic evaluation. The scoring process may take up to three hours. Once the evaluation is complete, the results are displayed in the user's account and remain private unless the user opts to publish them. In this case, the submission undergoes expert verification, which includes checking the log files automatically generated by the evaluation script and reviewing the provided submission information. Once approved, the model's score is publicly visible on the leaderboard, while the specific outputs remain private. The user process for the benchmark is illustrated in Figure~\ref{fig:submit} of the Appendix.

\section{Evaluation results and analysis}
\label{sec:eval}

\input{table/mera_code_results}

We evaluate both code-oriented and general-purpose models of various sizes. Table~\ref{tab:mera_results} shows the results of the models evaluation. Detailed results for public and private tasks are presented in the Appendix \ref{sec:full_metrics}. The snapshot of the leaderboard was taken on November 15, 2025.


\textbf{Task complexity.}
The most difficult task is \rucodereviewer which requires the largest number of skills according to our taxonomy. Moreover, the task presumes strong capabilities in Programming and Natural languages simultaneously, making it even harder.
The easiest task is \codecorrectness, probably due to its simple format (binary classification).
In general, current benchmark results show modest model performance, with substantial headroom due to task difficulty.

\textbf{Impact of natural languages.}
Since all prompts in MERA Code are formulated in Russian, a strong proficiency in the language is required for the models to get higher scores on some benchmark tasks. Hence, relatively small models, RuadaptQwen3-8B-Hybrid and A-vibe, targeted at the Russian language, appear at the top of the leaderboard. Particularly, A-vibe demonstrates superiority (0.307) on the \strucom (generating commentaries in Russian) task.

\textbf{Influence of the pretraining data.}
The Table~\ref{tab:mera_results} demonstrates that even smaller models may outperform larger ones from the same or ``close'' model families if they are trained specifically on the code tasks data. To be precise, Qwen2.5-Coder-32B-Instruct shows larger Total Score (0.3) compared to Qwen2.5-72B-Instruct (0.288). The same applies to the Qwen3-Coder-30B-A3B-Instruct result (0.301). 

\textbf{Model specialization.}
Our benchmark demonstrates that models lean toward certain specializations and programming languages. GPT-4.1 is efficient in Software Engineering, showing top performance on \realcode (0.418), \realcodejava (0.416), \javatestgen (0.344), and \rucodereviewer (0.096) --- the tasks that cover code completions, unit test generation, and code review. Gemini 2.5 flash achieves top scores on \ruhumaneval (0.604) and \rucodeeval (0.610), which are designed for the evaluation of Python code generation abilities. DeepSeek-Coder-V2-Instruct is the best on the \yabloco benchmark (0.149), outperforming other models in C++ and Long Context Comprehension. Seed-Coder-8b-Instruct is proficient in Python style guides (\codelintereval: 0.655). Such a distribution of skills among models confirms the complexity and diversity of our benchmark.

\section{Conclusion.}
In this work, we present the MERA Code, a comprehensive evaluation tool designed for coding tasks and programming languages across eight different languages. MERA provides an open-source evaluation toolkit, a structured scoring system, and a public leaderboard, enabling standardized assessments of coding skills while addressing limitations in non-English contexts. This framework enhances our understanding of model capabilities and helps identify strengths and weaknesses based on community tests. We are releasing MERA to the research community to encourage innovation, establish standards, and promote reproducible evaluation practices. 

\section*{Limitations}
\label{sec:limitations}
While the proposed benchmark advances the evaluation of LLMs' coding abilities, several limitations remain.

\paragraph{Limited Representativeness} Despite efforts to ensure coverage, our datasets cannot comprehensively represent the entire landscape of programming problems, especially as real-world coding tasks are highly diverse and domain-specific. As a result, while our benchmark captures a broad range of difficulty levels and programming domains, certain edge cases and emerging languages or paradigms may not be well represented.  Technical challenges related to non-Python languages complicate our ability to cover this variety.

\paragraph{Code Quality Assessment} 


We employ various types of evaluations across the datasets, including n-gram matching, LM-as-Judge, and pass@k. While many datasets utilize the LM-as-Judge technique for evaluation, this method does not always adequately assess deeper aspects of code quality, such as readability, efficiency, maintainability, security, and adherence to best practices. Additionally, significant challenges arise from the parsing of generated answers and the models' instruction-following capabilities, which can affect the accuracy of the results. These qualitative factors often require more nuanced human judgment, which cannot be easily scaled for everyday benchmarking.

\paragraph{Testing Conditions} 

The benchmark currently assumes that coding prompts are well-specified and unambiguous, which is often not the case in practice. The ability of large language models (LLMs) to handle unspecified, noisy, or ambiguous real-world requirements remains a challenge. Additionally, our evaluation infrastructure assumes that the generated code can be executed in isolated environments, which may not accurately reflect the complexity and constraints of production development settings. 

There are instances where Docker containers cannot be built due to issues arising from the network setups of several datasets (such as RealCode, RealCodeJava, and JavaTestGen) that are designed to reproduce the conditions found in their respective repositories. These containers may not always compile successfully. While we strive for accuracy in our deployment, we cannot guarantee that it will always score everything perfectly. However, any discrepancies are typically within 4\%, which accounts for about 0.2\% of the total, and should not significantly impact the overall ranking.

\paragraph{Data Contamination} 
While we have implemented measures to minimize data contamination (e.g., by removing public problems found in known training sets), there remains the possibility that LLMs — particularly when trained on vast internet corpora — may have encountered similar or even identical problems during training, potentially inflating performance results.
Additionally, some datasets were created from repositories that might contribute to the risk of data contamination.

\paragraph{Scoring optimization} The scoring system in the benchmark takes time to produce results, often requiring up to 2 hours due to the complexity of running environments and libraries for various programming languages. We plan to address code optimization and testing environments in the future.

\paragraph{Technical Constraints} As with any rapidly evolving technology, improvements in LLM architectures and training data may quickly impact the relevance of our benchmark. Therefore, regular updates and community engagement will be necessary to maintain its utility as a robust evaluation tool.

\section*{Ethical Statement}
\label{ethical}
As with any research that advances the development and evaluation of LLMs, introducing a new coding benchmark necessitates careful consideration of ethical implications.

Firstly, the benchmark has been constructed mindful of data privacy and intellectual property rights and released under the MIT licence. All programming tasks and datasets utilized are either original or derived from public domain resources, and are used with proper attribution and permissions. We have taken precautions to ensure that example problems do not contain proprietary or sensitive information and that datasets do not inadvertently leak private user data.

Secondly, the public release of such a benchmark may facilitate broader research into improving LLMs for software development. While this has clear benefits in promoting open scientific progress, it may also be misused, for example, to enhance automated systems capable of generating malicious code or exploiting software vulnerabilities. We strongly discourage the use of this benchmark, or any resulting models, for unethical or harmful purposes and urge the community to adhere to responsible usage guidelines.

Thirdly, although LLMs have the potential to democratize programming by lowering barriers to software development, concerns exist regarding their impact on the workforce, code quality, and reliance on automated tools. Our benchmark is intended strictly for research and evaluation purposes. We encourage users to complement model assessment with human oversight and to remain vigilant for biases, errors, or unsafe behavior that may emerge from automatic code generation.

Lastly, by releasing the benchmark and associated evaluation code, we commit to transparency and reproducibility in our research. We invite feedback from the broader community, particularly regarding unintended biases, fairness across programming languages, and the benchmark's accessibility for a global and diverse audience.

\noindent\textit{AI-assistants Help} We improve and proofread the text of this article using Writefull assistant integrated into Overleaf (Writefull's/Open AI GPT models), Grammarly\footnote{\href{https://app.grammarly.com/}{https://app.grammarly.com/}} to correct grammatical, spelling, and style errors and paraphrase sentences.We want to clarify that these tools are used exclusively to improve the quality of English writing, in full accordance with ACL policies regarding the responsible use of AI writing assistance. However, some parts of our publication may potentially be identified as AI-generated, AI-edited, or a combination of human and AI contributions.

\section*{Acknowledgments}

We gratefully acknowledge the AI Alliance Russia\footnote{\url{https://a-ai.ru/?lang=en}} for organizing and supporting our work, offering legal guidance, and providing computational resources essential for the development and maintenance of the MERA Code benchmark.

We also extend our sincere thanks to all our industrial and academic partners for their valuable contributions to the MERA Code — through their shared expertise and datasets, they have made this benchmark possible.

We would like to express our deep appreciation to Ekaterina Morgunova, Egor Nizamov, Ivan Bondarenko, Georgy Mkrtchyan, Ivan Kharkevich, Nikolay Bushkov, Irina Shakhova, Denis Kokosinsky, Kirill Pikhtovnikov, Anton Bykov, and Oleg Sedukhin for their contributions and generous support throughout this work.

In the following list, we mention all authors of MERA Code with a description of their contribution:
\begin{itemize}[nosep]
\item Codebase techlead \& Contributor of ruCodeEval and ruHumanEval \& Baseline evaluation \& Deployment: \textit{Artem Chervyakov},
\item Methodological Review \& Taxonomy of MERA Code \& Code review \& Baseline evaluation \& Deployment: \textit{Alexander Kharitonov},
\item Contributors of RealCode, RealCodeJava, JavaTestGen: \textit{Pavel Zadorozhny,  Pavel Adamenko, Rodion Levichev, Dmitrii Vorobev, Dmitrii Salikhov} ,
\item Contributor of YABLoCo: \textit{Aidar Valeev}, 
\item Contributor of UnitTests: \textit{Alena Pestova},
\item Contributor of StRuCom: \textit{Maria Dziuba},
\item Contributors of ruCodeReviewer: \textit{Ilseyar Alimova, Artem Zavgorodnev, Aleksandr Medvedev, Stanislav Moiseev},
\item Contributors of CodeCorrectness: \textit{Elena Bruches, Daniil Grebenkin, Roman Derunets},
\item Research advisor \& Contributor of CodeLinterEval: \textit{Vikulov Vladimir},
\item Baseline evaluation \& Deployment: \textit{Anton Emelyanov},
\item Research advisor \& Contributor of RealCode, RealCodeJava, JavaTestGen: \textit{Dmitrii Babaev}, 
\item Research advisor \& Contributor of YABLoCo: \textit{Vladimir V. Ivanov},
\item Research advisor \& Contributor of StRuCom and UnitTests: \textit{Valentin Malykh}, 
\item Administration \& Ideology \& Contributor of ruCodeEval and ruHumanEval \& Senior research advisor: \textit{Alena Fenogenova}
\end{itemize}

\bibliography{custom}

\appendix
\section{Metrics description}
\label{sec:metric_desc}

\hspace{4mm}\textbf{Pass@k}~\cite{chen2021evaluating} evaluates the functional correctness of the generated code.

\textbf{Compile@k} evaluates the correctness of the compilation of the generated code.

\textbf{chrF}~\cite{popovic2015chrf} is used in code-to-text tasks since its enhanced sensitivity to Russian morphological complexity.

\textbf{BLEU}~\cite{papineni-etal-2002-bleu} measures the n-grams level similarity of the prediction and gold answer.

\textbf{CodeBLEU (CBLEU)}~\cite{ren2020codebleu} combines regular BLEU with the measure of similarity of code syntax via abstract syntax trees (AST) and code semantics via data-flow.

\textbf{Exact Match (EM)} is the rate at which the predictions exactly match the true references.

\textbf{Judge@k} measures whether one of the top-k predictions matches with the gold answer via LLM-as-a-Judge~\cite{zheng2023judging}. The details about the validation of LLM-as-a-Judge could be found in Appendix \ref{sec:rucodereview}.

\section{Prompts language analysis}
\label{sec:prompts_en}
\input{table/en_prompts}

We hypothesize that the language of the instructions does not affect the benchmark task scores. To test this, the original Russian prompts from MERA Code were translated into English using GPT-4.1, followed by LLM-as-a-Judge validation and manual corrections. 
We evaluated ten LLMs of varying scales, selected from a public leaderboard (see Table ~\ref{tab:mera_results}), encompassing both models specifically trained on Russian and general multilingual architectures.
As we research the impact of the language of the task instructions, 4 out of 11 tasks (\rucodeeval, \rucodereviewer, \ruhumaneval, \strucom) have been excluded: they contain natural language as part of either the input data or the golden answer, which does not fit into the hypotheses. 

The Table \ref{tab:en_prompts_models} reports per-model and per-task score deltas between the Russian and English prompts (Russian -- English). For most models and tasks, the differences remain close to zero (typically a few hundredths) and vary in sign, indicating no systematic advantage for either language. This presumably reflects the ``English-default'' nature of programming data: LLMs are widely trained on code paired with English documentation, comments, and instructions, so switching the instruction language often does not change the core code-centric signal and, therefore, has little effect on scores. The other explanation is that most tasks are dominated by the code and formal constraints, so models mainly focus on the programming content rather than the natural-language surface form. However, these two ideas are not contradictory.

Notable deviations are rare but informative. Yi-Coder-9B-Chat shows the largest gains from Russian on pass@1 tasks, especially \realcodejava (+0.208) and \javatestgen (+0.216), while \unittests is one of the most mixed, with some models improving under Russian (e.g., GigaChat-2-Max +0.145) and others favoring English (e.g., Yi-Coder-9B-Chat -0.14). Overall, the benchmark appears robust to English-Russian prompt changes, with language effects emerging mainly for specific model-task combinations.



\section{Datasets examples}
\label{sec:appendix_ds_exp}


\paragraph{ruHumanEval}
\label{sec:ruhumaneval}
\ruhumaneval is the Russian counterpart of the HumanEval dataset~\cite{chen2021evaluating}, which assesses models' abilities to generate solutions for straightforward programming problems in Python.
The dataset contains the translated into Russian and manually verified tasks of the original dataset~\footnote{\href{https://huggingface.co/datasets/openai_humaneval}{https://huggingface.co/datasets/openai\_humaneval}}, including the test cases, which were taken from~\citet{liu2023your} (10 test cases per task).

\paragraph{ruCodeEval}
\label{sec:rucodeeval}
\rucodeeval is created from scratch by assembling various programming tasks of the same difficulty level as the ruHumanEval test part and manually writing the test cases and documentation strings.
All tasks were verified to ensure that no repetition (with \ruhumaneval) occurred in the test samples. 

These tasks evaluate the functional correctness of code generation by providing input information, including a textual function description (docstring) and examples of expected results for different test cases.

An example of \rucodeeval and \ruhumaneval tasks:
\input{data_samples/rucodeeval}




\paragraph{JavaTestGen}
\label{sec:javatestgen}

To construct \textsc{JavaTestGen}, we first collected 5,000 of the most starred and recently updated Java repositories on GitHub. Repositories without an open-source license, larger than 100MB, or those that failed to compile and pass their tests using Maven were excluded. From the remaining pool, we selected repositories with at least two passing test cases and fully green test suites, yielding 500 high-quality Java projects.

Using static analysis, we extracted focal classes and their corresponding test classes, retaining only self-contained test modules (i.e., those depending solely on the focal class’s module). We further filtered class pairs with the following criteria: (i) source file length between 1,000 and 5,000 characters; (ii) not an enum or interface; (iii) at least two public methods in the source class; and (iv) test class with at least two assertions. This resulted in a curated set of 227 focal–test pairs.

At inference time, the model is given only the focal class and must generate the corresponding JUnit~5 test class. For evaluation, we replace the original test file with the generated one and run \texttt{mvn clean test} in a controlled Dockerized environment. We report two metrics: \texttt{compile@1}, the proportion of generated tests that compile, and \texttt{pass@1}, the proportion that pass all assertions.

An example of \javatestgen task:
\input{data_samples/javatestgen}

\paragraph{RealCode}
\label{sec:realcode}

Recent work has demonstrated that widely used surface-level metrics for code generation, such as BLEU~\cite{10.3115/1073083.1073135} and CodeBLEU~\cite{ren2020codebleu}, correlate poorly with functional correctness and real-world utility~\cite{Chen2021EvaluatingLL, 10.1145/3695991}. These metrics tend to reward superficial similarity to reference implementations, often failing to capture whether the generated code satisfies the intended behavioral requirements~\cite{10.1145/3695991}. Consequently, leading benchmarks such as HumanEval~\cite{Chen2021EvaluatingLL}, MBPP~\cite{Austin2021ProgramSW}, and AlphaCode (ClassEval)~\cite{Li2022CompetitionlevelCG} rely on execution-based evaluation, using \textit{pass@k} (a task is considered solved if any of $k$ samples passes all tests) as the primary correctness metric. However, these benchmarks are typically limited to synthetic or isolated tasks, reducing their ability to reflect practical code completion and maintenance scenarios~\cite{zhuo2025bigcodebench}.

To address these limitations, we introduce \textbf{RealCode}, a benchmark designed to evaluate code generation models on realistic code completion tasks within authentic, actively maintained Python repositories. We curated projects from public GitHub repositories created in 2024, filtering for open-source licenses, a minimum of three GitHub stars and one fork, and repository size under 30MB to ensure real-world relevance and manageable computational overhead. Additionally, we excluded repositories where function-level docstrings were consistently missing or empty, as documentation is often essential for guiding meaningful code generation. Each repository was automatically cloned, built, and tested using a rule-based pipeline in a Dockerized environment, retaining only those that successfully built and passed their test suites.

We then identified all functions within each project that were covered by at least one test. For these functions, we replaced the body with a mock implementation (e.g., a \texttt{pass} statement or a default return value) and reran the project's test suite. We retained only those functions for which the tests failed with the mock but passed with the original implementation, ensuring that solving the task requires synthesizing semantically meaningful and test-sensitive code.

Each task is defined by the left-side context (i.e., the beginning of the file and the function signature up to the body), with the target being the original function body. The model is expected to generate only the function body, which is inserted back into the codebase during evaluation.

The generated completion is then integrated into the original repository and evaluated using \texttt{pytest} with the project's existing test suite. The primary evaluation metric is \texttt{pass@1}, indicating whether the generated code passes all relevant tests. All steps, from project preparation to test execution, are fully automated and reproducible via our open-source \texttt{repotest} library.

\textbf{Prompting Considerations.} One of the challenges in code generation benchmarks is enforcing consistent formatting in model outputs. Generated completions must be syntactically valid and correctly indented to be parsable and executable in the original project. This is especially important in languages like Python, where whitespace is semantically meaningful. Our prompt format ensures that models produce completions that integrate cleanly into the original context. The second major challenge is semantic correctness: even if the format is valid, the model must synthesize correct logic that satisfies the tests based on limited surrounding context.

An example of \realcode task:
\input{data_samples/realcode}

\paragraph{RealCodeJava}
\label{sec:realcodejava}
The benchmark aims to evaluate the code generation model's ability to produce compilable and functional Java code in the context of genuine open-source Java repositories.

Tasks were gathered from public GitHub repositories created in 2024--2025, which were thoroughly filtered to include only repositories with open-source licenses and those that met popularity and utility criteria (minimum 3 stars on GitHub). We then selected projects that used the Maven build system and had unit tests within. We left only those that could be successfully built and that had passed the tests.

Each task consists of a function, which is covered by at least one test. We verify the test's capability to detect non-working code by replacing the function body with an auto-generated mock implementation. Specifically in Java, our mocks return null values, empty collections, or some predefined constants depending on the function's signature. Mocks are expected to pass the compilation stage but fail during the test execution phase. We only keep samples that meet this criterion. Finally, all tasks are automatically evaluated for perceived target function complexity and ranked accordingly. We approximate the perceived complexity of a function by the presence of certain patterns in the code, such as cross-file API calls, usage of complex lambda expressions, or safeguard checks.

For each task, inputs include left-side context (the beginning of the file and the function signature with opening curly brace "\{"). From there, the model is expected to generate the body of the function, namely the lines of code up to the last closing curly brace "\}", including it. The resulting answer is then inserted back into the codebase and verified.

Generated completions are evaluated using the original project's existing test suite and our open-source \texttt{repotest} library. We use \texttt{}{pass@1} as the main metric to measure the model's ability to provide a functional solution to the given task.

An example of \realcodejava task:
\input{data_samples/realcode_java}


\paragraph{RuCodeReviewer}
\label{sec:rucodereview}
The first benchmark for automated code review comment generation in Russian, designed to address the limitations of existing English-centric and often noisy datasets. The creation of this dataset is motivated by the need for a high-quality, reproducible evaluation framework for models' abilities in the code review task. 

The data for RuCodeReviewer originates from Backend Academy, an industry-oriented educational program where student-submitted code across four languages (Java, Python, Go, and Scala) is manually reviewed by professional software developers. Utilizing the GitHub API, 1,300 merge requests containing 2,859 review comments were collected. A rigorous two-stage pipeline was used to clean the data. First, an LLM-assisted filtering stage, based on GPT-4o, identifies comments that are self-contained and reproducible solely from the provided code diff. A manually validated subsample of 1,300 examples marked as "non-reproducible" at this stage contained no misclassified cases, demonstrating the high quality of the proposed approach. Second, two independent human annotators, both experienced software developers, verify each "reproducible" candidate, ensuring that comments are actionable and fully understandable without external context. This human validation step results in a substantial inter-annotator agreement, with a Cohen's kappa coefficient of 0.78. The final dataset is consists of 689 merge-request diff-comment pairs prioritizing quality over size.

To overcome the ``one-to-many'' challenge in evaluation of the generation tasks, where multiple valid outputs exist for a single input, a novel evaluation methodology centered on an LLM-as-a-Judge was proposed. This framework leverages a large language model to assess the semantic equivalence between model-generated and human reference answers. Performance is reported using a \texttt{pass@k} metric, which credits a model if at least one of its $k$ generated candidates is deemed valid. The Qwen2.5-Coder-32B~\footnote{https://huggingface.co/Qwen/Qwen2.5-Coder-32B} model was used as the judge model, so its performance measurements on this benchmark may be biased.

\textbf{LLM-Judge Validation} 
To validate the automatic judge for the RuCodeReviewer task, we first generate candidate review comments with GPT-4o and manually annotate 300 generated–reference pairs. The goal of this step is to check whether the generated comment preserves the original semantic intent. Based on this annotated set, we refine the prompt for our LLM-based judge. After tuning, we test the chosen model \textbf{Qwen2.5-Coder-32B} on 100 Russian comment pairs. The model reaches a \textbf{Spearman correlation of 0.831} with human ratings, which shows that it can reliably assess semantic equivalence.  

For a broader check of consistency, we compare GPT-4o outputs using three different judges: Qwen2.5-Coder-32B, GPT-4o, and Claude-Sonnet-3.7. Their scores remain within each other’s standard deviation ranges, giving us confidence that Qwen2.5-Coder-32B is a solid choice. Since it is also faster and can be deployed locally, we adopt it as our default judge.

\input{data_samples/rucodereviewer}

\input{data_samples/rucodereviewer_judge}

\paragraph{UnitTests}
\label{sec:unittests}
UnitTests is a multilingual dataset for evaluating models on generating unit tests in Python, Java, Go, JavaScript, and C\#. Each example includes a function, its context, and the corresponding test with its context. 

The data is sourced from open-licensed GitHub code. First, a list of repositories for each language was obtained. We chose the repositories with permissive licenses only and with the number of stars more than 10. We also filtered out fork repositories. The list of licenses used in the dataset: MIT License, Apache License 2.0, The Unlicense, Mozilla Public License 2.0, BSD 2-Clause "Simplified" License, BSD 3-Clause "New" or "Revised" License, EPL 1.0 license, EPL 2.0 license, MPL 2.0 License, Unlicense License, 0BSD license.

Then, the repositories were downloaded and parsed using syntax parsers. For all languages (except Python) the tree-sitter was used for code parsing, specifically, for searching and parsing functions/methods and classes, identifying calls, etc. For Python, we use the built-in ast library.

Next, methods and functions, along with their unit tests, were mapped using a method adopted from the paper~\citet{tufano2021unit}. For Java, the mapping procedure was identical to the method by~\citet{tufano2021unit}. For Python, Go, and C\#, we develop similar mapping methods based on functions names and unique method invocation. For Javascript, the tests were mapped to functions by the last local method/function invocation in the test because test functions do not have identifiers when declared in it() and test().

When building the dataset, the same filtering rules for all languages were used:
(i) Empty tests are removed. 
 (ii) No more than 200 method-test pairs were collected from one repository. If there were more pairs, they were sampled randomly. 
 (iii) The test case should be less than 5000 characters. This limit is set to remove overly long tests from the data. 
 (iv) Maximum input length (focal function with context) should be less than 70000 characters.
(v) Maximum number of assertions (the word "assert" in the test case) is 20.
 (vi) For Python and Java, there was additional filtering for tests with syntax errors (using ast and javalang libraries correspondingly).
 (vii) The training data was filtered for duplicates of test cases both within a set, and possible overlaps with the validation and test data were removed.

For the benchmark, the data was sampled for each language so as to cover different cases as evenly as possible in terms of the length of the focal function, the test function, and the entire context; and in terms of the type of both functions (function/method). The benchmark comprises a total of 2,500 samples, with 500 samples for each programming language. For evaluation, we use the CodeBLEU~\cite{ren2020codebleumethodautomaticevaluation} metric to compare the original test from the repository and the generated test.

\input{data_samples/unittests}

\paragraph{CodeLinterEval}
\label{sec:codelintereval}
CodeLinterEval is a dataset for evaluating model abilities for generating and correcting code based on linter errors in the Python language. The benchmark evaluates understanding linter errors (the ability to correctly interpret messages like E111, E231, etc.), correct code refactoring (the ability to make corrections while preserving the logic of the program), following the code style (PEP 8) that includes correct indents, spaces, formatting, and contextual understanding of the code - the model should not break the logic when fixing the style. The benchmark contains 110 tasks: source code with errors, feedback – a list of errors with description from the linter, instruction – explicit instruction to correct the code based on feedback, and reference canonical code.

Explicit indication of errors allows us to evaluate the model's ability to correct code according to the linter, rather than "guess" errors.
The canonical solution provides a clear ground truth for evaluation.
The instruction explicitly specifies the task so that the model does not deviate from the goal.

If the model corrects the code according to the feedback and the linter does not detect errors after checking the generation results, then the result is correct. If errors persist or new ones appear, the model does not solve the problem. The metric is \texttt{pass@k}, determined based on the success of the linter check relative to the total size of the dataset.

An example of \codelintereval task:
\input{data_samples/codelintereval}

\paragraph{CodeCorrectness}
\label{sec:CodeCorrectness}
CodeCorrectness is a novel benchmark designed to evaluate LLMs’ ability to identify whether a unit test is \textit{correct} (compilable and executable) or \textit{incorrect} (failing compilation or execution). Although advanced tools are being developed to assess generated code correctness~\cite{bui2025correctness}, dedicated benchmarks for evaluating LLMs' capabilities in code assessment remain an underexplored research area.

The dataset comprises 1,361 samples across Java, Python, and Go, constructed through a multi-stage process. Focal file (the code under test) – test file pairs were first automatically collected from GitHub repositories using filters for permissive licenses (e.g., MIT, Apache-2.0), repository popularity (stars), and recency (last commit date). Exclusion criteria removed pairs with minimal test or focal code lines, additional imports from the project, imports from specific libraries, or file operations in test cases. The resulting samples include both 466 original human-written tests that passed all filters, compiled and executed successfully, alongside 895 LLM-generated tests. A subset of these LLM-generated tests contains errors, failing during compilation or execution; samples exhibiting only syntax errors were filtered out. This process ensured sufficient context for the task and provided real-world complexity.

We employ \textbf{exact match} as the evaluation metric for this task. Each sample presents a focal file code and a corresponding test file code containing test cases. Models are tasked with predicting a binary label: compilation and execution success or failure. Ground truth labels were rigorously verified using actual execution within isolated environments, confirming that a specific failure reason occurred for failing tests. 

\input{data_samples/codecorrectness}

\paragraph{StRuCom}
\label{sec:StRuCom}
StRuCom \cite{dziuba2025strucom} addresses a critical challenge in AI-assisted code documentation by introducing the first large-scale dataset (153K examples) of structured Russian comments for Python, Java, JavaScript, C\#, and Go. This resource enables robust evaluation of documentation generation models for Russian-speaking developers, filling the void of standardized benchmarks for non-English code explanations.

To construct StRuCom, the authors developed a hybrid sourcing methodology combining human-authored and AI-generated content. Human-written comments were extracted from over $150$K Russian GitHub repositories, identified through repository metadata analysis and license filtering. To ensure representation across all target languages, the corpus was supplemented with synthetic examples that were either created by Qwen2.5-Coder-32B \cite{hui2024qwen2} or enhanced by Miqu-70B \footnote{\url{https://huggingface.co/miqudev/miqu-1-70b}}. Every comment underwent automated validation against language-specific docstring conventions (GoogleDoc \footnote{\url{https://google.github.io/styleguide/pyguide.html}}, JSDoc \footnote{\url{https://jsdoc.app}}, etc.) using custom verification tools. The validation protocol enforced strict structural completeness, requiring all comments to comprehensively describe parameters, return values, exceptions, and types (if needed). For this benchmark, we curated a balanced subset comprising $7 500$ comments for training ($1 500$ per language) and $500$ for testing ($100$ per language).

For evaluation, the chrF metric \cite{popovic2015chrf} is employed due to its enhanced sensitivity to Russian morphological complexity. Unlike BLEU \cite{papineni-etal-2002-bleu} -- which often fails to capture inflectional nuances -- chrF operates through character n-gram weighting. This approach effectively detects subtle morphological shifts in case endings, verb conjugations, and derivational suffixes that are essential for assessing documentation fluency in Russian. The metric's design aligns with the language's rich inflectional system, providing a more reliable quality assessment for generated comments.

An example of StRuCom task:
\input{data_samples/strucom.tex}

\paragraph{YABLoCo}
\label{sec:YABLoCo}
\par The YABLoCo benchmark evaluates the proficiency of modern large language models (LLMs) in completing code functions based on textual descriptions from the user. While previous research has explored this capability for standalone functions \cite{jiang2024surveylargelanguagemodels},
there has been limited investigation into functions intended for integration within larger code repositories. YABLoCo consists of 4 large repositories, with sizes ranging from 200,000 to 2,000,000 lines of code. The authors of the benchmark generated a call graph for each repository. For example, the LLVM repository contains approximately 175,000 functions and 571,000 inter-function calls. Next, the functions that do not correspond to the following criteria were filtered out from the benchmark: low test coverage, lack of developer comments, and excessively long function bodies. From the remaining candidates, the authors of YABLoCo manually selected 208 functions that featured clear docstrings aligned with their respective function bodies. These docstrings serve as the descriptive prompts for the LLMs being tested.

The main challenge for an LLM in generating a function for a specific code repository lies in the context size of the existing codebase. The new function must not only precisely follow the text description but also effectively utilize other functions and classes within the repository. This necessitates a careful selection of context that is both relevant for generation and not excessively lengthy. The authors of YABLoCo~\cite{valeev2025yabloco}
discovered that an effective context for this purpose consists of the functions that are called by the target function. This context is referred to as the "oracle," as it is typically unavailable to the new function during generation, although it is known within the benchmark. Simply incorporating the oracle functions, without any modifications to the LLMs, resulted in an appreciable improvement on YABLoCo in the overall \texttt{pass@10} metric, increasing it by 14 points (from 22.4 to 36.1).

An example of YABLoCo task:
\input{data_samples/yabloco}

\begin{figure*}[htbp]
    \centering
    \includegraphics[width=0.95\textwidth]{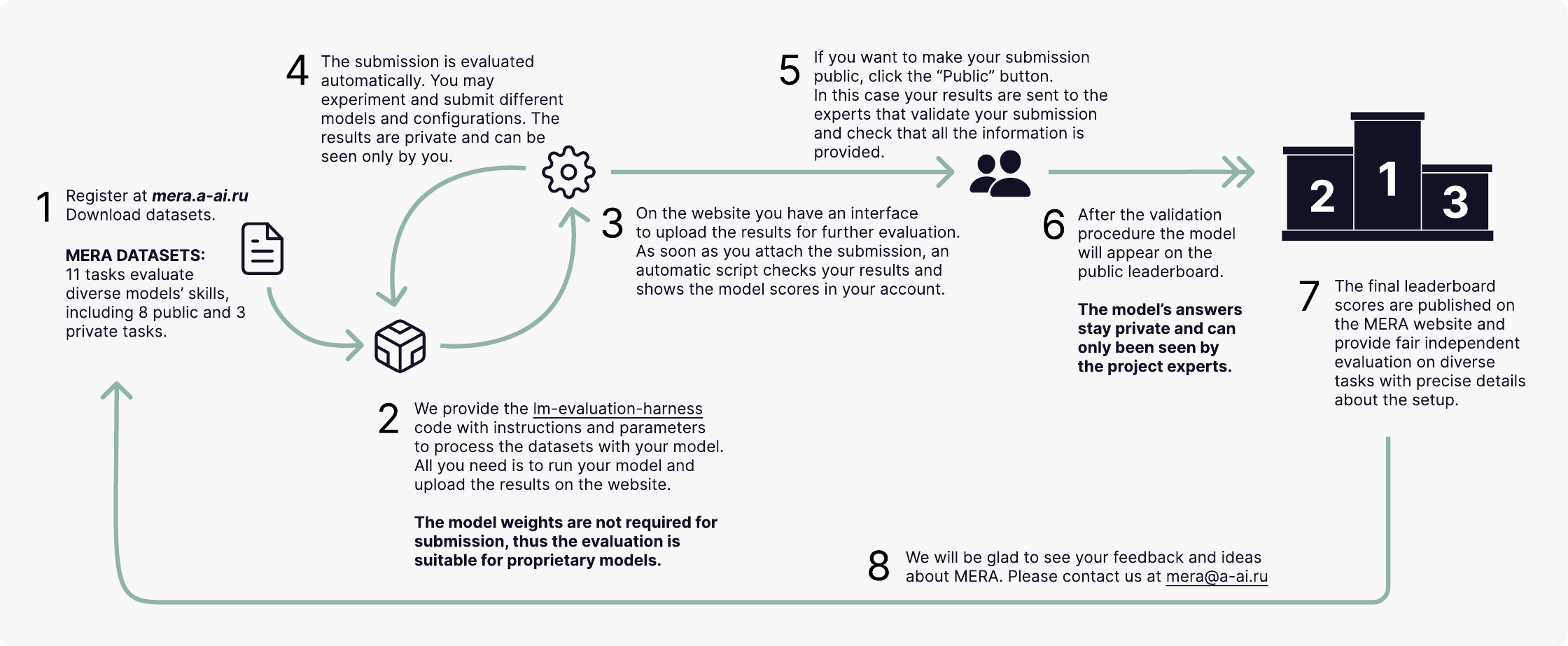}
    \caption{The user path for the submission process on the MERA Code evaluation platform.}
    \label{fig:submit}
\end{figure*}

\section{Taxonomy details}
\label{sec:appendx_taxonomy}
In the Table~\ref{tab:taxonomy_info} we provide an explanation of skills presented in the MERA Code taxonomy, starting from the second level of taxonomy. 
\input{table/taxonomy_examples}

\section{Evaluation results}
\label{sec:full_metrics}

This section provides two tables: Table~\ref{tab:private_results} shows all metrics of the baseline models on private tasks of MERA Code, Table~\ref{tab:public_results} provides the metrics for the same models on public tasks of the benchmark.

\input{table/private_results}

\input{table/public_results}

\end{document}

%% file: table/short_en_prompts_per_task.tex
\begin{table}[ht]
\centering
\small
\begin{tabular}{lc}
\toprule
\textbf{Task} & \textbf{Difference} \\
\midrule
CodeCorrectness (EM) & 0.01 [-0.02, 0.04] \\
CodeLinterEval (pass@1) & 0.03 [-0.01, 0.05] \\
JavaTestGen (pass@1) & -0.01 [-0.02, 0.01] \\
RealCode (pass@1) & 0.0 [-0.01, 0.04] \\
RealCodeJava (pass@1) & -0.02 [-0.05, 0.01] \\
UnitTests (CodeBLEU) & 0.0 [-0.01, 0.0] \\
YABLoCo (pass@1) & 0.06 [0.01, 0.07] \\
\bottomrule
\end{tabular}
\caption{The statistics on the difference between the scores of the models with the Russian prompts and the English ones. The difference is displayed as "Median [Q1, Q3]". Negative values mean that the score with the English prompts is larger than that with the Russian prompts.}
\label{tab:en_prompts_tasks}
\end{table}

%% file: table/datasets.tex
\begin{table*}[ht!]
    \setlength{\tabcolsep}{2pt}
    \centering
    \scriptsize
    \renewcommand\arraystretch{1.3}
    \begin{tabularx}{\textwidth}{
        @{}c|
        >{\raggedright\arraybackslash}p{0.14\textwidth}
        >{\raggedright\arraybackslash}p{0.14\textwidth}
        >{\raggedright\arraybackslash}p{0.14\textwidth}
        >{\raggedright\arraybackslash}p{0.06\textwidth}
        >{\raggedright\arraybackslash}p{0.06\textwidth}
        >{\raggedright\arraybackslash}p{0.4\textwidth}
    }
        \toprule
         & \textbf{Task name} &  
         \textbf{Language} &
         \textbf{Metrics} & 
         \textbf{Size} &
         \textbf{Prompts} &
         \makecell[c]{\textbf{Skills}} \\
        \midrule
        \multirow{5}{*}{\rotatebox[origin=c]{90}{\textbf{Private}}} & \hyperref[sec:rucodeeval]{\rucodeeval} & Python & pass@k & 164 & 10 & \makecell[tl]{Instruction Following, Code Perception, Completion,\\ Algorithms \& Data Structures} \\
         & \hyperref[sec:rucodereview]{\rucodereviewer}  & \makecell[tl]{Java, Scala, Go,\\ Python} & \makecell[tl]{Judge@k, \\BLEU, chrF} & 689 & 10 & Instruction Following, Code Perception, Review, Simulation, Explanation, Programming Patters, Style Guides\\
         & \hyperref[sec:codelintereval]{\codelintereval} & Python & pass@k & 110 & 10 & Instruction Following, Code Perception, Style Guides, Review, Editing \\
        \midrule
        \multirow{11}{*}{\rotatebox[origin=c]{90}{\textbf{Public}}} 
         & \hyperref[sec:ruhumaneval]{\ruhumaneval} & Python & pass@k & 164 & 10 & Instruction Following, Code Perception, Completion  \\
         & \hyperref[sec:StRuCom]{\strucom} & \makecell[tl]{Python, Java, Go,\\ C\#, JavaScript} & chrF & 500 & 10 & Instruction Following, Code Perception, Simulation, Documentation \\
         & \hyperref[sec:unittests]{\unittests} & \makecell[tl]{Python, Java, Go,\\ C\#, JavaScript} & CodeBLEU & 2500 & 20 & Instruction Following, Code Perception, Synthesis, Testing, \makecell[tl]{Long Context Comprehension} \\
         & \hyperref[sec:CodeCorrectness]{\codecorrectness} & Python, Java, Go & EM & 1361 & 11 & \makecell[tl]{Instruction Following, Code Perception, Simulation,\\ Error Classification} \\
        & \hyperref[sec:realcode]{\realcode} & Python & pass@k & 802 & 10 & Instruction Following, Code Perception, Completion \\
        & \hyperref[sec:realcodejava]{\realcodejava} & Java & pass@k & 298 & 10 & Instruction Following, Code Perception, Completion \\
        & \hyperref[sec:javatestgen]{\javatestgen} & Java & pass@k, compile@k & 227 & 10 & Instruction Following, Code Perception, Completion, Testing \\
        & \hyperref[sec:YABLoCo]{\yabloco} & C, C++ & pass@k, EM & 208 & 11 & Instruction Following, Code Perception, Completion, Long Context Comprehension \\
        \bottomrule
    \end{tabularx}
    \caption{The MERA Code tasks outline. \textbf{Size} is the number of test samples. \textbf{Prompts} is the number of unique prompts used for each task. \textbf{Skills} lists skills from taxonomy which are necessary for particular task.}
    \label{tab:tasks_info}
\end{table*}

%% file: table/mera_code_results.tex
\begin{table*}[ht!]
\centering
\tiny
\addtolength{\tabcolsep}{-0.65em}
\begin{tabular}{lcc|c|c|ccccclccccc}
\multirow{2}*{\textbf{Model}} & \multirow{2}*{\textbf{Citation}} & \multirow{2}*{\textbf{\begin{tabular}{@{}c@{}}\textbf{Params,} \\ \textbf{B}\end{tabular}}} & \multirow{2}*{\textbf{\begin{tabular}{@{}c@{}}\textbf{Total} \\ \textbf{Score}\end{tabular}}} & \multirow{2}*{\textbf{\begin{tabular}{@{}c@{}}\textbf{Private} \\ \textbf{Score}\end{tabular}}} & \textbf{\rot{90}{YABLoCo}} & \textbf{\rot{90}{RealCode}} & \textbf{\rot{90}{RealCodeJava}} & \textbf{\rot{90}{ruCodeEval}} & \textbf{\rot{90}{ruHumanEval}} & \textbf{\rot{90}{JavaTestGen}} & \textbf{\rot{90}{CodeLinterEval}} & \textbf{\rot{90}{ruCodeReviewer}} & \textbf{\rot{90}{UnitTests}} & \textbf{\rot{90}{StRuCom}} & \textbf{\rot{90}{CodeCorrectness}} \\

\cmidrule(lr){6-12} \cmidrule(lr){13-13} \cmidrule(lr){14-14} \cmidrule(lr){15-15} \cmidrule(lr){16-16} 

 & & & & & \multicolumn{7}{c}{\textbf{pass@1}} & \textbf{Judge@1} & \textbf{CBLEU} & \textbf{chrF} & \textbf{EM} \\

\toprule

GPT-4.1 & \citealp{gpt4.1} & - & \textbf{0.381} & 0.382 & 0.144 & \textbf{0.418} & \textbf{0.416} & 0.443 & 0.450 & \textbf{0.344} & 0.555 & \textbf{0.096} & 0.201 & 0.297 & 0.660 \\
Gemini 2.5 flash & \citealp{gemini2025} & - & 0.360 & \textbf{0.427} & 0.120 & 0.388 & 0.386 & \textbf{0.610} & \textbf{0.604} & 0.211 & 0.496 & 0.064 & 0.212 & 0.217 & 0.404 \\
DeepSeek-Coder-V2-Inst & \citealp{deepseekcoderv2} & 236 & 0.350 & 0.360 & \textbf{0.149} & 0.364 & 0.386 & 0.433 & 0.392 & 0.269 & 0.494 & 0.060 & 0.188 & 0.200 & 0.714 \\
GigaChat-2-Max & \citealp{mamedov-etal-2025-gigachat} & - & 0.338 & 0.361 & 0.091 & 0.318 & 0.262 & 0.534 & 0.524 & 0.145 & 0.425 & 0.055 & 0.223 & 0.291 & 0.686 \\
RuadaptQwen3-8B-Hybrid & \citealp{Tikhomirov_Chernyshov_2024} & 8 & 0.326 & 0.380 & 0.115 & 0.191 & 0.245 & 0.419 & 0.427 & 0.093 & 0.500 & 0.042 & 0.218 & 0.244 & \textbf{0.744} \\
Qwen3-Coder-480B-A35B-Inst & \citealp{yang2025qwen3technicalreport} & 480 & 0.326 & 0.329 & 0.082 & 0.411 & 0.342 & 0.463 & 0.421 & 0.220 & 0.430 & 0.058 & 0.231 & 0.202 & 0.540 \\
Kimi-K2-Inst-0905 & \citealp{kimiteam2025kimik2openagentic} & 1000 & 0.325 & 0.380 & 0.077 & 0.188 & 0.342 & 0.448 & 0.448 & 0.233 & 0.423 & 0.074 & 0.153 & 0.232 & 0.571 \\
A-vibe & \href{https://huggingface.co/AvitoTech/avibe}{HF Page} & 8 & 0.317 & 0.367 & 0.115 & 0.157 & 0.279 & 0.263 & 0.257 & 0.119 & 0.614 & 0.023 & \textbf{0.232} & \textbf{0.307} & 0.724 \\
Qwen3-Coder-30B-A3B-Inst & \citealp{yang2025qwen3technicalreport} & 30 & 0.301 & 0.373 & 0.072 & 0.188 & 0.309 & 0.391 & 0.350 & 0.189 & 0.621 & 0.058 & 0.231 & 0.202 & 0.540 \\
Qwen2.5-Coder-32B-Inst & \citealp{hui2024qwen25codertechnicalreport} & 32 & 0.300 & 0.306 & 0.111 & 0.323 & 0.386 & 0.311 & 0.289 & 0.220 & 0.466 & 0.065 & 0.206 & 0.213 & 0.519 \\
GigaCode 1.4 & \href{https://gigacode.ru}{Web page} & 32 & 0.293 & 0.166 & 0.135 & 0.322 & 0.352 & 0.357 & 0.305 & 0.313 & 0.027 & 0.064 & 0.230 & 0.276 & 0.676 \\
Qwen2.5-72B-Inst & \citealp{hui2024qwen25codertechnicalreport} & 72 & 0.288 & 0.254 & 0.144 & 0.362 & 0.349 & 0.174 & 0.157 & 0.189 & 0.481 & 0.048 & 0.160 & 0.252 & 0.702 \\
Seed-Coder-8B-Inst & \citealp{seed2025seedcoderletcodemodel} & 8 & 0.271 & 0.345 & 0.106 & 0.106 & 0.305 & 0.317 & 0.210 & 0.264 & \textbf{0.655} & 0.035 & 0.161 & 0.237 & 0.403 \\
DeepSeek-V3.1 & \citealp{deepseekai2025deepseekv3technicalreport} & 685 & 0.248 & 0.259 & 0.106 & 0.084 & 0.275 & 0.345 & 0.282 & 0.189 & 0.230 & 0.065 & 0.159 & 0.293 & 0.372 \\
GigaChat-2-Lite & \citealp{mamedov-etal-2025-gigachat} & - & 0.229 & 0.269 & 0.048 & 0.099 & 0.148 & 0.309 & 0.251 & 0.035 & 0.316 & 0.025 & 0.107 & 0.144 & 0.712 \\
Yi-Coder-9B-Chat & \citealp{ai2025yiopenfoundationmodels} & 9 & 0.206 & 0.181 & 0.135 & 0.067 & 0.255 & 0.350 & 0.173 & 0.229 & 0.145 & 0.016 & 0.168 & 0.192 & 0.364 \\
Mixtral-8x22B-Inst & \citealp{jiang2024mixtralexperts} & 141 & 0.182 & 0.028 & 0.106 & 0.180 & 0.366 & 0.000 & 0.000 & 0.229 & 0.027 & 0.016 & 0.194 & 0.152 & 0.597 \\
Kodify Nano & \href{https://mts.ai/ru/product/kodify}{Web page} & 1.5 & 0.181 & 0.226 & 0.082 & 0.041 & 0.104 & 0.191 & 0.173 & 0.053 & 0.235 & 0.003 & 0.112 & 0.092 & 0.475 \\
CodeLlama70b-Inst & \citealp{roziere2023code} & 70 & 0.159 & 0.310 & 0.120 & 0.039 & 0.232 & 0.040 & 0.035 & 0.167 & 0.507 & 0.006 & 0.035 & 0.043 & 0.025 \\
gpt-oss-120b & \citealp{openai2025gptoss120bgptoss20bmodel} & 120 & 0.155 & 0.106 & 0.096 & 0.350 & 0.309 & 0.004 & 0.012 & 0.273 & 0.208 & 0.000 & 0.107 & 0.055 & 0.093 \\
DeepSeek-Coder-V2-Lite-Inst & \citealp{deepseekcoderv2} & 16 & 0.146 & 0.018 & 0.072 & 0.096 & 0.255 & 0.000 & 0.000 & 0.137 & 0.000 & 0.022 & 0.127 & 0.189 & 0.539 \\
CodeLlama34b-Inst & \citealp{roziere2023code} & 34 & 0.141 & 0.097 & 0.120 & 0.019 & 0.195 & 0.000 & 0.000 & 0.088 & 0.144 & 0.023 & 0.176 & 0.120 & 0.472 \\
CodeLlama7b-Inst & \citealp{roziere2023code} & 7 & 0.140 & 0.176 & 0.096 & 0.000 & 0.151 & 0.000 & 0.000 & 0.040 & 0.315 & 0.000 & 0.106 & 0.098 & 0.455 \\
gpt-oss-20b & \citealp{openai2025gptoss120bgptoss20bmodel} & 20 & 0.095 & 0.019 & 0.038 & 0.305 & 0.285 & 0.000 & 0.000 & 0.013 & 0.044 & 0.001 & 0.148 & 0.062 & 0.115 \\



\bottomrule
\end{tabular}
\caption{MERA Code benchmark results with model specifications. The private tasks are CodeLinterEval, ruCodeEval and ruCodeReviewer. The board is captured on the 15.11.2025.}
\label{tab:mera_results}
\end{table*}

%% file: table/en_prompts.tex

\begin{table*}
\centering
\scriptsize
\begin{tabular}{l|ccccccc}
\multirow{2}*{\textbf{Model}} & \textbf{\rot{90}{YABLoCo}} & \textbf{\rot{90}{RealCode}} & \textbf{\rot{90}{RealCodeJava}} & \textbf{\rot{90}{JavaTestGen}} & \textbf{\rot{90}{CodeLinterEval}} & \textbf{\rot{90}{UnitTests}} & \textbf{\rot{90}{CodeCorrectness}} \\

\cmidrule(lr){2-6} \cmidrule(lr){7-7} \cmidrule(lr){8-8}

 & \multicolumn{5}{c}{\textbf{pass@1}} & \textbf{CBLEU} & \textbf{EM} \\

\toprule

DeepSeek-Coder-V2-Instruct & 0.077 & 0.054 & 0.003 & -0.013 & 0.039 & -0.028 & 0.001 \\
Qwen3-Coder-30B-A3B-Instruct & 0.0 & -0.014 & -0.094 & -0.035 & -0.006 & 0.007 & -0.02 \\
Qwen2.5-Coder-32B-Instruct & 0.072 & 0.0 & 0.01 & 0.035 & 0.076 & 0.06 & 0.0 \\
Qwen2.5-72B-Instruct & 0.091 & 0.035 & 0.0 & -0.035 & -0.003 & 0.083 & -0.009 \\
GigaChat-2-Max & 0.053 & 0.007 & -0.07 & -0.004 & 0.042 & 0.145 & -0.002 \\
GigaChat-2-Lite & 0.01 & 0.04 & 0.013 & -0.009 & -0.023 & 0.02 & 0.002 \\
A-vibe & 0.058 & -0.009 & -0.04 & -0.018 & 0.014 & 0.04 & 0.043 \\
Kodify-Nano & 0.058 & -0.035 & -0.044 & -0.004 & -0.035 & -0.018 & -0.011 \\
Seed-Coder-8B-Instruct & 0.067 & -0.005 & 0.007 & 0.04 & 0.039 & -0.06 & 0.014 \\
Yi-Coder-9B-Chat & 0.091 & 0.062 & 0.208 & 0.216 & 0.1 & -0.14 & 0.032 \\

\bottomrule
\end{tabular}
\caption{Detailed results on difference between scores with the Russian prompts and the English ones applied to the benchmark tasks. The difference is always ``Russian --- English''. Negative values mean that the score with the English prompts is larger than that with the Russian prompts.}
\label{tab:en_prompts_models}
\end{table*}


%% file: data_samples/rucodeeval.tex
\begin{itemize}
\item \textbf{instruction:} \texttt{The input represents a function with a description in the form of a docstring. Given the input function, you need to implement it based on the template: ``\{function\}''.}

\item \textbf{function:} \texttt{\\def gcd(a: int, b: int) -> int:\\ """Returns the greatest common divisor of two integers a and b.\\ Examples:\\ gcd(3, 5)\\ 1\\ gcd(25, 15)\\ 5"""}

\item \textbf{tests:} \texttt{"[\{'a': 3, 'b': 7\}, \{'a': 10, 'b': 15\}, \{'a': 49, 'b': 14\}, \{'a': 144, 'b': 60\}]"}

\item \textbf{outputs} (golden answer): \texttt{[1, 5, 7, 12]}
\end{itemize}

%% file: data_samples/javatestgen.tex









\label{sec:javagen-example}

\hrule
\vspace{0.7em}
{\ttfamily\small
\noindent Context code:\\
\hrule
\vspace{0.5em}
\begin{verbatim}
package com.github.quiram.course;

import java.util.List;

import static java.lang.String.join;
import static java.util.Arrays.asList;

public class ReverseCommand extends 
Command {
    @Override
    protected boolean 
    safelySupports(String input) {
    ....
}
\end{verbatim}
}
\hrule
\vspace{0.7em}
\noindent\textit{Prompt Template (translated)}\\
\hrule
\vspace{0.5em}
\noindent{\small *This prompt was used during data generation and translated from Russian:}

\begin{quote}
You are given the implementation of a class \{\texttt{class}\}

\{\texttt{context code}\}

Your task is to write a unit test class \{\texttt{test\_class}\} for the class above.

Use Junit5. Ensure full coverage of all test scenarios even if the code doesn't contain corresponding branches.  
Write tests for happy path, edge cases, illegal arguments and other cases.  
Ensure only one assert in each test method. Test method names should be meaningful.  
Add necessary imports and annotations.
\end{quote}

%% file: data_samples/realcode.tex
\label{sec:realcodepython-example}

\hrule
\vspace{0.7em}
{\ttfamily\small
\noindent Left Context:\\
\hrule
\vspace{0.5em}
\begin{verbatim}
from dataclasses import dataclass
from typing import Self

from pysphinx.const import SECURITY_PARAMETER
from pysphinx.crypto import compute_hmac_sha256

@dataclass
class IntegrityHmac:
    """
    This class represents a HMAC-SHA256 
    that can be used 
    for integrity authentication.
    """

    value: bytes

    SIZE: int = SECURITY_PARAMETER

    def __init__(self, value: bytes):
        """Override the default constructor
        to check the size of value"""
\end{verbatim}
\hrule
\vspace{0.7em}
\noindent Ground Truth Body:\\
\hrule
\vspace{0.5em}
\begin{verbatim}
        if len(value) != self.SIZE:
            raise ValueError("invalid length of HMAC", 
                            len(value))

        self.value = value
\end{verbatim}
}
\hrule
\vspace{0.7em}
\noindent\textit{Prompt Template (translated)}\\
\hrule
\vspace{0.5em}
\noindent{\small *This prompt was used during data generation and translated from Russian:}

\begin{quote}
Use the following code:  
\texttt{\{left\_context\}}

Now — only implement the body of a single function. Wrap your answer in a block:  

\verb|```python|\\
\texttt{<code>}\\ 
\verb|```|

Indentation must be correct. Do not include the function signature. Do not write other functions. Your answer will be inserted into the function and tested.
\end{quote}







%% file: data_samples/realcode_java.tex
\label{sec:realcodejava-example}

\hrule
\vspace{0.7em}
{\ttfamily\small
\noindent Left Context:\\
\hrule
\vspace{0.5em}
\begin{verbatim}
package com.comp301.a06image;

import java.awt.Color;

/**
 * The SolidColorImage class is 
 * an Image implementation
 * that represents an image with a solid color.
 * The image is defined by 
 * a width, a height, and a color.
 * The color of a pixel can be retrieved by
 * calling the getPixelColor method.
 */
public class SolidColorImage implements Image {
  private int width;
  private int height;
  private Color color;

  /**
    * Returns the color of the pixel at the specified 
   * coordinates.
   *
   * @param x The x coordinate of the pixel
   * @param y The y coordinate of the pixel
   * @return The color of the pixel
   */
  @Override
  public Color getPixelColor(int x, int y) {
\end{verbatim}
\hrule
\vspace{0.7em}
\noindent Ground Truth Body:\\
\hrule
\vspace{0.5em}
\begin{verbatim}
    if (x < 0) {
        throw new IllegalArgumentException(
                "Pixel x-coordinate must be non-
                negative.");
    }

    if (y < 0) {
        throw new IllegalArgumentException(
                "Pixel y-coordinate must be non-
                negative.");
    }

    if (x >= this.width) {
        throw new IllegalArgumentException(
                "Pixel x-coordinate must be less than 
             the image width.");
    }

    if (y >= this.height) {
        throw new IllegalArgumentException(
                "Pixel y-coordinate must be less than 
             the image height.");
    }
    // Return the solid color of the image
    return this.color;
  }    
\end{verbatim}
}
\hrule
\vspace{0.7em}
\noindent\textit{Prompt Template (translated)}\\
\hrule
\vspace{0.5em}
\noindent{\small *This prompt was used during data generation and translated from Russian:}

\begin{quote}
Use the following code:  
\texttt{\{left\_context\}}

Write the contents of the last function after the header with arguments. Do not invent new functions and classes, but you can use existing ones in the code. The answer consists of one function. Place the answer in the block:

\verb|```java|\\
\texttt{function body}\\ 
\verb|```|

Apply the indents and formatting as in the example.
\end{quote}

%% file: data_samples/rucodereviewer.tex
\paragraph{Enum values}
\hrule
\vspace{0.5em}
\begin{itemize}
\item \textbf{Diff Block} (Original): \texttt{\\+CategoryType = tuple[tuple[str], tuple[str]]\\
+\\
+class LetterStatus(Enum):\\
+\ \ \ NO\_USED:\ int = 0\\
+\ \ \ USED\_IN\_WORD:\ int = 1\\
+\ \ \ USED\_NOT\_IN\_WORD:\ int = 2}

\item \textbf{Gold comment} (Translated): \texttt{NO\_USED = enum.auto()}

\end{itemize}



\paragraph{Go benchmark loop}
\hrule
\vspace{0.5em}
\begin{itemize}
\item \textbf{Diff Block} (Original): \texttt{\\+func BenchmarkSummation(b \*testing.B) \{\\
+\ \ sizes := []int\{config.HundredThousands,\ ...\}\\
+\ \ for \_, size := range sizes \{\\
+\ \ \ \ data := make([]int,\ size)\\
+\ \ \ \ b.Run(fmt.Sprintf("chan\_size\_
+\ \ \ \ b.Run(fmt.Sprintf("mu\_size\_
+\ \ \ \ b.Run(fmt.Sprintf("wg\_size\_
+\ \ \}\\
+\}}

\item \textbf{Gold comment} (Translated): \texttt{Add a \textit{for range b.N} loop — for very fast functions, a single run might be insufficient.}

\end{itemize}



\paragraph{Java null checks}
\hrule
\vspace{0.5em}
\begin{itemize}
\item \textbf{Diff Block} (Original): \texttt{\\+import static backend.academy.Values.RANDOM\_COLOR;\\
+import static backend.academy.Values.VARIATIONS;\\
+\\
+@Getter\\
+public class Transformations \{\\
+\ \ private Color lastColor;\\
+\ \ private final Coefficients[] coefficients;\\
+\ \ private final double[] weights;\\
+\ \ private final Color[] colors;\\
+\ \ private final PointFunction[] variations;\\
+\ \ \dots\\}

\item \textbf{Gold comment} (Translated): \texttt{\\- Missing \texttt{null} checks for arrays\\
- \texttt{variations} is not validated\\}

\end{itemize}



\paragraph{Python nesting}
\hrule
\vspace{0.5em}
\begin{itemize}
\item \textbf{Diff Block} (Original): \texttt{\\+\ \ \ \ self.\_bytes\_sum = 0\\
+\ \ \ \ self.\_filter\_name = filter\_name\\
+\\
+\ \ def parse(self) -> LogReport:\\
+\\
+\ \ \ \ """\\
+\ \ \ \ Method to begin parsing data\\
+\ \ \ \ \dots\\
+\ \ \ \ """\\
+\\
+\ \ \ \ files = glob.glob(self.\_path)\\
+\ \ \ \ \dots\\}

\item \textbf{Gold comment} (Translated): \texttt{\\Excessive nesting. Split checks into separate functions and simplify the block;\\
currently it’s a loop → try/except → context manager → another loop with branching — it’s too much.\\}

\end{itemize}



%% file: data_samples/rucodereviewer_judge.tex
\textbf{RuCodeReview Judge Prompt}

\hrule
\vspace{0.5em}
{\ttfamily\small
This prompt, translated from the original Russian formulation, is designed for an LLM-as-a-Judge:\\
"""\\
You are a \textbf{Judge}.

You will be provided with a code change snippet and two code comments describing a problem.
Your task is to determine whether both comments describe the same problem.

If the two comments describe the same problem and proposed solution, answer: correct.
If the comments describe different problems and solutions, answer: wrong.

Write your answer, correct or wrong, without quotation marks or other extraneous characters.

\textbf{Input data:}\\
You'll get the following data as input:\\
- Code block difference: \textcolor{blue}{\textbf{\{diff\_block\}}}\\
- Comment 1: \textcolor{blue}{\textbf{\{comment1\}}}\\
- Comment 2: \textcolor{blue}{\textbf{\{comment2\}}}\\
Answer:\\
"""
}
\vspace{0.5em}
\hrule

%% file: data_samples/unittests.tex
\begin{itemize}
\item \textbf{instruction:} \texttt{Write a test for the following go code from the file 'lists/mergesorted.go'.\\
You need to write the test function on go. The test will be placed to the file 'lists/mergesorted\_test.go'.\\
You can use the following entities imported or declared in the test file:\\
package lists\\
import (\\
	"math/rand"\\
	"reflect"\\
	"sort"\\
	"testing"\\
)\\
Pay attention to the following code when writing the test:\\
\#lists/mergesorted.go\\
package lists\\
\#focal function/method here\\
\\
Code for testing:\\
func MergeSorted(l, f *List) (m *List, ok bool) {\\
	m = new(List)\\
	for l.Len() > 0 || f.Len() > 0 {\\
		vl, nl, okl := PopInt(l)\\
		vf, nf, okf := PopInt(f)\\
		if !okl || !okf {\\
			return m, false\\
		}\\
\\
		ll, n := l, nl // The assumption is: vl <= vf.\\
		switch {\\
		case l.Len() == 0:\\
			ll, n = f, nf\\
		case f.Len() == 0:\\
			ll, n = l, nl\\
		case vl > vf:\\
			ll, n = f, nf\\
		}\\
\\
		m.Insert(ll.Remove(n))\\
	}\\
	return m, true\\
}\\
Write only the test function without any explanations or comments.\\
Your answer should be formatted using markdown as follows:\\
\\\textasciigrave\textasciigrave\textasciigrave go\\
<your code>
\\\textasciigrave\textasciigrave\textasciigrave
}

\end{itemize}

%% file: data_samples/codelintereval.tex
\begin{itemize}
\item \textbf{instruction:} \textit{Rewrite the code based on the errors received from the linter. Errors indicate critical weaknesses: potential bugs, security vulnerabilities and violations of clean code principles. Fix ALL these errors without exceptions, keep the original logic of the program, strictly adhere to PEP-8 for Python, do not add comments and explanations. Errors and warning from the linter: ``\{feedback\}'' Code: ``\{code\}''. Provide the response in the format corresponding to the template for the response:\\\textasciigrave\textasciigrave\textasciigrave\
python\\<code>\\\textasciigrave\textasciigrave\textasciigrave\
}

\item \textbf{code:}
\begin{verbatim}
def first_repeated_char(str1):
  for index,c in enumerate(str1):
    if str1[:index+1].count(c) > 1:
      return c
\end{verbatim}

\item \textbf{feedback:} \texttt{E111: indentation is not a multiple of 4 in 2 line. E231: missing whitespace after ',' in 2 line. E111: indentation is not a multiple of 4 in 4 line. W292: no newline at end of file in 4 line\\}

\end{itemize}

%% file: data_samples/codecorrectness.tex
\makeatletter \renewcommand\verbatim@font{\normalfont\itshape}
\begin{itemize}
  \item \textbf{instruction:} \\
        \textit{Below are the focal and test files. Please check if the test will run correctly.}\\ 
        \textit{Focal file:}\\
        \texttt{\textasciigrave\textasciigrave\textasciigrave}\{\textit{lang}\}\\
        \{\textit{focal}\}\\\texttt{\textasciigrave\textasciigrave\textasciigrave}\\
        \textit{Test file:}\\
        \texttt{\textasciigrave\textasciigrave\textasciigrave}\{\textit{lang}\}\\
        \{\textit{test}\}\\\texttt{\textasciigrave\textasciigrave\textasciigrave}\\ 
        \textit{Your answer: if the test runs without errors, write} \texttt{<<success>>} \textit{otherwise---} \texttt{<<failed>>}.
  \item \textbf{language:} \textit{Go}
  \item \textbf{focal file (snippet):}
\begin{verbatim}
type Queue[V core.Value] struct {
    items []core.Node[V]
}
...
\end{verbatim}
  \item \textbf{test file (snippet):}
\begin{verbatim}
func TestQueue_WithLargeNumberOfItems
                    (t *testing.T) {
    q := NewQueue[int]()
    ...
    assert.True(t, q.IsEmpty())
}
\end{verbatim}
\item \textbf{outputs:} \textit{success}
\end{itemize}

%% file: data_samples/strucom.tex
\begin{itemize}
\item \textbf{instruction:} \texttt{Write the Russian-language documentation for the function.}

\item \textbf{inputs:} \texttt{\\private void button15\_Click(object sender, EventArgs e) {
    label12.Text = "";
    richTextBox1.Clear();
    richTextBox2.Clear();
    textBox1.Clear();
    textBox2.Clear();
    textBox4.Clear();
}}

\item \textbf{outputs} (golden answer): \texttt{\\/// <summary>\\
/// Handler for the button click event button15.\\
/// When calling this function, the text in various controls of the form is cleared:\\
/// - The text is deleted from the label label12.\\
/// - The contents of the multiline text field richTextBox1 are cleared.\\
/// - The contents of the multiline text field richTextBox2 are cleared.\\
/// - The text field is cleared TextBox1.\\
/// - The textBox2 text field is cleared.\\
/// - The textbox textbox is being cleared./// </summary>\\
/// <param name="sender">The object that triggered the event (in this case, the button15 button).\\</param>\\
/// <param name="e">Event parameters that contain additional information about the event.</param>"}
\end{itemize}



%% file: data_samples/yabloco.tex
\begin{itemize}
\item \textbf{instruction:} \texttt{Generate code on C.}

\item \textbf{inputs:} \texttt{\\Function signature:\\
void *UI\_add\_user\_data(UI *ui, void *user\_data).\\
\\
Function description:\\
The following function is used to store a pointer to user-specific data.\\
\\
Use this context:\\
LONG\_CONTEXT\\
}

\item \textbf{outputs} (golden answer): \texttt{\\
void *UI\_add\_user\_data(UI *ui, void *user\_data)\\
\{\\
    void *old\_data = ui->user\_data;\\
    ui->user\_data = user\_data;\\
    ui->flags = $\sim$UI\_FLAG\_DUPL\_DATA;\\
    return old\_data;\\
\}
}

\end{itemize}




%% file: table/taxonomy_examples.tex
\begin{table*}[ht!]
    \setlength{\tabcolsep}{3pt}
    \centering
    \scriptsize
    \renewcommand\arraystretch{0.95}
    \begin{tabularx}{\textwidth}{
    @{}c|
        >{\raggedright\arraybackslash}p{0.15\textwidth}
        >{\raggedright\arraybackslash}p{0.25\textwidth}
        >{\raggedright\arraybackslash}p{0.3\textwidth}
    }
        \toprule 
        \textbf{Taxon} & \textbf{Base Taxon} & \textbf{Description} & \textbf{Example of task}\\
        \midrule
        Text & Perception & Ability to take as input text modality & MERA Code: StRuCom, UnitTests \\
        Image & Perception & Ability to take as input image modality & UML Diagrams, Block Schemes \\
        Audio & Perception & Ability to take as input audio modality & Vibe Coding \\
        Video & Perception & Ability to take as input video modality & Programming tutorial captioning \\
        Code & Perception & Ability to take as input code files & HTML, Markdown, Pseudo-code, bash scripts \\
        Tools & Perception & Ability to take information from various tools & Calculator, Web-Search, Interpreter \\
        Long Context Comprehension & Perception & Ability to take as input sequences with length at least 32000 tokens.\\
        \midrule
        Style Guides & Knowledge & Standards of code style for certain programming language & MERA Code: CodeLinterEval(PEP8) \\
        Programming Patterns & Knowledge & Programming patterns and best practices for developing a software project & GoF \\
        Programming Languages & Knowledge & Syntax of specific programming language & Python, Java, C++ \\
        Information \& Cyber Security & Knowledge & Awareness of model about threats in information systems  & Malware, hacking, phishing, ransomware \\
        Algorithms \& Data Structures & Knowledge & Theoretical knowledge in Computer Science & Sorting algorithms, Asymptotic complexity \\
        \midrule
        Comparison & Reasoning & Compare several code snippets with each other and derive their differences and commonalities & Commit message diff \\
        Side-by-Side Evaluation & Reasoning & Score a pair of code snippets according to certain criteria & Code Review \\
        Code-to-Code Search & Reasoning & Identify similar code snippets based on their syntactical and semantical similarity & Duplicates identification \\
        Knowledge Extraction & Reasoning & Find information in a given context relevant to a certain user query & Question answering \\
        Logs analysis & Reasoning & Extract particular events or summarize logs of any program execution & Identify the reason of code crashing \\
        Text-to-Code Search & Reasoning & Semantic search of the code snippet based on a given query & Question answering \\
        Dependencies understanding & Reasoning & Understanding of dependencies required to run certain application  & Requirements list creation \\
        Classification & Reasoning & Classify the code snippet according to a specific labels & MERA Code: Code correctness classification \\
        Detection & Reasoning & Highlight and classify the span of code snippet & Vulnerability detection \\
        Segmentation & Reasoning & Classify every unit of input message(token, pixel, etc) to a specific labels & OCR \\
        Simulation & Reasoning & Simulation the behavior of code snippet after running & MERA Code: StRuCom \\
        Testing & Reasoning & Design a test suit for a program with the fully possible test coverage & MERA Code: JavaTestGen \\
        Review & Reasoning & Critical analysis of provided code snippet & MERA Code: ruCodeReviewer \\
        Instruction Following & Reasoning & Understanding of a provided instruction and adjustment of the answer according to it & MERA Code: ruCodeReviewer \\
        Planning & Reasoning & Plan future actions to achieve the certain goal & End-to-End software development \\
        \midrule
        Synthesis & Generation & Generation of code from scratch by a provided text prompt. &  MERA Code: Unit Tests generation \\
        Completion & Generation & Generation of code based on provided code snippet and optional text instruction &  MERA Code: RealCode, RealCodeJava, ruCodeEval \\
        Editing & Generation & Edit a provided code snippet accordingly to text instruction &  MERA Code: CodeLinterEval \\
        Translation & Generation & Translate code from one programming language to another. & Rewriting the code-base to a new programming language \\
        Text Generation & Generation & Generation of text based on a ginen code &  Summarization of code snippet into natural language \\
        Explanation & Generation & Textual description and explanation of code snippet & README creation \\
        Documentation & Generation & Generate structured explanation of code snippet & MERA Code: StRuCom \\
        Diagram Generation & Generation & Create visual description of software project & UML Diagram creation \\
        \bottomrule
    \end{tabularx}
    \caption{The MERA Code taxonomy overview. It represents skills beginning from the second layer of taxonomy. \textbf{Taxon} - name of skill, \textbf{Base Taxon} - name of one of four grounding skills which is a predecessor of certain skill, \textbf{Description} - characteristic of skill, \textbf{Example of task} - real-world task where a mentioned skill could be used. Examples include tasks from MERA Code, if there are exists. MERA Team works on the extension of current benchmark to cover all skills, presented in the taxonomy.}
    \label{tab:taxonomy_info}
\end{table*}

%% file: table/private_results.tex
\begin{table*}
\centering
\scriptsize
\addtolength{\tabcolsep}{-0.4em}
\begin{tabular}{l|c|ccccccccccc}  

\multirow{2}*{\textbf{Model}} & \multirow{2}*{\textbf{\begin{tabular}{@{}c@{}}\textbf{Private} \\ \textbf{Score}\end{tabular}}} & \multicolumn{3}{c}{\textbf{{ruCodeEval}}} & \multicolumn{3}{c}{\textbf{{CodeLinterEval}}} & \multicolumn{5}{c}{\textbf{{ruCodeReviewer}}} \\

\cmidrule(lr){3-5} \cmidrule(lr){6-8} \cmidrule(lr){9-13} 

 & & \rot{90}{pass@1} & \rot{90}{pass@5} & \rot{90}{pass@10} & \rot{90}{pass@1} & \rot{90}{pass@5} & \rot{90}{pass@10} & \rot{90}{BLEU} & \rot{90}{chrF} & \rot{90}{Judge@1} & \rot{90}{Judge@5} & \rot{90}{Judge@10} \\

\toprule

Gemini 2.5 flash & \textbf{0.427} & \textbf{0.610} & \textbf{0.645} & \textbf{0.652} & 0.496 & 0.538 & 0.545 & 0.031 & 0.152 & 0.064 & 0.145 & 0.193 \\
GPT-4.1 & 0.382 & 0.443 & 0.484 & 0.494 & 0.555 & 0.585 & 0.600 & 0.017 & 0.136 & \textbf{0.096} & 0.102 & 0.102 \\
RuadaptQwen3-8B-Hybrid & 0.380 & 0.419 & 0.548 & 0.598 & 0.500 & 0.537 & 0.545 & 0.034 & 0.181 & 0.042 & 0.078 & 0.118 \\
Kimi-K2-Inst-0905 & 0.380 & 0.448 & 0.552 & 0.579 & 0.423 & 0.548 & 0.573 & 0.022 & 0.132 & 0.074 & 0.129 & 0.144 \\
Qwen3-Coder-30B-A3B-Inst & 0.373 & 0.391 & 0.418 & 0.433 & 0.621 & 0.632 & 0.636 & 0.018 & 0.139 & 0.058 & 0.074 & 0.083 \\
A-vibe & 0.367 & 0.263 & 0.405 & 0.439 & 0.614 & 0.664 & 0.673 & \textbf{0.035} & 0.182 & 0.023 & 0.071 & 0.102 \\
GigaChat-2-Max & 0.361 & 0.534 & 0.581 & 0.591 & 0.425 & 0.458 & 0.464 & 0.016 & 0.139 & 0.055 & 0.055 & 0.060 \\
DeepSeek-Coder-V2-Inst & 0.360 & 0.433 & 0.450 & 0.457 & 0.494 & 0.590 & 0.618 & 0.015 & 0.135 & 0.060 & 0.061 & 0.062 \\
Seed-Coder-8B-Inst & 0.345 & 0.317 & 0.317 & 0.317 & \textbf{0.655} & 0.655 & 0.655 & 0.019 & 0.156 & 0.035 & 0.052 & 0.060 \\
Qwen3-Coder-480B-A35B-Inst & 0.329 & 0.463 & 0.466 & 0.467 & 0.430 & 0.445 & 0.464 & 0.018 & 0.139 & 0.058 & 0.074 & 0.083 \\
CodeLlama70b-Inst & 0.310 & 0.040 & 0.120 & 0.171 & 0.507 & \textbf{0.908} & \textbf{0.973} & 0.007 & 0.051 & 0.006 & 0.023 & 0.025 \\
Qwen2.5-Coder-32B-Inst & 0.306 & 0.311 & 0.311 & 0.311 & 0.466 & 0.472 & 0.473 & 0.034 & \textbf{0.187} & 0.065 & \textbf{0.174} & \textbf{0.222} \\
GigaChat-2-Lite & 0.269 & 0.309 & 0.395 & 0.427 & 0.316 & 0.407 & 0.455 & 0.012 & 0.094 & 0.025 & 0.026 & 0.026 \\
DeepSeek-V3.1 & 0.259 & 0.345 & 0.417 & 0.433 & 0.230 & 0.314 & 0.355 & 0.018 & 0.133 & 0.065 & 0.086 & 0.093 \\
Qwen2.5-72B-Inst & 0.254 & 0.174 & 0.177 & 0.177 & 0.481 & 0.497 & 0.500 & 0.023 & 0.158 & 0.048 & 0.104 & 0.136 \\
Kodify Nano & 0.226 & 0.191 & 0.313 & 0.354 & 0.235 & 0.432 & 0.473 & 0.005 & 0.043 & 0.003 & 0.003 & 0.003 \\
Yi-Coder-9B-Chat & 0.181 & 0.350 & 0.362 & 0.372 & 0.145 & 0.157 & 0.164 & 0.009 & 0.078 & 0.016 & 0.016 & 0.016 \\
CodeLlama7b-Inst & 0.176 & 0.000 & 0.000 & 0.000 & 0.315 & 0.602 & 0.664 & 0.000 & 0.001 & 0.000 & 0.000 & 0.000 \\
GigaCode 1.4 & 0.166 & 0.357 & 0.364 & 0.366 & 0.027 & 0.027 & 0.027 & 0.029 & 0.182 & 0.064 & 0.120 & 0.145 \\
gpt-oss-120b & 0.106 & 0.004 & 0.016 & 0.024 & 0.208 & 0.328 & 0.355 & 0.003 & 0.025 & 0.000 & 0.000 & 0.000 \\
CodeLlama34b-Inst & 0.097 & 0.000 & 0.000 & 0.000 & 0.144 & 0.271 & 0.327 & 0.016 & 0.127 & 0.023 & 0.023 & 0.023 \\
CodeLlama13b-Inst & 0.071 & 0.000 & 0.000 & 0.000 & 0.058 & 0.225 & 0.355 & 0.000 & 0.000 & 0.000 & 0.000 & 0.000 \\
Mixtral-8x22B-Inst & 0.028 & 0.000 & 0.000 & 0.000 & 0.027 & 0.045 & 0.045 & 0.017 & 0.135 & 0.016 & 0.025 & 0.025 \\
gpt-oss-20b & 0.019 & 0.000 & 0.000 & 0.000 & 0.044 & 0.061 & 0.064 & 0.001 & 0.006 & 0.001 & 0.001 & 0.001 \\
DeepSeek-Coder-V2-Lite-Inst & 0.018 & 0.000 & 0.000 & 0.000 & 0.000 & 0.000 & 0.000 & 0.019 & 0.146 & 0.022 & 0.038 & 0.044 \\

\bottomrule
\end{tabular}
\caption{MERA Code benchmark results on private tasks. The best results are highlighted in bold.}
\label{tab:private_results}
\end{table*}

%% file: table/public_results.tex
\begin{table*}
\centering
\scriptsize
\addtolength{\tabcolsep}{-0.48em}
\begin{tabular}{l|c|cccccccccccc}  
\multirow{2}*{\textbf{Model}} & \multirow{2}*{\textbf{\begin{tabular}{@{}c@{}}\textbf{Total} \\ \textbf{Score}\end{tabular}}} & \multicolumn{2}{c}{\textbf{\rot{70}{YABLoCo}}} & \multicolumn{1}{c}{\textbf{\rot{70}{stRuCom}}} & \multicolumn{1}{c}{\textbf{\rot{70}{RealCode}}} & \multicolumn{1}{c}{\textbf{\rot{70}{UnitTests}}} & \multicolumn{2}{c}{\textbf{\rot{70}{JavaTestGen}}} & \multicolumn{3}{c}{\textbf{\rot{70}{ruHumanEval}}} & \multicolumn{1}{c}{\textbf{\rot{70}{RealCodeJava}}} & \multicolumn{1}{c}{\textbf{\rot{70}{CodeCorrectness}}} \\

\cmidrule(lr){3-4} \cmidrule(lr){5-5} \cmidrule(lr){6-6} \cmidrule(lr){7-7} \cmidrule(lr){8-9} \cmidrule(lr){10-12} \cmidrule(lr){13-13} \cmidrule(lr){14-14}      

 & & \rot{90}{pass@1} & \rot{90}{EM} & \rot{90}{chrF} & \rot{90}{pass@1} & \rot{90}{CBLEU} & \rot{90}{pass@1} & \rot{90}{compile@1} & \rot{90}{pass@1} & \rot{90}{pass@5} & \rot{90}{pass@10} & \rot{90}{pass@1} & \rot{90}{EM} \\

\toprule

GPT-4.1 & \textbf{0.381} & 0.144 & 0.034 & 0.297 & \textbf{0.418} & 0.201 & \textbf{0.344} & 0.639 & 0.450 & 0.480 & 0.494 & \textbf{0.416} & 0.660 \\
Gemini 2.5 flash & 0.360 & 0.120 & 0.029 & 0.217 & 0.388 & 0.212 & 0.211 & 0.502 & \textbf{0.604} & \textbf{0.654} & \textbf{0.665} & 0.386 & 0.404 \\
DeepSeek-Coder-V2-Inst & 0.350 & \textbf{0.149} & 0.034 & 0.200 & 0.364 & 0.188 & 0.269 & 0.581 & 0.392 & 0.411 & 0.415 & 0.386 & 0.714 \\
GigaChat-2-Max & 0.338 & 0.091 & 0.014 & 0.291 & 0.318 & 0.223 & 0.145 & 0.361 & 0.524 & 0.559 & 0.567 & 0.262 & 0.686 \\
RuadaptQwen3-8B-Hybrid & 0.326 & 0.115 & 0.014 & 0.244 & 0.191 & 0.218 & 0.093 & 0.326 & 0.427 & 0.570 & 0.604 & 0.245 & \textbf{0.744} \\
Qwen3-Coder-480B-A35B-Inst & 0.326 & 0.082 & 0.019 & 0.202 & 0.411 & 0.231 & 0.220 & 0.573 & 0.421 & 0.424 & 0.427 & 0.342 & 0.540 \\
Kimi-K2-Instruct-0905 & 0.325 & 0.077 & 0.029 & 0.232 & 0.188 & 0.153 & 0.233 & 0.507 & 0.448 & 0.554 & 0.579 & 0.342 & 0.571 \\
A-vibe & 0.317 & 0.115 & 0.024 & \textbf{0.307} & 0.157 & \textbf{0.232} & 0.119 & 0.348 & 0.257 & 0.411 & 0.470 & 0.279 & 0.724 \\
Qwen3-Coder-30B-A3B-Inst & 0.301 & 0.072 & 0.019 & 0.202 & 0.188 & 0.231 & 0.189 & 0.441 & 0.350 & 0.368 & 0.378 & 0.309 & 0.540 \\
Qwen2.5-Coder-32B-Inst & 0.300 & 0.111 & 0.019 & 0.213 & 0.323 & 0.206 & 0.220 & 0.529 & 0.289 & 0.293 & 0.293 & 0.386 & 0.519 \\
GigaCode 1.4 & 0.293 & 0.135 & \textbf{0.038} & 0.276 & 0.322 & 0.230 & 0.313 & 0.639 & 0.305 & 0.305 & 0.305 & 0.352 & 0.676 \\
Qwen2.5-72B-Inst & 0.288 & 0.144 & 0.024 & 0.252 & 0.362 & 0.160 & 0.189 & 0.476 & 0.157 & 0.163 & 0.165 & 0.349 & 0.702 \\
Seed-Coder-8B-Inst & 0.271 & 0.106 & 0.010 & 0.237 & 0.106 & 0.161 & 0.264 & \textbf{0.643} & 0.210 & 0.219 & 0.220 & 0.305 & 0.403 \\
DeepSeek-V3.1 & 0.248 & 0.106 & 0.024 & 0.293 & 0.084 & 0.159 & 0.189 & 0.520 & 0.282 & 0.372 & 0.390 & 0.275 & 0.372 \\
GigaChat-2-Lite & 0.229 & 0.048 & 0.005 & 0.144 & 0.099 & 0.107 & 0.035 & 0.278 & 0.251 & 0.344 & 0.372 & 0.148 & 0.712 \\
Yi-Coder-9B-Chat & 0.206 & 0.135 & 0.024 & 0.192 & 0.067 & 0.168 & 0.229 & 0.590 & 0.173 & 0.197 & 0.201 & 0.255 & 0.364 \\
Mixtral-8x22B-Inst & 0.182 & 0.106 & 0.019 & 0.152 & 0.180 & 0.194 & 0.229 & 0.502 & 0.000 & 0.000 & 0.000 & 0.366 & 0.597 \\
Kodify Nano & 0.181 & 0.082 & 0.000 & 0.092 & 0.041 & 0.112 & 0.053 & 0.300 & 0.173 & 0.299 & 0.341 & 0.104 & 0.475 \\
CodeLlama70b-Inst & 0.159 & 0.120 & \textbf{0.038} & 0.043 & 0.039 & 0.035 & 0.167 & 0.348 & 0.035 & 0.116 & 0.165 & 0.232 & 0.025 \\
gpt-oss-120b & 0.155 & 0.096 & 0.019 & 0.055 & 0.350 & 0.107 & 0.273 & 0.511 & 0.012 & 0.024 & 0.024 & 0.309 & 0.093 \\
DeepSeek-Coder-V2-Lite-Inst & 0.146 & 0.072 & 0.005 & 0.189 & 0.096 & 0.127 & 0.137 & 0.471 & 0.000 & 0.000 & 0.000 & 0.255 & 0.539 \\
CodeLlama34b-Inst & 0.141 & 0.120 & 0.029 & 0.120 & 0.019 & 0.176 & 0.088 & 0.317 & 0.000 & 0.000 & 0.000 & 0.195 & 0.472 \\
CodeLlama7b-Inst & 0.140 & 0.096 & 0.014 & 0.098 & 0.000 & 0.106 & 0.040 & 0.251 & 0.000 & 0.000 & 0.000 & 0.151 & 0.455 \\
CodeLlama13b-Inst & 0.127 & 0.082 & 0.024 & 0.143 & 0.016 & 0.156 & 0.093 & 0.361 & 0.001 & 0.003 & 0.006 & 0.171 & 0.419 \\
gpt-oss-20b & 0.095 & 0.038 & 0.000 & 0.062 & 0.305 & 0.148 & 0.013 & 0.084 & 0.000 & 0.000 & 0.000 & 0.285 & 0.115 \\

\bottomrule
\end{tabular}
\caption{MERA Code benchmark results on public tasks. The best results are highlighted in bold.}
\label{tab:public_results}
\end{table*}

%% file: custom.bib
@article{ren2020codebleu,
  title={Codebleu: a method for automatic evaluation of code synthesis},
  author={Ren, Shuo and Guo, Daya and Lu, Shuai and Zhou, Long and Liu, Shujie and Tang, Duyu and Sundaresan, Neel and Zhou, Ming and Blanco, Ambrosio and Ma, Shuai},
  journal={arXiv preprint arXiv:2009.10297},
  year={2020}
}

@article{liu2023your,
  title={Is your code generated by chatgpt really correct? rigorous evaluation of large language models for code generation},
  author={Liu, Jiawei and Xia, Chunqiu Steven and Wang, Yuyao and Zhang, Lingming},
  journal={Advances in Neural Information Processing Systems},
  volume={36},
  pages={21558--21572},
  year={2023}
}

@misc{eval-harness,
  author       = {Gao, Leo and Tow, Jonathan and Abbasi, Baber and Biderman, Stella and Black, Sid and DiPofi, Anthony and Foster, Charles and Golding, Laurence and Hsu, Jeffrey and Le Noac'h, Alain and Li, Haonan and McDonell, Kyle and Muennighoff, Niklas and Ociepa, Chris and Phang, Jason and Reynolds, Laria and Schoelkopf, Hailey and Skowron, Aviya and Sutawika, Lintang and Tang, Eric and Thite, Anish and Wang, Ben and Wang, Kevin and Zou, Andy},
  title        = {The Language Model Evaluation Harness},
  month        = 07,
  year         = 2024,
  publisher    = {Zenodo},
  version      = {v0.4.3},
  doi          = {10.5281/zenodo.12608602},
  url          = {https://zenodo.org/records/12608602}
}

@article{zhang2023unifying,
  title={Unifying the perspectives of nlp and software engineering: A survey on language models for code},
  author={Zhang, Ziyin and Chen, Chaoyu and Liu, Bingchang and Liao, Cong and Gong, Zi and Yu, Hang and Li, Jianguo and Wang, Rui},
  journal={arXiv preprint arXiv:2311.07989},
  year={2023}
}

@inproceedings{ruf2015classification,
  title={Classification of programming tasks according to required skills and knowledge representation},
  author={Ruf, Alexander and Berges, Marc and Hubwieser, Peter},
  booktitle={Informatics in Schools. Curricula, Competences, and Competitions: 8th International Conference on Informatics in Schools: Situation, Evolution, and Perspectives, ISSEP 2015, Ljubljana, Slovenia, September 28-October 1, 2015, Proceedings 8},
  pages={57--68},
  year={2015},
  organization={Springer}
}

@misc{ren2020codebleumethodautomaticevaluation,
      title={CodeBLEU: a Method for Automatic Evaluation of Code Synthesis}, 
      author={Shuo Ren and Daya Guo and Shuai Lu and Long Zhou and Shujie Liu and Duyu Tang and Neel Sundaresan and Ming Zhou and Ambrosio Blanco and Shuai Ma},
      year={2020},
      eprint={2009.10297},
      archivePrefix={arXiv},
      primaryClass={cs.SE},
      url={https://arxiv.org/abs/2009.10297}, 
}

@misc{tufano2021unit,
      title={Unit Test Case Generation with Transformers and Focal Context}, 
      author={Michele Tufano and Dawn Drain and Alexey Svyatkovskiy and Shao Kun Deng and Neel Sundaresan},
      year={2021},
      eprint={2009.05617},
      archivePrefix={arXiv},
      primaryClass={cs.SE}
}

@article{bui2025correctness,
  title={Correctness Assessment of Code Generated by Large Language Models Using Internal Representations},
  author={Bui, Tuan-Dung and Vu, Thanh Trong and Nguyen, Thu-Trang and Nguyen, Son and Vo, Hieu Dinh},
  journal={arXiv preprint arXiv:2501.12934},
  year={2025}
}

@article{dziuba2025strucom,
  title={Strucom: A novel dataset of structured code comments in russian},
  author={Dziuba, Maria and Malykh, Valentin},
  journal={arXiv preprint arXiv:2505.11026},
  year={2025}
}

@article{hui2024qwen2,
  title={Qwen2. 5-coder technical report},
  author={Hui, Binyuan and Yang, Jian and Cui, Zeyu and Yang, Jiaxi and Liu, Dayiheng and Zhang, Lei and Liu, Tianyu and Zhang, Jiajun and Yu, Bowen and Lu, Keming and others},
  journal={arXiv preprint arXiv:2409.12186},
  year={2024}
}

@inproceedings{popovic2015chrf,
  title={chrF: character n-gram F-score for automatic MT evaluation},
  author={Popovi{\'c}, Maja},
  booktitle={Proceedings of the tenth workshop on statistical machine translation},
  pages={392--395},
  year={2015}
}

@inproceedings{papineni-etal-2002-bleu,
    title = "{B}leu: a Method for Automatic Evaluation of Machine Translation",
    author = "Papineni, Kishore  and
      Roukos, Salim  and
      Ward, Todd  and
      Zhu, Wei-Jing",
    editor = "Isabelle, Pierre  and
      Charniak, Eugene  and
      Lin, Dekang",
    booktitle = "Proceedings of the 40th Annual Meeting of the Association for Computational Linguistics",
    month = jul,
    year = "2002",
    address = "Philadelphia, Pennsylvania, USA",
    publisher = "Association for Computational Linguistics",
    url = "https://aclanthology.org/P02-1040/",
    doi = "10.3115/1073083.1073135",
    pages = "311--318"
}

@article{chen2021evaluating,
  title={Evaluating large language models trained on code},
  author={Chen, Mark and Tworek, Jerry and Jun, Heewoo and Yuan, Qiming and Pinto, Henrique Ponde De Oliveira and Kaplan, Jared and Edwards, Harri and Burda, Yuri and Joseph, Nicholas and Brockman, Greg and others},
  journal={arXiv preprint arXiv:2107.03374},
  year={2021}
}

@article{austin2021program,
  title={Program Synthesis with Large Language Models},
  author={Austin, Jacob and Odena, Augustus and Nye, Maxwell and Bosma, Maarten and Michalewski, Henryk and Dohan, David and Jiang, Ellen and Cai, Carrie and Terry, Michael and Le, Quoc and others},
  journal={arXiv preprint arXiv:2108.07732},
  year={2021}
}

@article{du2023classeval,
  title={ClassEval: A Manually-Crafted Benchmark for Evaluating LLMs on Class-level Code Generation},
  author={Du, Xueying and Liu, Mingwei and Wang, Kaixin and Wang, Hanlin and Liu, Junwei and Chen, Yixuan and Feng, Jiayi and Sha, Chaofeng and Peng, Xin and Lou, Yiling},
  journal={CoRR},
  year={2023}
}

@inproceedings{liurepobench,
  title={RepoBench: Benchmarking Repository-Level Code Auto-Completion Systems},
  author={Liu, Tianyang and Xu, Canwen and McAuley, Julian},
  booktitle={The Twelfth International Conference on Learning Representations},
  year={2023} 
}

@inproceedings{jimenez2024swe,
  title={SWE-BENCH: CAN LANGUAGE MODELS RESOLVE REAL-WORLD GITHUB ISSUES?},
  author={Jimenez, Carlos E and Yang, John and Wettig, Alexander and Yao, Shunyu and Pei, Kexin and Press, Ofir and Narasimhan, Karthik},
  booktitle={12th International Conference on Learning Representations, ICLR 2024},
  year={2024}
}

@inproceedings{jainlivecodebench,
  title={LiveCodeBench: Holistic and Contamination Free Evaluation of Large Language Models for Code},
  author={Jain, Naman and Han, King and Gu, Alex and Li, Wen-Ding and Yan, Fanjia and Zhang, Tianjun and Wang, Sida and Solar-Lezama, Armando and Sen, Koushik and Stoica, Ion},
  booktitle={The Thirteenth International Conference on Learning Representations},
  year={2024}
}

@article{bogomolov2024long,
  title={Long Code Arena: a Set of Benchmarks for Long-Context Code Models},
  author={Bogomolov, Egor and Eliseeva, Aleksandra and Galimzyanov, Timur and Glukhov, Evgeniy and Shapkin, Anton and Tigina, Maria and Golubev, Yaroslav and Kovrigin, Alexander and van Deursen, Arie and Izadi, Maliheh and others},
  journal={CoRR},
  year={2024}
}

@misc{valeev2025yabloco,
      title={YABLoCo: Yet Another Benchmark for Long Context Code Generation}, 
      author={Aidar Valeev and Roman Garaev and Vadim Lomshakov and Irina Piontkovskaya and Vladimir Ivanov and Israel Adewuyi},
      year={2025},
      eprint={2505.04406},
      archivePrefix={arXiv},
      primaryClass={cs.CL},
      url={https://arxiv.org/abs/2505.04406}, 
}

@misc{jiang2024surveylargelanguagemodels,
      title={A Survey on Large Language Models for Code Generation}, 
      author={Juyong Jiang and Fan Wang and Jiasi Shen and Sungju Kim and Sunghun Kim},
      year={2024},
      eprint={2406.00515},
      archivePrefix={arXiv},
      primaryClass={cs.CL},
      url={https://arxiv.org/abs/2406.00515}, 
}

@article{lu2021codexglue,
  title={CodeXGLUE: {A} Machine Learning Benchmark Dataset for Code Understanding},
  author={Lu, Shuai and Guo, Daya and Ren, Shuo and Huang, Junjie and Svyatkovskiy, Alexey and Blanco, Ambrosio and Clement, Colin and Drain, Dawn and Jiang, Daxin and Tang, Duyu and others},
  journal={CoRR},
  year={2021}
}

@article{hendrycksapps2021,
  title={Measuring Coding Challenge Competence With APPS},
  author={Dan Hendrycks and Steven Basart and Saurav Kadavath and Mantas Mazeika and Akul Arora and Ethan Guo and Collin Burns and Samir Puranik and Horace He and Dawn Song and Jacob Steinhardt},
  journal={NeurIPS},
  year={2021}
}

@article{li2022competition,
  title={Competition-level code generation with alphacode},
  author={Li, Yujia and Choi, David and Chung, Junyoung and Kushman, Nate and Schrittwieser, Julian and Leblond, R{\'e}mi and Eccles, Tom and Keeling, James and Gimeno, Felix and Dal Lago, Agustin and others},
  journal={Science},
  volume={378},
  number={6624},
  pages={1092--1097},
  year={2022},
  publisher={American Association for the Advancement of Science}
}

@article{quan2025codeelo,
  title={CodeElo: Benchmarking Competition-level Code Generation of LLMs with Human-comparable Elo Ratings},
  author={Quan, Shanghaoran and Yang, Jiaxi and Yu, Bowen and Zheng, Bo and Liu, Dayiheng and Yang, An and Ren, Xuancheng and Gao, Bofei and Miao, Yibo and Feng, Yunlong and others},
  journal={arXiv preprint arXiv:2501.01257},
  year={2025}
}

@article{cassano2022multipl,
  title={Multipl-e: A scalable and extensible approach to benchmarking neural code generation},
  author={Cassano, Federico and Gouwar, John and Nguyen, Daniel and Nguyen, Sydney and Phipps-Costin, Luna and Pinckney, Donald and Yee, Ming-Ho and Zi, Yangtian and Anderson, Carolyn Jane and Feldman, Molly Q and others},
  journal={arXiv preprint arXiv:2208.08227},
  year={2022}
}

@inproceedings{zheng2023codegeex,
  title={Codegeex: A pre-trained model for code generation with multilingual benchmarking on humaneval-x},
  author={Zheng, Qinkai and Xia, Xiao and Zou, Xu and Dong, Yuxiao and Wang, Shan and Xue, Yufei and Shen, Lei and Wang, Zihan and Wang, Andi and Li, Yang and others},
  booktitle={Proceedings of the 29th ACM SIGKDD Conference on Knowledge Discovery and Data Mining},
  pages={5673--5684},
  year={2023}
}

@inproceedings{athiwaratkun2023multi,
  title={Multi-lingual Evaluation of Code Generation Models},
  author={Athiwaratkun, Ben and Gouda, Sanjay Krishna and Wang, Zijian and Li, Xiaopeng and Tian, Yuchen and Tan, Ming and Ahmad, Wasi Uddin and Wang, Shiqi and Sun, Qing and Shang, Mingyue and others},
  booktitle={ICLR},
  year={2023}
}

@article{Chen2021EvaluatingLL,
  title={Evaluating Large Language Models Trained on Code},
  author={Mark Chen and Jerry Tworek and Heewoo Jun and Qiming Yuan and Henrique Pond{\'e} and Jared Kaplan and Harrison Edwards and Yura Burda and Nicholas Joseph and Greg Brockman and Alex Ray and Raul Puri and Gretchen Krueger and Michael Petrov and Heidy Khlaaf and Girish Sastry and Pamela Mishkin and Brooke Chan and Scott Gray and Nick Ryder and Mikhail Pavlov and Alethea Power and Lukasz Kaiser and Mo Bavarian and Clemens Winter and Phil Tillet and Felipe Petroski Such and David W. Cummings and Matthias Plappert and Fotios Chantzis and Elizabeth Barnes and Ariel Herbert-Voss and William H. Guss and Alex Nichol and Igor Babuschkin and Suchir Balaji and Shantanu Jain and Andrew Carr and Jan Leike and Josh Achiam and Vedant Misra and Evan Morikawa and Alec Radford and Matthew M. Knight and Miles Brundage and Mira Murati and Katie Mayer and Peter Welinder and Bob McGrew and Dario Amodei and Sam McCandlish and Ilya Sutskever and Wojciech Zaremba},
  journal={ArXiv},
  year={2021},
  volume={abs/2107.03374},
  url={https://api.semanticscholar.org/CorpusID:235755472}
}

@article{Austin2021ProgramSW,
  title={Program Synthesis with Large Language Models},
  author={Jacob Austin and Augustus Odena and Maxwell Nye and Maarten Bosma and Henryk Michalewski and David Dohan and Ellen Jiang and Carrie J. Cai and Michael Terry and Quoc V. Le and Charles Sutton},
  journal={ArXiv},
  year={2021},
  volume={abs/2108.07732},
  url={https://api.semanticscholar.org/CorpusID:237142385}
}

@article{Li2022CompetitionlevelCG,
  title={Competition-level code generation with AlphaCode},
  author={Yujia Li and David Choi and Junyoung Chung and Nate Kushman and Julian Schrittwieser and R{\'e}mi Leblond and Tom and Eccles and James Keeling and Felix Gimeno and Agustin Dal Lago and Thomas Hubert and Peter Choy and Cyprien de and Masson d’Autume and Igor Babuschkin and Xinyun Chen and Po-Sen Huang and Johannes Welbl and Sven Gowal and Alexey and Cherepanov and James Molloy and Daniel Jaymin Mankowitz and Esme Sutherland Robson and Pushmeet Kohli and Nando de and Freitas and Koray Kavukcuoglu and Oriol Vinyals},
  journal={Science},
  year={2022},
  volume={378},
  pages={1092 - 1097},
  url={https://api.semanticscholar.org/CorpusID:246527904}
}

@article{10.1145/3695991,
author = {Dong, Yihong and Ding, Jiazheng and Jiang, Xue and Li, Ge and Li, Zhuo and Jin, Zhi},
title = {CodeScore: Evaluating Code Generation by Learning Code Execution},
year = {2025},
issue_date = {March 2025},
publisher = {Association for Computing Machinery},
address = {New York, NY, USA},
volume = {34},
number = {3},
issn = {1049-331X},
url = {https://doi.org/10.1145/3695991},
doi = {10.1145/3695991},
journal = {ACM Trans. Softw. Eng. Methodol.},
month = feb,
articleno = {77},
numpages = {22}
}

@inproceedings{
zhuo2025bigcodebench,
title={BigCodeBench: Benchmarking Code Generation with Diverse Function Calls and Complex Instructions},
author={Terry Yue Zhuo and Vu Minh Chien and Jenny Chim and Han Hu and Wenhao Yu and Ratnadira Widyasari and Imam Nur Bani Yusuf and Haolan Zhan and Junda He and Indraneil Paul and Simon Brunner and Chen GONG and James Hoang and Armel Randy Zebaze and Xiaoheng Hong and Wen-Ding Li and Jean Kaddour and Ming Xu and Zhihan Zhang and Prateek Yadav and Naman Jain and Alex Gu and Zhoujun Cheng and Jiawei Liu and Qian Liu and Zijian Wang and David Lo and Binyuan Hui and Niklas Muennighoff and Daniel Fried and Xiaoning Du and Harm de Vries and Leandro Von Werra},
booktitle={The Thirteenth International Conference on Learning Representations},
year={2025},
url={https://openreview.net/forum?id=YrycTjllL0}
}

@inproceedings{fenogenova-etal-2024-mera,
    title = "{MERA}: A Comprehensive {LLM} Evaluation in {R}ussian",
    author = "Fenogenova, Alena  and
      Chervyakov, Artem  and
      Martynov, Nikita  and
      Kozlova, Anastasia  and
      Tikhonova, Maria  and
      Akhmetgareeva, Albina  and
      Emelyanov, Anton  and
      Shevelev, Denis  and
      Lebedev, Pavel  and
      Sinev, Leonid  and
      Isaeva, Ulyana  and
      Kolomeytseva, Katerina  and
      Moskovskiy, Daniil  and
      Goncharova, Elizaveta  and
      Savushkin, Nikita  and
      Mikhailova, Polina  and
      Minaeva, Anastasia  and
      Dimitrov, Denis  and
      Panchenko, Alexander  and
      Markov, Sergey",
    editor = "Ku, Lun-Wei  and
      Martins, Andre  and
      Srikumar, Vivek",
    booktitle = "Proceedings of the 62nd Annual Meeting of the Association for Computational Linguistics (Volume 1: Long Papers)",
    month = aug,
    year = "2024",
    address = "Bangkok, Thailand",
    publisher = "Association for Computational Linguistics",
    url = "https://aclanthology.org/2024.acl-long.534/",
    doi = "10.18653/v1/2024.acl-long.534",
    pages = "9920--9948"
}

@inproceedings{gao-etal-2021-making,
    title = "Making Pre-trained Language Models Better Few-shot Learners",
    author = "Gao, Tianyu  and
      Fisch, Adam  and
      Chen, Danqi",
    editor = "Zong, Chengqing  and
      Xia, Fei  and
      Li, Wenjie  and
      Navigli, Roberto",
    booktitle = "Proceedings of the 59th Annual Meeting of the Association for Computational Linguistics and the 11th International Joint Conference on Natural Language Processing (Volume 1: Long Papers)",
    month = aug,
    year = "2021",
    address = "Online",
    publisher = "Association for Computational Linguistics",
    url = "https://aclanthology.org/2021.acl-long.295/",
    doi = "10.18653/v1/2021.acl-long.295",
    pages = "3816--3830"
}

@inproceedings{shin-etal-2020-autoprompt,
    title = "{A}uto{P}rompt: {E}liciting {K}nowledge from {L}anguage {M}odels with {A}utomatically {G}enerated {P}rompts",
    author = "Shin, Taylor  and
      Razeghi, Yasaman  and
      Logan IV, Robert L.  and
      Wallace, Eric  and
      Singh, Sameer",
    editor = "Webber, Bonnie  and
      Cohn, Trevor  and
      He, Yulan  and
      Liu, Yang",
    booktitle = "Proceedings of the 2020 Conference on Empirical Methods in Natural Language Processing (EMNLP)",
    month = nov,
    year = "2020",
    address = "Online",
    publisher = "Association for Computational Linguistics",
    url = "https://aclanthology.org/2020.emnlp-main.346/",
    doi = "10.18653/v1/2020.emnlp-main.346",
    pages = "4222--4235"
}

@inproceedings{10.3115/1073083.1073135,
author = {Papineni, Kishore and Roukos, Salim and Ward, Todd and Zhu, Wei-Jing},
title = {BLEU: a method for automatic evaluation of machine translation},
year = {2002},
publisher = {Association for Computational Linguistics},
address = {USA},
url = {https://doi.org/10.3115/1073083.1073135},
doi = {10.3115/1073083.1073135},
booktitle = {Proceedings of the 40th Annual Meeting on Association for Computational Linguistics},
pages = {311–318},
numpages = {8},
location = {Philadelphia, Pennsylvania},
series = {ACL '02}
}

@article{zheng2023judging,
  title={Judging llm-as-a-judge with mt-bench and chatbot arena},
  author={Zheng, Lianmin and Chiang, Wei-Lin and Sheng, Ying and Zhuang, Siyuan and Wu, Zhanghao and Zhuang, Yonghao and Lin, Zi and Li, Zhuohan and Li, Dacheng and Xing, Eric and others},
  journal={Advances in Neural Information Processing Systems},
  volume={36},
  pages={46595--46623},
  year={2023}
}

@misc{gpt4.1,
      title={GPT-4 Technical Report}, 
      author={OpenAI and Josh Achiam and Steven Adler and Sandhini Agarwal and Lama Ahmad and Ilge Akkaya and Florencia Leoni Aleman and Diogo Almeida and Janko Altenschmidt and Sam Altman and Shyamal Anadkat and Red Avila and Igor Babuschkin and Suchir Balaji and Valerie Balcom and Paul Baltescu and Haiming Bao and Mohammad Bavarian and Jeff Belgum and Irwan Bello and Jake Berdine and Gabriel Bernadett-Shapiro and Christopher Berner and Lenny Bogdonoff and Oleg Boiko and Madelaine Boyd and Anna-Luisa Brakman and Greg Brockman and Tim Brooks and Miles Brundage and Kevin Button and Trevor Cai and Rosie Campbell and Andrew Cann and Brittany Carey and Chelsea Carlson and Rory Carmichael and Brooke Chan and Che Chang and Fotis Chantzis and Derek Chen and Sully Chen and Ruby Chen and Jason Chen and Mark Chen and Ben Chess and Chester Cho and Casey Chu and Hyung Won Chung and Dave Cummings and Jeremiah Currier and Yunxing Dai and Cory Decareaux and Thomas Degry and Noah Deutsch and Damien Deville and Arka Dhar and David Dohan and Steve Dowling and Sheila Dunning and Adrien Ecoffet and Atty Eleti and Tyna Eloundou and David Farhi and Liam Fedus and Niko Felix and Simón Posada Fishman and Juston Forte and Isabella Fulford and Leo Gao and Elie Georges and Christian Gibson and Vik Goel and Tarun Gogineni and Gabriel Goh and Rapha Gontijo-Lopes and Jonathan Gordon and Morgan Grafstein and Scott Gray and Ryan Greene and Joshua Gross and Shixiang Shane Gu and Yufei Guo and Chris Hallacy and Jesse Han and Jeff Harris and Yuchen He and Mike Heaton and Johannes Heidecke and Chris Hesse and Alan Hickey and Wade Hickey and Peter Hoeschele and Brandon Houghton and Kenny Hsu and Shengli Hu and Xin Hu and Joost Huizinga and Shantanu Jain and Shawn Jain and Joanne Jang and Angela Jiang and Roger Jiang and Haozhun Jin and Denny Jin and Shino Jomoto and Billie Jonn and Heewoo Jun and Tomer Kaftan and Łukasz Kaiser and Ali Kamali and Ingmar Kanitscheider and Nitish Shirish Keskar and Tabarak Khan and Logan Kilpatrick and Jong Wook Kim and Christina Kim and Yongjik Kim and Jan Hendrik Kirchner and Jamie Kiros and Matt Knight and Daniel Kokotajlo and Łukasz Kondraciuk and Andrew Kondrich and Aris Konstantinidis and Kyle Kosic and Gretchen Krueger and Vishal Kuo and Michael Lampe and Ikai Lan and Teddy Lee and Jan Leike and Jade Leung and Daniel Levy and Chak Ming Li and Rachel Lim and Molly Lin and Stephanie Lin and Mateusz Litwin and Theresa Lopez and Ryan Lowe and Patricia Lue and Anna Makanju and Kim Malfacini and Sam Manning and Todor Markov and Yaniv Markovski and Bianca Martin and Katie Mayer and Andrew Mayne and Bob McGrew and Scott Mayer McKinney and Christine McLeavey and Paul McMillan and Jake McNeil and David Medina and Aalok Mehta and Jacob Menick and Luke Metz and Andrey Mishchenko and Pamela Mishkin and Vinnie Monaco and Evan Morikawa and Daniel Mossing and Tong Mu and Mira Murati and Oleg Murk and David Mély and Ashvin Nair and Reiichiro Nakano and Rajeev Nayak and Arvind Neelakantan and Richard Ngo and Hyeonwoo Noh and Long Ouyang and Cullen O'Keefe and Jakub Pachocki and Alex Paino and Joe Palermo and Ashley Pantuliano and Giambattista Parascandolo and Joel Parish and Emy Parparita and Alex Passos and Mikhail Pavlov and Andrew Peng and Adam Perelman and Filipe de Avila Belbute Peres and Michael Petrov and Henrique Ponde de Oliveira Pinto and Michael and Pokorny and Michelle Pokrass and Vitchyr H. Pong and Tolly Powell and Alethea Power and Boris Power and Elizabeth Proehl and Raul Puri and Alec Radford and Jack Rae and Aditya Ramesh and Cameron Raymond and Francis Real and Kendra Rimbach and Carl Ross and Bob Rotsted and Henri Roussez and Nick Ryder and Mario Saltarelli and Ted Sanders and Shibani Santurkar and Girish Sastry and Heather Schmidt and David Schnurr and John Schulman and Daniel Selsam and Kyla Sheppard and Toki Sherbakov and Jessica Shieh and Sarah Shoker and Pranav Shyam and Szymon Sidor and Eric Sigler and Maddie Simens and Jordan Sitkin and Katarina Slama and Ian Sohl and Benjamin Sokolowsky and Yang Song and Natalie Staudacher and Felipe Petroski Such and Natalie Summers and Ilya Sutskever and Jie Tang and Nikolas Tezak and Madeleine B. Thompson and Phil Tillet and Amin Tootoonchian and Elizabeth Tseng and Preston Tuggle and Nick Turley and Jerry Tworek and Juan Felipe Cerón Uribe and Andrea Vallone and Arun Vijayvergiya and Chelsea Voss and Carroll Wainwright and Justin Jay Wang and Alvin Wang and Ben Wang and Jonathan Ward and Jason Wei and CJ Weinmann and Akila Welihinda and Peter Welinder and Jiayi Weng and Lilian Weng and Matt Wiethoff and Dave Willner and Clemens Winter and Samuel Wolrich and Hannah Wong and Lauren Workman and Sherwin Wu and Jeff Wu and Michael Wu and Kai Xiao and Tao Xu and Sarah Yoo and Kevin Yu and Qiming Yuan and Wojciech Zaremba and Rowan Zellers and Chong Zhang and Marvin Zhang and Shengjia Zhao and Tianhao Zheng and Juntang Zhuang and William Zhuk and Barret Zoph},
      year={2024},
      eprint={2303.08774},
      archivePrefix={arXiv},
      primaryClass={cs.CL},
      url={https://arxiv.org/abs/2303.08774}, 
}

@misc{gemini2025,
      title={Gemini 2.5: Pushing the Frontier with Advanced Reasoning, Multimodality, Long Context, and Next Generation Agentic Capabilities}, 
      author={Gheorghe Comanici and Eric Bieber and Mike Schaekermann and Ice Pasupat and Noveen Sachdeva and Inderjit Dhillon and Marcel Blistein and Ori Ram and Dan Zhang and Evan Rosen and Luke Marris and Sam Petulla and Colin Gaffney and Asaf Aharoni and Nathan Lintz and Tiago Cardal Pais and Henrik Jacobsson and Idan Szpektor and Nan-Jiang Jiang and Krishna Haridasan and Ahmed Omran and Nikunj Saunshi and Dara Bahri and Gaurav Mishra and Eric Chu and Toby Boyd and Brad Hekman and Aaron Parisi and Chaoyi Zhang and Kornraphop Kawintiranon and Tania Bedrax-Weiss and Oliver Wang and Ya Xu and Ollie Purkiss and Uri Mendlovic and Ilaï Deutel and Nam Nguyen and Adam Langley and Flip Korn and Lucia Rossazza and Alexandre Ramé and Sagar Waghmare and Helen Miller and Nathan Byrd and Ashrith Sheshan and Raia Hadsell and Sangnie Bhardwaj and Pawel Janus and Tero Rissa and Dan Horgan and Alvin Abdagic and Lior Belenki and James Allingham and Anima Singh and Theo Guidroz and Srivatsan Srinivasan and Herman Schmit and Kristen Chiafullo and Andre Elisseeff and Nilpa Jha and Prateek Kolhar and Leonard Berrada and Frank Ding and Xiance Si and Shrestha Basu Mallick and Franz Och and Sofia Erell and Eric Ni and Tejasi Latkar and Sherry Yang and Petar Sirkovic and Ziqiang Feng and Robert Leland and Rachel Hornung and Gang Wu and Charles Blundell and Hamidreza Alvari and Po-Sen Huang and Cathy Yip and Sanja Deur and Li Liu and Gabriela Surita and Pablo Duque and Dima Damen and Johnson Jia and Arthur Guez and Markus Mircea and Animesh Sinha and Alberto Magni and Paweł Stradomski and Tal Marian and Vlado Galić and Wenhu Chen and Hisham Husain and Achintya Singhal and Dominik Grewe and François-Xavier Aubet and Shuang Song and Lorenzo Blanco and Leland Rechis and Lewis Ho and Rich Munoz and Kelvin Zheng and Jessica Hamrick and Kevin Mather and Hagai Taitelbaum and Eliza Rutherford and Yun Lei and Kuangyuan Chen and Anand Shukla and Erica Moreira and Eric Doi and Berivan Isik and Nir Shabat and Dominika Rogozińska and Kashyap Kolipaka and Jason Chang and Eugen Vušak and Srinivasan Venkatachary and Shadi Noghabi and Tarun Bharti and Younghoon Jun and Aleksandr Zaks and Simon Green and Jeshwanth Challagundla and William Wong and Muqthar Mohammad and Dean Hirsch and Yong Cheng and Iftekhar Naim and Lev Proleev and Damien Vincent and Aayush Singh and Maxim Krikun and Dilip Krishnan and Zoubin Ghahramani and Aviel Atias and Rajeev Aggarwal and Christo Kirov and Dimitrios Vytiniotis and Christy Koh and Alexandra Chronopoulou and Pawan Dogra and Vlad-Doru Ion and Gladys Tyen and Jason Lee and Felix Weissenberger and Trevor Strohman and Ashwin Balakrishna and Jack Rae and Marko Velic and Raoul de Liedekerke and Oded Elyada and Wentao Yuan and Canoee Liu and Lior Shani and Sergey Kishchenko and Bea Alessio and Yandong Li and Richard Song and Sam Kwei and Orion Jankowski and Aneesh Pappu and Youhei Namiki and Yenai Ma and Nilesh Tripuraneni and Colin Cherry and Marissa Ikonomidis and Yu-Cheng Ling and Colin Ji and Beka Westberg and Auriel Wright and Da Yu and David Parkinson and Swaroop Ramaswamy and Jerome Connor and Soheil Hassas Yeganeh and Snchit Grover and George Kenwright and Lubo Litchev and Chris Apps and Alex Tomala and Felix Halim and Alex Castro-Ros and Zefei Li and Anudhyan Boral and Pauline Sho and Michal Yarom and Eric Malmi and David Klinghoffer and Rebecca Lin and Alan Ansell and Pradeep Kumar S and Shubin Zhao and Siqi Zuo and Adam Santoro and Heng-Tze Cheng and Solomon Demmessie and Yuchi Liu and Nicole Brichtova and Allie Culp and Nathaniel Braun and Dan Graur and Will Ng and Nikhil Mehta and Aaron Phillips and Patrik Sundberg and Varun Godbole and Fangyu Liu and Yash Katariya and David Rim and Mojtaba Seyedhosseini and Sean Ammirati and Jonas Valfridsson and Mahan Malihi and Timothy Knight and Andeep Toor and Thomas Lampe and Abe Ittycheriah and Lewis Chiang and Chak Yeung and Alexandre Fréchette and Jinmeng Rao and Huisheng Wang and Himanshu Srivastava and Richard Zhang and Rocky Rhodes and Ariel Brand and Dean Weesner and Ilya Figotin and Felix Gimeno and Rachana Fellinger and Pierre Marcenac and José Leal and Eyal Marcus and Victor Cotruta and Rodrigo Cabrera and Sheryl Luo and Dan Garrette and Vera Axelrod and Sorin Baltateanu and David Barker and Dongkai Chen and Horia Toma and Ben Ingram and Jason Riesa and Chinmay Kulkarni and Yujing Zhang and Hongbin Liu and Chao Wang and Martin Polacek and Will Wu and Kai Hui and Adrian N Reyes and Yi Su and Megan Barnes and Ishaan Malhi and Anfal Siddiqui and Qixuan Feng and Mihai Damaschin and Daniele Pighin and Andreas Steiner and Samuel Yang and Ramya Sree Boppana and Simeon Ivanov and Arun Kandoor and Aditya Shah and Asier Mujika and Da Huang and Christopher A. Choquette-Choo and Mohak Patel and Tianhe Yu and Toni Creswell and Jerry and Liu and Catarina Barros and Yasaman Razeghi and Aurko Roy and Phil Culliton and Binbin Xiong and Jiaqi Pan and Thomas Strohmann and Tolly Powell and Babi Seal and Doug DeCarlo and Pranav Shyam and Kaan Katircioglu and Xuezhi Wang and Cassidy Hardin and Immanuel Odisho and Josef Broder and Oscar Chang and Arun Nair and Artem Shtefan and Maura O'Brien and Manu Agarwal and Sahitya Potluri and Siddharth Goyal and Amit Jhindal and Saksham Thakur and Yury Stuken and James Lyon and Kristina Toutanova and Fangxiaoyu Feng and Austin Wu and Ben Horn and Alek Wang and Alex Cullum and Gabe Taubman and Disha Shrivastava and Chongyang Shi and Hamish Tomlinson and Roma Patel and Tao Tu and Ada Maksutaj Oflazer and Francesco Pongetti and Mingyao Yang and Adrien Ali Taïga and Vincent Perot and Nuo Wang Pierse and Feng Han and Yoel Drori and Iñaki Iturrate and Ayan Chakrabarti and Legg Yeung and Dave Dopson and Yi-ting Chen and Apoorv Kulshreshtha and Tongfei Guo and Philip Pham and Tal Schuster and Junquan Chen and Alex Polozov and Jinwei Xing and Huanjie Zhou and Praneeth Kacham and Doron Kukliansky and Antoine Miech and Sergey Yaroshenko and Ed Chi and Sholto Douglas and Hongliang Fei and Mathieu Blondel and Preethi Myla and Lior Madmoni and Xing Wu and Daniel Keysers and Kristian Kjems and Isabela Albuquerque and Lijun Yu and Joel D'sa and Michelle Plantan and Vlad Ionescu and Jaume Sanchez Elias and Abhirut Gupta and Manish Reddy Vuyyuru and Fred Alcober and Tong Zhou and Kaiyang Ji and Florian Hartmann and Subha Puttagunta and Hugo Song and Ehsan Amid and Anca Stefanoiu and Andrew Lee and Paul Pucciarelli and Emma Wang and Amit Raul and Slav Petrov and Isaac Tian and Valentin Anklin and Nana Nti and Victor Gomes and Max Schumacher and Grace Vesom and Alex Panagopoulos and Konstantinos Bousmalis and Daniel Andor and Josh Jacob and Yuan Zhang and Bill Rosgen and Matija Kecman and Matthew Tung and Alexandra Belias and Noah Goodman and Paul Covington and Brian Wieder and Nikita Saxena and Elnaz Davoodi and Muhuan Huang and Sharath Maddineni and Vincent Roulet and Folawiyo Campbell-Ajala and Pier Giuseppe Sessa and Xintian and Wu and Guangda Lai and Paul Collins and Alex Haig and Vytenis Sakenas and Xiaowei Xu and Marissa Giustina and Laurent El Shafey and Pichi Charoenpanit and Shefali Garg and Joshua Ainslie and Boone Severson and Montse Gonzalez Arenas and Shreya Pathak and Sujee Rajayogam and Jie Feng and Michiel Bakker and Sheng Li and Nevan Wichers and Jamie Rogers and Xinyang Geng and Yeqing Li and Rolf Jagerman and Chao Jia and Nadav Olmert and David Sharon and Matthew Mauger and Sandeep Mariserla and Hongxu Ma and Megha Mohabey and Kyuyeun Kim and Alek Andreev and Scott Pollom and Juliette Love and Vihan Jain and Priyanka Agrawal and Yannick Schroecker and Alisa Fortin and Manfred Warmuth and Ji Liu and Andrew Leach and Irina Blok and Ganesh Poomal Girirajan and Roee Aharoni and Benigno Uria and Andrei Sozanschi and Dan Goldberg and Lucian Ionita and Marco Tulio Ribeiro and Martin Zlocha and Vighnesh Birodkar and Sami Lachgar and Liangzhe Yuan and Himadri Choudhury and Matt Ginsberg and Fei Zheng and Gregory Dibb and Emily Graves and Swachhand Lokhande and Gabriel Rasskin and George-Cristian Muraru and Corbin Quick and Sandeep Tata and Pierre Sermanet and Aditya Chawla and Itay Karo and Yan Wang and Susan Zhang and Orgad Keller and Anca Dragan and Guolong Su and Ian Chou and Xi Liu and Yiqing Tao and Shruthi Prabhakara and Marc Wilson and Ruibo Liu and Shibo Wang and Georgie Evans and David Du and Alfonso Castaño and Gautam Prasad and Mona El Mahdy and Sebastian Gerlach and Machel Reid and Jarrod Kahn and Amir Zait and Thanumalayan Sankaranarayana Pillai and Thatcher Ulrich and Guanyu Wang and Jan Wassenberg and Efrat Farkash and Kiran Yalasangi and Congchao Wang and Maria Bauza and Simon Bucher and Ting Liu and Jun Yan and Gary Leung and Vikas Sindhwani and Parker Barnes and Avi Singh and Ivan Jurin and Jichuan Chang and Niket Kumar Bhumihar and Sivan Eiger and Gui Citovsky and Ben Withbroe and Zhang Li and Siyang Xue and Niccolò Dal Santo and Georgi Stoyanov and Yves Raimond and Steven Zheng and Yilin Gao and Vít Listík and Sławek Kwasiborski and Rachel Saputro and Adnan Ozturel and Ganesh Mallya and Kushal Majmundar and Ross West and Paul Caron and Jinliang Wei and Lluis Castrejon and Sharad Vikram and Deepak Ramachandran and Nikhil Dhawan and Jiho Park and Sara Smoot and George van den Driessche and Yochai Blau and Chase Malik and Wei Liang and Roy Hirsch and Cicero Nogueira dos Santos and Eugene Weinstein and Aäron van den Oord and Sid Lall and Nicholas FitzGerald and Zixuan Jiang and Xuan Yang and Dale Webster and Ali Elqursh and Aedan Pope and Georges Rotival and David Raposo and Wanzheng Zhu and Jeff Dean and Sami Alabed and Dustin Tran and Arushi Gupta and Zach Gleicher and Jessica Austin and Edouard Rosseel and Megh Umekar and Dipanjan Das and Yinghao Sun and Kai Chen and Karolis Misiunas and Xiang Zhou and Yixian Di and Alyssa Loo and Josh Newlan and Bo Li and Vinay Ramasesh and Ying Xu and Alex Chen and Sudeep Gandhe and Radu Soricut and Nikita Gupta and Shuguang Hu and Seliem El-Sayed and Xavier Garcia and Idan Brusilovsky and Pu-Chin Chen and Andrew Bolt and Lu Huang and Alex Gurney and Zhiying Zhang and Alexander Pritzel and Jarek Wilkiewicz and Bryan Seybold and Bhargav Kanagal Shamanna and Felix Fischer and Josef Dean and Karan Gill and Ross Mcilroy and Abhishek Bhowmick and Jeremy Selier and Antoine Yang and Derek Cheng and Vladimir Magay and Jie Tan and Dhriti Varma and Christian Walder and Tomas Kocisky and Ryo Nakashima and Paul Natsev and Mike Kwong and Ionel Gog and Chiyuan Zhang and Sander Dieleman and Thomas Jimma and Andrey Ryabtsev and Siddhartha Brahma and David Steiner and Dayou Du and Ante Žužul and Mislav Žanić and Mukund Raghavachari and Willi Gierke and Zeyu Zheng and Dessie Petrova and Yann Dauphin and Yuchuan Liu and Ido Kessler and Steven Hand and Chris Duvarney and Seokhwan Kim and Hyo Lee and Léonard Hussenot and Jeffrey Hui and Josh Smith and Deepali Jain and Jiawei Xia and Gaurav Singh Tomar and Keyvan Amiri and Du Phan and Fabian Fuchs and Tobias Weyand and Nenad Tomasev and Alexandra Cordell and Xin Liu and Jonathan Mallinson and Pankaj Joshi and Andy Crawford and Arun Suggala and Steve Chien and Nick Fernando and Mariella Sanchez-Vargas and Duncan Williams and Phil Crone and Xiyang Luo and Igor Karpov and Jyn Shan and Terry Thurk and Robin Strudel and Paul Voigtlaender and Piyush Patil and Tim Dozat and Ali Khodaei and Sahil Singla and Piotr Ambroszczyk and Qiyin Wu and Yifan Chang and Brian Roark and Chaitra Hegde and Tianli Ding and Angelos Filos and Zhongru Wu and André Susano Pinto and Shuang Liu and Saarthak Khanna and Aditya Pandey and Siobhan Mcloughlin and Qiujia Li and Sam Haves and Allan Zhou and Elena Buchatskaya and Isabel Leal and Peter de Boursac and Nami Akazawa and Nina Anderson and Terry Chen and Krishna Somandepalli and Chen Liang and Sheela Goenka and Stephanie Winkler and Alexander Grushetsky and Yifan Ding and Jamie Smith and Fan Ye and Jordi Pont-Tuset and Eric Li and Ruichao Li and Tomer Golany and Dawid Wegner and Tao Jiang and Omer Barak and Yuan Shangguan and Eszter Vértes and Renee Wong and Jörg Bornschein and Alex Tudor and Michele Bevilacqua and Tom Schaul and Ankit Singh Rawat and Yang Zhao and Kyriakos Axiotis and Lei Meng and Cory McLean and Jonathan Lai and Jennifer Beattie and Nate Kushman and Yaxin Liu and Blair Kutzman and Fiona Lang and Jingchen Ye and Praneeth Netrapalli and Pushkar Mishra and Myriam Khan and Megha Goel and Rob Willoughby and David Tian and Honglei Zhuang and JD Chen and Zak Tsai and Tasos Kementsietsidis and Arjun Khare and James Keeling and Keyang Xu and Nathan Waters and Florent Altché and Ashok Popat and Bhavishya Mittal and David Saxton and Dalia El Badawy and Michael Mathieu and Zheng Zheng and Hao Zhou and Nishant Ranka and Richard Shin and Qingnan Duan and Tim Salimans and Ioana Mihailescu and Uri Shaham and Ming-Wei Chang and Yannis Assael and Nishanth Dikkala and Martin Izzard and Vincent Cohen-Addad and Cat Graves and Vlad Feinberg and Grace Chung and DJ Strouse and Danny Karmon and Sahand Sharifzadeh and Zoe Ashwood and Khiem Pham and Jon Blanton and Alex Vasiloff and Jarred Barber and Mark Geller and Aurick Zhou and Fedir Zubach and Tzu-Kuo Huang and Lei Zhang and Himanshu Gupta and Matt Young and Julia Proskurnia and Ronny Votel and Valentin Gabeur and Gabriel Barcik and Aditya Tripathi and Hongkun Yu and Geng Yan and Beer Changpinyo and Filip Pavetić and Amy Coyle and Yasuhisa Fujii and Jorge Gonzalez Mendez and Tianhao Zhou and Harish Rajamani and Blake Hechtman and Eddie Cao and Da-Cheng Juan and Yi-Xuan Tan and Valentin Dalibard and Yilun Du and Natalie Clay and Kaisheng Yao and Wenhao Jia and Dimple Vijaykumar and Yuxiang Zhou and Xinyi Bai and Wei-Chih Hung and Steven Pecht and Georgi Todorov and Nikhil Khadke and Pramod Gupta and Preethi Lahoti and Arnaud Autef and Karthik Duddu and James Lee-Thorp and Alexander Bykovsky and Tautvydas Misiunas and Sebastian Flennerhag and Santhosh Thangaraj and Jed McGiffin and Zack Nado and Markus Kunesch and Andreas Noever and Amir Hertz and Marco Liang and Victor Stone and Evan Palmer and Samira Daruki and Arijit Pramanik and Siim Põder and Austin Kyker and Mina Khan and Evgeny Sluzhaev and Marvin Ritter and Avraham Ruderman and Wenlei Zhou and Chirag Nagpal and Kiran Vodrahalli and George Necula and Paul Barham and Ellie Pavlick and Jay Hartford and Izhak Shafran and Long Zhao and Maciej Mikuła and Tom Eccles and Hidetoshi Shimokawa and Kanav Garg and Luke Vilnis and Hanwen Chen and Ilia Shumailov and Kuang-Huei Lee and Abdelrahman Abdelhamed and Meiyan Xie and Vered Cohen and Ester Hlavnova and Dan Malkin and Chawin Sitawarin and James Lottes and Pauline Coquinot and Tianli Yu and Sandeep Kumar and Jingwei Zhang and Aroma Mahendru and Zafarali Ahmed and James Martens and Tao Chen and Aviel Boag and Daiyi Peng and Coline Devin and Arseniy Klimovskiy and Mary Phuong and Danny Vainstein and Jin Xie and Bhuvana Ramabhadran and Nathan Howard and Xinxin Yu and Gitartha Goswami and Jingyu Cui and Sam Shleifer and Mario Pinto and Chih-Kuan Yeh and Ming-Hsuan Yang and Sara Javanmardi and Dan Ethier and Chace Lee and Jordi Orbay and Suyog Kotecha and Carla Bromberg and Pete Shaw and James Thornton and Adi Gerzi Rosenthal and Shane Gu and Matt Thomas and Ian Gemp and Aditya Ayyar and Asahi Ushio and Aarush Selvan and Joel Wee and Chenxi Liu and Maryam Majzoubi and Weiren Yu and Jake Abernethy and Tyler Liechty and Renke Pan and Hoang Nguyen and Qiong and Hu and Sarah Perrin and Abhinav Arora and Emily Pitler and Weiyi Wang and Kaushik Shivakumar and Flavien Prost and Ben Limonchik and Jing Wang and Yi Gao and Timothee Cour and Shyamal Buch and Huan Gui and Maria Ivanova and Philipp Neubeck and Kelvin Chan and Lucy Kim and Huizhong Chen and Naman Goyal and Da-Woon Chung and Lu Liu and Yao Su and Anastasia Petrushkina and Jiajun Shen and Armand Joulin and Yuanzhong Xu and Stein Xudong Lin and Yana Kulizhskaya and Ciprian Chelba and Shobha Vasudevan and Eli Collins and Vasilisa Bashlovkina and Tony Lu and Doug Fritz and Jongbin Park and Yanqi Zhou and Chen Su and Richard Tanburn and Mikhail Sushkov and Mitchelle Rasquinha and Jinning Li and Jennifer Prendki and Yiming Li and Pallavi LV and Shriya Sharma and Hen Fitoussi and Hui Huang and Andrew Dai and Phuong Dao and Mike Burrows and Henry Prior and Danfeng Qin and Golan Pundak and Lars Lowe Sjoesund and Art Khurshudov and Zhenkai Zhu and Albert Webson and Elizabeth Kemp and Tat Tan and Saurabh Agrawal and Susie Sargsyan and Liqun Cheng and Jim Stephan and Tom Kwiatkowski and David Reid and Arunkumar Byravan and Assaf Hurwitz Michaely and Nicolas Heess and Luowei Zhou and Sonam Goenka and Viral Carpenter and Anselm Levskaya and Bo Wang and Reed Roberts and Rémi Leblond and Sharat Chikkerur and Stav Ginzburg and Max Chang and Robert Riachi and Chuqiao and Xu and Zalán Borsos and Michael Pliskin and Julia Pawar and Morgane Lustman and Hannah Kirkwood and Ankit Anand and Aditi Chaudhary and Norbert Kalb and Kieran Milan and Sean Augenstein and Anna Goldie and Laurel Prince and Karthik Raman and Yanhua Sun and Vivian Xia and Aaron Cohen and Zhouyuan Huo and Josh Camp and Seher Ellis and Lukas Zilka and David Vilar Torres and Lisa Patel and Sho Arora and Betty Chan and Jonas Adler and Kareem Ayoub and Jacky Liang and Fayaz Jamil and Jiepu Jiang and Simon Baumgartner and Haitian Sun and Yael Karov and Yaroslav Akulov and Hui Zheng and Irene Cai and Claudio Fantacci and James Rubin and Alex Rav Acha and Mengchao Wang and Nina D'Souza and Rohit Sathyanarayana and Shengyang Dai and Simon Rowe and Andrey Simanovsky and Omer Goldman and Yuheng Kuang and Xiaoyue Pan and Andrew Rosenberg and Tania Rojas-Esponda and Praneet Dutta and Amy Zeng and Irina Jurenka and Greg Farquhar and Yamini Bansal and Shariq Iqbal and Becca Roelofs and Ga-Young Joung and Parker Beak and Changwan Ryu and Ryan Poplin and Yan Wu and Jean-Baptiste Alayrac and Senaka Buthpitiya and Olaf Ronneberger and Caleb Habtegebriel and Wei Li and Paul Cavallaro and Aurora Wei and Guy Bensky and Timo Denk and Harish Ganapathy and Jeff Stanway and Pratik Joshi and Francesco Bertolini and Jessica Lo and Olivia Ma and Zachary Charles and Geta Sampemane and Himanshu Sahni and Xu Chen and Harry Askham and David Gaddy and Peter Young and Jiewen Tan and Matan Eyal and Arthur Bražinskas and Li Zhong and Zhichun Wu and Mark Epstein and Kai Bailey and Andrew Hard and Kamyu Lee and Sasha Goldshtein and Alex Ruiz and Mohammed Badawi and Matthias Lochbrunner and JK Kearns and Ashley Brown and Fabio Pardo and Theophane Weber and Haichuan Yang and Pan-Pan Jiang and Berkin Akin and Zhao Fu and Marcus Wainwright and Chi Zou and Meenu Gaba and Pierre-Antoine Manzagol and Wendy Kan and Yang Song and Karina Zainullina and Rui Lin and Jeongwoo Ko and Salil Deshmukh and Apoorv Jindal and James Svensson and Divya Tyam and Heri Zhao and Christine Kaeser-Chen and Scott Baird and Pooya Moradi and Jamie Hall and Qiuchen Guo and Vincent Tsang and Bowen Liang and Fernando Pereira and Suhas Ganesh and Ivan Korotkov and Jakub Adamek and Sridhar Thiagarajan and Vinh Tran and Charles Chen and Chris Tar and Sanil Jain and Ishita Dasgupta and Taylan Bilal and David Reitter and Kai Zhao and Giulia Vezzani and Yasmin Gehman and Pulkit Mehta and Lauren Beltrone and Xerxes Dotiwalla and Sergio Guadarrama and Zaheer Abbas and Stefani Karp and Petko Georgiev and Chun-Sung Ferng and Marc Brockschmidt and Liqian Peng and Christoph Hirnschall and Vikas Verma and Yingying Bi and Ying Xiao and Avigail Dabush and Kelvin Xu and Phil Wallis and Randall Parker and Qifei Wang and Yang Xu and Ilkin Safarli and Dinesh Tewari and Yin Zhang and Seungyeon Kim and Andrea Gesmundo and Mackenzie Thomas and Sergey Levi and Ahmed Chowdhury and Kanishka Rao and Peter Garst and Sam Conway-Rahman and Helen Ran and Kay McKinney and Zhisheng Xiao and Wenhao Yu and Rohan Agrawal and Axel Stjerngren and Catalin Ionescu and Jingjing Chen and Vivek Sharma and Justin Chiu and Fei Liu and Ken Franko and Clayton Sanford and Xingyu Cai and Paul Michel and Sanjay Ganapathy and Jane Labanowski and Zachary Garrett and Ben Vargas and Sean Sun and Bryan Gale and Thomas Buschmann and Guillaume Desjardins and Nimesh Ghelani and Palak Jain and Mudit Verma and Chulayuth Asawaroengchai and Julian Eisenschlos and Jitendra Harlalka and Hideto Kazawa and Don Metzler and Joshua Howland and Ying Jian and Jake Ades and Viral Shah and Tynan Gangwani and Seungji Lee and Roman Ring and Steven M. Hernandez and Dean Reich and Amer Sinha and Ashutosh Sathe and Joe Kovac and Ashleah Gill and Ajay Kannan and Andrea D'olimpio and Martin Sevenich and Jay Whang and Been Kim and Khe Chai Sim and Jilin Chen and Jiageng Zhang and Shuba Lall and Yossi Matias and Bill Jia and Abe Friesen and Sara Nasso and Ashish Thapliyal and Bryan Perozzi and Ting Yu and Anna Shekhawat and Safeen Huda and Peter Grabowski and Eric Wang and Ashwin Sreevatsa and Hilal Dib and Mehadi Hassen and Parker Schuh and Vedrana Milutinovic and Chris Welty and Michael Quinn and Ali Shah and Bangju Wang and Gabe Barth-Maron and Justin Frye and Natalie Axelsson and Tao Zhu and Yukun Ma and Irene Giannoumis and Hanie Sedghi and Chang Ye and Yi Luan and Kevin Aydin and Bilva Chandra and Vivek Sampathkumar and Ronny Huang and Victor Lavrenko and Ahmed Eleryan and Zhi Hong and Steven Hansen and Sara Mc Carthy and Bidisha Samanta and Domagoj Ćevid and Xin Wang and Fangtao Li and Michael Voznesensky and Matt Hoffman and Andreas Terzis and Vikash Sehwag and Gil Fidel and Luheng He and Mu Cai and Yanzhang He and Alex Feng and Martin Nikoltchev and Samrat Phatale and Jason Chase and Rory Lawton and Ming Zhang and Tom Ouyang and Manuel Tragut and Mehdi Hafezi Manshadi and Arjun Narayanan and Jiaming Shen and Xu Gao and Tolga Bolukbasi and Nick Roy and Xin Li and Daniel Golovin and Liviu Panait and Zhen Qin and Guangxing Han and Thomas Anthony and Sneha Kudugunta and Viorica Patraucean and Aniket Ray and Xinyun Chen and Xiaochen Yang and Tanuj Bhatia and Pranav Talluri and Alex Morris and Andrija Ražnatović and Bethanie Brownfield and James An and Sheng Peng and Patrick Kane and Ce Zheng and Nico Duduta and Joshua Kessinger and James Noraky and Siqi Liu and Keran Rong and Petar Veličković and Keith Rush and Alex Goldin and Fanny Wei and Shiva Mohan Reddy Garlapati and Caroline Pantofaru and Okwan Kwon and Jianmo Ni and Eric Noland and Julia Di Trapani and Françoise Beaufays and Abhijit Guha Roy and Yinlam Chow and Aybuke Turker and Geoffrey Cideron and Lantao Mei and Jon Clark and Qingyun Dou and Matko Bošnjak and Ralph Leith and Yuqing Du and Amir Yazdanbakhsh and Milad Nasr and Chester Kwak and Suraj Satishkumar Sheth and Alex Kaskasoli and Ankesh Anand and Balaji Lakshminarayanan and Sammy Jerome and David Bieber and Chun-Te Chu and Alexandre Senges and Tianxiao Shen and Mukund Sridhar and Ndaba Ndebele and Benjamin Beyret and Shakir Mohamed and Mia Chen and Markus Freitag and Jiaxian Guo and Luyang Liu and Paul Roit and Heng Chen and Shen Yan and Tom Stone and JD Co-Reyes and Jeremy Cole and Salvatore Scellato and Shekoofeh Azizi and Hadi Hashemi and Alicia Jin and Anand Iyer and Marcella Valentine and András György and Arun Ahuja and Daniel Hernandez Diaz and Chen-Yu Lee and Nathan Clement and Weize Kong and Drew Garmon and Ishaan Watts and Kush Bhatia and Khyatti Gupta and Matt Miecnikowski and Hugo Vallet and Ankur Taly and Edward Loper and Saket Joshi and James Atwood and Jo Chick and Mark Collier and Fotis Iliopoulos and Ryan Trostle and Beliz Gunel and Ramiro Leal-Cavazos and Arnar Mar Hrafnkelsson and Michael Guzman and Xiaoen Ju and Andy Forbes and Jesse Emond and Kushal Chauhan and Ben Caine and Li Xiao and Wenjun Zeng and Alexandre Moufarek and Daniel Murphy and Maya Meng and Nitish Gupta and Felix Riedel and Anil Das and Elijah Lawal and Shashi Narayan and Tiberiu Sosea and James Swirhun and Linda Friso and Behnam Neyshabur and Jing Lu and Sertan Girgin and Michael Wunder and Edouard Yvinec and Aroonalok Pyne and Victor Carbune and Shruti Rijhwani and Yang Guo and Tulsee Doshi and Anton Briukhov and Max Bain and Ayal Hitron and Xuanhui Wang and Ashish Gupta and Ke Chen and Cosmo Du and Weiyang Zhang and Dhruv Shah and Arjun Akula and Max Dylla and Ashyana Kachra and Weicheng Kuo and Tingting Zou and Lily Wang and Luyao Xu and Jifan Zhu and Justin Snyder and Sachit Menon and Orhan Firat and Igor Mordatch and Yuan Yuan and Natalia Ponomareva and Rory Blevins and Lawrence Moore and Weijun Wang and Phil Chen and Martin Scholz and Artur Dwornik and Jason Lin and Sicheng Li and Diego Antognini and Te I and Xiaodan Song and Matt Miller and Uday Kalra and Adam Raveret and Oscar Akerlund and Felix Wu and Andrew Nystrom and Namrata Godbole and Tianqi Liu and Hannah DeBalsi and Jewel Zhao and Buhuang Liu and Avi Caciularu and Lauren Lax and Urvashi Khandelwal and Victoria Langston and Eric Bailey and Silvio Lattanzi and Yufei Wang and Neel Kovelamudi and Sneha Mondal and Guru Guruganesh and Nan Hua and Ofir Roval and Paweł Wesołowski and Rishikesh Ingale and Jonathan Halcrow and Tim Sohn and Christof Angermueller and Bahram Raad and Eli Stickgold and Eva Lu and Alec Kosik and Jing Xie and Timothy Lillicrap and Austin Huang and Lydia Lihui Zhang and Dominik Paulus and Clement Farabet and Alex Wertheim and Bing Wang and Rishabh Joshi and Chu-ling Ko and Yonghui Wu and Shubham Agrawal and Lily Lin and XiangHai Sheng and Peter Sung and Tyler Breland-King and Christina Butterfield and Swapnil Gawde and Sumeet Singh and Qiao Zhang and Raj Apte and Shilpa Shetty and Adrian Hutter and Tao Li and Elizabeth Salesky and Federico Lebron and Jonni Kanerva and Michela Paganini and Arthur Nguyen and Rohith Vallu and Jan-Thorsten Peter and Sarmishta Velury and David Kao and Jay Hoover and Anna Bortsova and Colton Bishop and Shoshana Jakobovits and Alessandro Agostini and Alekh Agarwal and Chang Liu and Charles Kwong and Sasan Tavakkol and Ioana Bica and Alex Greve and Anirudh GP and Jake Marcus and Le Hou and Tom Duerig and Rivka Moroshko and Dave Lacey and Andy Davis and Julien Amelot and Guohui Wang and Frank Kim and Theofilos Strinopoulos and Hui Wan and Charline Le Lan and Shankar Krishnan and Haotian Tang and Peter Humphreys and Junwen Bai and Idan Heimlich Shtacher and Diego Machado and Chenxi Pang and Ken Burke and Dangyi Liu and Renga Aravamudhan and Yue Song and Ed Hirst and Abhimanyu Singh and Brendan Jou and Liang Bai and Francesco Piccinno and Chuyuan Kelly Fu and Robin Alazard and Barak Meiri and Daniel Winter and Charlie Chen and Mingda Zhang and Jens Heitkaemper and John Lambert and Jinhyuk Lee and Alexander Frömmgen and Sergey Rogulenko and Pranav Nair and Paul Niemczyk and Anton Bulyenov and Bibo Xu and Hadar Shemtov and Morteza Zadimoghaddam and Serge Toropov and Mateo Wirth and Hanjun Dai and Sreenivas Gollapudi and Daniel Zheng and Alex Kurakin and Chansoo Lee and Kalesha Bullard and Nicolas Serrano and Ivana Balazevic and Yang Li and Johan Schalkwyk and Mark Murphy and Mingyang Zhang and Kevin Sequeira and Romina Datta and Nishant Agrawal and Charles Sutton and Nithya Attaluri and Mencher Chiang and Wael Farhan and Gregory Thornton and Kate Lin and Travis Choma and Hung Nguyen and Kingshuk Dasgupta and Dirk Robinson and Iulia Comşa and Michael Riley and Arjun Pillai and Basil Mustafa and Ben Golan and Amir Zandieh and Jean-Baptiste Lespiau and Billy Porter and David Ross and Sujeevan Rajayogam and Mohit Agarwal and Subhashini Venugopalan and Bobak Shahriari and Qiqi Yan and Hao Xu and Taylor Tobin and Pavel Dubov and Hongzhi Shi and Adrià Recasens and Anton Kovsharov and Sebastian Borgeaud and Lucio Dery and Shanthal Vasanth and Elena Gribovskaya and Linhai Qiu and Mahdis Mahdieh and Wojtek Skut and Elizabeth Nielsen and CJ Zheng and Adams Yu and Carrie Grimes Bostock and Shaleen Gupta and Aaron Archer and Chris Rawles and Elinor Davies and Alexey Svyatkovskiy and Tomy Tsai and Yoni Halpern and Christian Reisswig and Bartek Wydrowski and Bo Chang and Joan Puigcerver and Mor Hazan Taege and Jian Li and Eva Schnider and Xinjian Li and Dragos Dena and Yunhan Xu and Umesh Telang and Tianze Shi and Heiga Zen and Kyle Kastner and Yeongil Ko and Neesha Subramaniam and Aviral Kumar and Pete Blois and Zhuyun Dai and John Wieting and Yifeng Lu and Yoel Zeldes and Tian Xie and Anja Hauth and Alexandru Ţifrea and Yuqi Li and Sam El-Husseini and Dan Abolafia and Howard Zhou and Wen Ding and Sahra Ghalebikesabi and Carlos Guía and Andrii Maksai and Ágoston Weisz and Sercan Arik and Nick Sukhanov and Aga Świetlik and Xuhui Jia and Luo Yu and Weiyue Wang and Mark Brand and Dawn Bloxwich and Sean Kirmani and Zhe Chen and Alec Go and Pablo Sprechmann and Nithish Kannen and Alen Carin and Paramjit Sandhu and Isabel Edkins and Leslie Nooteboom and Jai Gupta and Loren Maggiore and Javad Azizi and Yael Pritch and Pengcheng Yin and Mansi Gupta and Danny Tarlow and Duncan Smith and Desi Ivanov and Mohammad Babaeizadeh and Ankita Goel and Satish Kambala and Grace Chu and Matej Kastelic and Michelle Liu and Hagen Soltau and Austin Stone and Shivani Agrawal and Min Kim and Kedar Soparkar and Srinivas Tadepalli and Oskar Bunyan and Rachel Soh and Arvind Kannan and DY Kim and Blake JianHang Chen and Afief Halumi and Sudeshna Roy and Yulong Wang and Olcan Sercinoglu and Gena Gibson and Sijal Bhatnagar and Motoki Sano and Daniel von Dincklage and Qingchun Ren and Blagoj Mitrevski and Mirek Olšák and Jennifer She and Carl Doersch and Jilei and Wang and Bingyuan Liu and Qijun Tan and Tamar Yakar and Tris Warkentin and Alex Ramirez and Carl Lebsack and Josh Dillon and Rajiv Mathews and Tom Cobley and Zelin Wu and Zhuoyuan Chen and Jon Simon and Swaroop Nath and Tara Sainath and Alexei Bendebury and Ryan Julian and Bharath Mankalale and Daria Ćurko and Paulo Zacchello and Adam R. Brown and Kiranbir Sodhia and Heidi Howard and Sergi Caelles and Abhinav Gupta and Gareth Evans and Anna Bulanova and Lesley Katzen and Roman Goldenberg and Anton Tsitsulin and Joe Stanton and Benoit Schillings and Vitaly Kovalev and Corey Fry and Rushin Shah and Kuo Lin and Shyam Upadhyay and Cheng Li and Soroush Radpour and Marcello Maggioni and Jing Xiong and Lukas Haas and Jenny Brennan and Aishwarya Kamath and Nikolay Savinov and Arsha Nagrani and Trevor Yacovone and Ryan Kappedal and Kostas Andriopoulos and Li Lao and YaGuang Li and Grigory Rozhdestvenskiy and Kazuma Hashimoto and Andrew Audibert and Sophia Austin and Daniel Rodriguez and Anian Ruoss and Garrett Honke and Deep Karkhanis and Xi Xiong and Qing Wei and James Huang and Zhaoqi Leng and Vittal Premachandran and Stan Bileschi and Georgios Evangelopoulos and Thomas Mensink and Jay Pavagadhi and Denis Teplyashin and Paul Chang and Linting Xue and Garrett Tanzer and Sally Goldman and Kaushal Patel and Shixin Li and Jeremy Wiesner and Ivy Zheng and Ian Stewart-Binks and Jie Han and Zhi Li and Liangchen Luo and Karel Lenc and Mario Lučić and Fuzhao Xue and Ryan Mullins and Alexey Guseynov and Chung-Ching Chang and Isaac Galatzer-Levy and Adam Zhang and Garrett Bingham and Grace Hu and Ale Hartman and Yue Ma and Jordan Griffith and Alex Irpan and Carey Radebaugh and Summer Yue and Lijie Fan and Victor Ungureanu and Christina Sorokin and Hannah Teufel and Peiran Li and Rohan Anil and Dimitris Paparas and Todd Wang and Chu-Cheng Lin and Hui Peng and Megan Shum and Goran Petrovic and Demetra Brady and Richard Nguyen and Klaus Macherey and Zhihao Li and Harman Singh and Madhavi Yenugula and Mariko Iinuma and Xinyi Chen and Kavya Kopparapu and Alexey Stern and Shachi Dave and Chandu Thekkath and Florence Perot and Anurag Kumar and Fangda Li and Yang Xiao and Matthew Bilotti and Mohammad Hossein Bateni and Isaac Noble and Lisa Lee and Amelio Vázquez-Reina and Julian Salazar and Xiaomeng Yang and Boyu Wang and Ela Gruzewska and Anand Rao and Sindhu Raghuram and Zheng Xu and Eyal Ben-David and Jieru Mei and Sid Dalmia and Zhaoyi Zhang and Yuchen Liu and Gagan Bansal and Helena Pankov and Steven Schwarcz and Andrea Burns and Christine Chan and Sumit Sanghai and Ricky Liang and Ethan Liang and Antoine He and Amy Stuart and Arun Narayanan and Yukun Zhu and Christian Frank and Bahar Fatemi and Amit Sabne and Oran Lang and Indro Bhattacharya and Shane Settle and Maria Wang and Brendan McMahan and Andrea Tacchetti and Livio Baldini Soares and Majid Hadian and Serkan Cabi and Timothy Chung and Nikita Putikhin and Gang Li and Jeremy Chen and Austin Tarango and Henryk Michalewski and Mehran Kazemi and Hussain Masoom and Hila Sheftel and Rakesh Shivanna and Archita Vadali and Ramona Comanescu and Doug Reid and Joss Moore and Arvind Neelakantan and Michaël Sander and Jonathan Herzig and Aviv Rosenberg and Mostafa Dehghani and JD Choi and Michael Fink and Reid Hayes and Eric Ge and Shitao Weng and Chia-Hua Ho and John Karro and Kalpesh Krishna and Lam Nguyen Thiet and Amy Skerry-Ryan and Daniel Eppens and Marco Andreetto and Navin Sarma and Silvano Bonacina and Burcu Karagol Ayan and Megha Nawhal and Zhihao Shan and Mike Dusenberry and Shantanu Thakoor and Sagar Gubbi and Duc Dung Nguyen and Reut Tsarfaty and Samuel Albanie and Jovana Mitrović and Meet Gandhi and Bo-Juen Chen and Alessandro Epasto and Georgi Stephanov and Ye Jin and Samuel Gehman and Aida Amini and Jack Weber and Feryal Behbahani and Shawn Xu and Miltos Allamanis and Xi Chen and Myle Ott and Claire Sha and Michal Jastrzebski and Hang Qi and David Greene and Xinyi Wu and Abodunrinwa Toki and Daniel Vlasic and Jane Shapiro and Ragha Kotikalapudi and Zhe Shen and Takaaki Saeki and Sirui Xie and Albin Cassirer and Shikhar Bharadwaj and Tatsuya Kiyono and Srinadh Bhojanapalli and Elan Rosenfeld and Sam Ritter and Jieming Mao and João Gabriel Oliveira and Zoltan Egyed and Bernd Bandemer and Emilio Parisotto and Keisuke Kinoshita and Juliette Pluto and Petros Maniatis and Steve Li and Yaohui Guo and Golnaz Ghiasi and Jean Tarbouriech and Srimon Chatterjee and Julie Jin and Katrina and Xu and Jennimaria Palomaki and Séb Arnold and Madhavi Sewak and Federico Piccinini and Mohit Sharma and Ben Albrecht and Sean Purser-haskell and Ashwin Vaswani and Chongyan Chen and Matheus Wisniewski and Qin Cao and John Aslanides and Nguyet Minh Phu and Maximilian Sieb and Lauren Agubuzu and Anne Zheng and Daniel Sohn and Marco Selvi and Anders Andreassen and Krishan Subudhi and Prem Eruvbetine and Oliver Woodman and Tomas Mery and Sebastian Krause and Xiaoqi Ren and Xiao Ma and Jincheng Luo and Dawn Chen and Wei Fan and Henry Griffiths and Christian Schuler and Alice Li and Shujian Zhang and Jean-Michel Sarr and Shixin Luo and Riccardo Patana and Matthew Watson and Dani Naboulsi and Michael Collins and Sailesh Sidhwani and Emiel Hoogeboom and Sharon Silver and Emily Caveness and Xiaokai Zhao and Mikel Rodriguez and Maxine Deines and Libin Bai and Patrick Griffin and Marco Tagliasacchi and Emily Xue and Spandana Raj Babbula and Bo Pang and Nan Ding and Gloria Shen and Elijah Peake and Remi Crocker and Shubha Srinivas Raghvendra and Danny Swisher and Woohyun Han and Richa Singh and Ling Wu and Vladimir Pchelin and Tsendsuren Munkhdalai and Dana Alon and Geoff Bacon and Efren Robles and Jannis Bulian and Melvin Johnson and George Powell and Felipe Tiengo Ferreira and Yaoyiran Li and Frederik Benzing and Mihajlo Velimirović and Hubert Soyer and William Kong and Tony and Nguyên and Zhen Yang and Jeremiah Liu and Joost van Amersfoort and Daniel Gillick and Baochen Sun and Nathalie Rauschmayr and Katie Zhang and Serena Zhan and Tao Zhou and Alexey Frolov and Chengrun Yang and Denis Vnukov and Louis Rouillard and Hongji Li and Amol Mandhane and Nova Fallen and Rajesh Venkataraman and Clara Huiyi Hu and Jennifer Brennan and Jenny Lee and Jerry Chang and Martin Sundermeyer and Zhufeng Pan and Rosemary Ke and Simon Tong and Alex Fabrikant and William Bono and Jindong Gu and Ryan Foley and Yiran Mao and Manolis Delakis and Dhruva Bhaswar and Roy Frostig and Nick Li and Avital Zipori and Cath Hope and Olga Kozlova and Swaroop Mishra and Josip Djolonga and Craig Schiff and Majd Al Merey and Eleftheria Briakou and Peter Morgan and Andy Wan and Avinatan Hassidim and RJ Skerry-Ryan and Kuntal Sengupta and Mary Jasarevic and Praveen Kallakuri and Paige Kunkle and Hannah Brennan and Tom Lieber and Hassan Mansoor and Julian Walker and Bing Zhang and Annie Xie and Goran Žužić and Adaeze Chukwuka and Alex Druinsky and Donghyun Cho and Rui Yao and Ferjad Naeem and Shiraz Butt and Eunyoung Kim and Zhipeng Jia and Mandy Jordan and Adam Lelkes and Mark Kurzeja and Sophie Wang and James Zhao and Andrew Over and Abhishek Chakladar and Marcel Prasetya and Neha Jha and Sriram Ganapathy and Yale Cong and Prakash Shroff and Carl Saroufim and Sobhan Miryoosefi and Mohamed Hammad and Tajwar Nasir and Weijuan Xi and Yang Gao and Young Maeng and Ben Hora and Chin-Yi Cheng and Parisa Haghani and Yoad Lewenberg and Caden Lu and Martin Matysiak and Naina Raisinghani and Huiyu Wang and Lexi Baugher and Rahul Sukthankar and Minh Giang and John Schultz and Noah Fiedel and Minmin Chen and Cheng-Chun Lee and Tapomay Dey and Hao Zheng and Shachi Paul and Celine Smith and Andy Ly and Yicheng Wang and Rishabh Bansal and Bartek Perz and Susanna Ricco and Stasha Blank and Vaishakh Keshava and Deepak Sharma and Marvin Chow and Kunal Lad and Komal Jalan and Simon Osindero and Craig Swanson and Jacob Scott and Anastasija Ilić and Xiaowei Li and Siddhartha Reddy Jonnalagadda and Afzal Shama Soudagar and Yan Xiong and Bat-Orgil Batsaikhan and Daniel Jarrett and Naveen Kumar and Maulik Shah and Matt Lawlor and Austin Waters and Mark Graham and Rhys May and Sabela Ramos and Sandra Lefdal and Zeynep Cankara and Nacho Cano and Brendan O'Donoghue and Jed Borovik and Frederick Liu and Jordan Grimstad and Mahmoud Alnahlawi and Katerina Tsihlas and Tom Hudson and Nikolai Grigorev and Yiling Jia and Terry Huang and Tobenna Peter Igwe and Sergei Lebedev and Xiaodan Tang and Igor Krivokon and Frankie Garcia and Melissa Tan and Eric Jia and Peter Stys and Shikhar Vashishth and Yu Liang and Balaji Venkatraman and Chenjie Gu and Anastasios Kementsietsidis and Chen Zhu and Junehyuk Jung and Yunfei Bai and Mohammad Javad Hosseini and Faruk Ahmed and Aditya Gupta and Xin Yuan and Shereen Ashraf and Shitij Nigam and Gautam Vasudevan and Pranjal Awasthi and Adi Mayrav Gilady and Zelda Mariet and Ramy Eskander and Haiguang Li and Hexiang Hu and Guillermo Garrido and Philippe Schlattner and George Zhang and Rohun Saxena and Petar Dević and Kritika Muralidharan and Ashwin Murthy and Yiqian Zhou and Min Choi and Arissa Wongpanich and Zhengdong Wang and Premal Shah and Yuntao Xu and Yiling Huang and Stephen Spencer and Alice Chen and James Cohan and Junjie Wang and Jonathan Tompson and Junru Wu and Ruba Haroun and Haiqiong Li and Blanca Huergo and Fan Yang and Tongxin Yin and James Wendt and Michael Bendersky and Rahma Chaabouni and Javier Snaider and Johan Ferret and Abhishek Jindal and Tara Thompson and Andrew Xue and Will Bishop and Shubham Milind Phal and Archit Sharma and Yunhsuan Sung and Prabakar Radhakrishnan and Mo Shomrat and Reeve Ingle and Roopali Vij and Justin Gilmer and Mihai Dorin Istin and Sam Sobell and Yang Lu and Emily Nottage and Dorsa Sadigh and Jeremiah Willcock and Tingnan Zhang and Steve Xu and Sasha Brown and Katherine Lee and Gary Wang and Yun Zhu and Yi Tay and Cheolmin Kim and Audrey Gutierrez and Abhanshu Sharma and Yongqin Xian and Sungyong Seo and Claire Cui and Elena Pochernina and Cip Baetu and Krzysztof Jastrzębski and Mimi Ly and Mohamed Elhawaty and Dan Suh and Eren Sezener and Pidong Wang and Nancy Yuen and George Tucker and Jiahao Cai and Zuguang Yang and Cindy Wang and Alex Muzio and Hai Qian and Jae Yoo and Derek Lockhart and Kevin R. McKee and Mandy Guo and Malika Mehrotra and Artur Mendonça and Sanket Vaibhav Mehta and Sherry Ben and Chetan Tekur and Jiaqi Mu and Muye Zhu and Victoria Krakovna and Hongrae Lee and AJ Maschinot and Sébastien Cevey and HyunJeong Choe and Aijun Bai and Hansa Srinivasan and Derek Gasaway and Nick Young and Patrick Siegler and Dan Holtmann-Rice and Vihari Piratla and Kate Baumli and Roey Yogev and Alex Hofer and Hado van Hasselt and Svetlana Grant and Yuri Chervonyi and David Silver and Andrew Hogue and Ayushi Agarwal and Kathie Wang and Preeti Singh and Four Flynn and Josh Lipschultz and Robert David and Lizzetth Bellot and Yao-Yuan Yang and Long Le and Filippo Graziano and Kate Olszewska and Kevin Hui and Akanksha Maurya and Nikos Parotsidis and Weijie Chen and Tayo Oguntebi and Joe Kelley and Anirudh Baddepudi and Johannes Mauerer and Gregory Shaw and Alex Siegman and Lin Yang and Shravya Shetty and Subhrajit Roy and Yunting Song and Wojciech Stokowiec and Ryan Burnell and Omkar Savant and Robert Busa-Fekete and Jin Miao and Samrat Ghosh and Liam MacDermed and Phillip Lippe and Mikhail Dektiarev and Zach Behrman and Fabian Mentzer and Kelvin Nguyen and Meng Wei and Siddharth Verma and Chris Knutsen and Sudeep Dasari and Zhipeng Yan and Petr Mitrichev and Xingyu Wang and Virat Shejwalkar and Jacob Austin and Srinivas Sunkara and Navneet Potti and Yan Virin and Christian Wright and Gaël Liu and Oriana Riva and Etienne Pot and Greg Kochanski and Quoc Le and Gargi Balasubramaniam and Arka Dhar and Yuguo Liao and Adam Bloniarz and Divyansh Shukla and Elizabeth Cole and Jong Lee and Sheng Zhang and Sushant Kafle and Siddharth Vashishtha and Parsa Mahmoudieh and Grace Chen and Raphael Hoffmann and Pranesh Srinivasan and Agustin Dal Lago and Yoav Ben Shalom and Zi Wang and Michael Elabd and Anuj Sharma and Junhyuk Oh and Suraj Kothawade and Maigo Le and Marianne Monteiro and Shentao Yang and Kaiz Alarakyia and Robert Geirhos and Diana Mincu and Håvard Garnes and Hayato Kobayashi and Soroosh Mariooryad and Kacper Krasowiak and Zhixin and Lai and Shibl Mourad and Mingqiu Wang and Fan Bu and Ophir Aharoni and Guanjie Chen and Abhimanyu Goyal and Vadim Zubov and Ankur Bapna and Elahe Dabir and Nisarg Kothari and Kay Lamerigts and Nicola De Cao and Jeremy Shar and Christopher Yew and Nitish Kulkarni and Dre Mahaarachchi and Mandar Joshi and Zhenhai Zhu and Jared Lichtarge and Yichao Zhou and Hannah Muckenhirn and Vittorio Selo and Oriol Vinyals and Peter Chen and Anthony Brohan and Vaibhav Mehta and Sarah Cogan and Ruth Wang and Ty Geri and Wei-Jen Ko and Wei Chen and Fabio Viola and Keshav Shivam and Lisa Wang and Madeleine Clare Elish and Raluca Ada Popa and Sébastien Pereira and Jianqiao Liu and Raphael Koster and Donnie Kim and Gufeng Zhang and Sayna Ebrahimi and Partha Talukdar and Yanyan Zheng and Petra Poklukar and Ales Mikhalap and Dale Johnson and Anitha Vijayakumar and Mark Omernick and Matt Dibb and Ayush Dubey and Qiong Hu and Apurv Suman and Vaibhav Aggarwal and Ilya Kornakov and Fei Xia and Wing Lowe and Alexey Kolganov and Ted Xiao and Vitaly Nikolaev and Steven Hemingray and Bonnie Li and Joana Iljazi and Mikołaj Rybiński and Ballie Sandhu and Peggy Lu and Thang Luong and Rodolphe Jenatton and Vineetha Govindaraj and Hui and Li and Gabriel Dulac-Arnold and Wonpyo Park and Henry Wang and Abhinit Modi and Jean Pouget-Abadie and Kristina Greller and Rahul Gupta and Robert Berry and Prajit Ramachandran and Jinyu Xie and Liam McCafferty and Jianling Wang and Kilol Gupta and Hyeontaek Lim and Blaž Bratanič and Andy Brock and Ilia Akolzin and Jim Sproch and Dan Karliner and Duhyeon Kim and Adrian Goedeckemeyer and Noam Shazeer and Cordelia Schmid and Daniele Calandriello and Parul Bhatia and Krzysztof Choromanski and Ceslee Montgomery and Dheeru Dua and Ana Ramalho and Helen King and Yue Gao and Lynn Nguyen and David Lindner and Divya Pitta and Oleaser Johnson and Khalid Salama and Diego Ardila and Michael Han and Erin Farnese and Seth Odoom and Ziyue Wang and Xiangzhuo Ding and Norman Rink and Ray Smith and Harshal Tushar Lehri and Eden Cohen and Neera Vats and Tong He and Parthasarathy Gopavarapu and Adam Paszke and Miteyan Patel and Wouter Van Gansbeke and Lucia Loher and Luis Castro and Maria Voitovich and Tamara von Glehn and Nelson George and Simon Niklaus and Zach Eaton-Rosen and Nemanja Rakićević and Erik Jue and Sagi Perel and Carrie Zhang and Yuval Bahat and Angéline Pouget and Zhi Xing and Fantine Huot and Ashish Shenoy and Taylor Bos and Vincent Coriou and Bryan Richter and Natasha Noy and Yaqing Wang and Santiago Ontanon and Siyang Qin and Gleb Makarchuk and Demis Hassabis and Zhuowan Li and Mandar Sharma and Kumaran Venkatesan and Iurii Kemaev and Roxanne Daniel and Shiyu Huang and Saloni Shah and Octavio Ponce and Warren and Chen and Manaal Faruqui and Jialin Wu and Slavica Andačić and Szabolcs Payrits and Daniel McDuff and Tom Hume and Yuan Cao and MH Tessler and Qingze Wang and Yinan Wang and Ivor Rendulic and Eirikur Agustsson and Matthew Johnson and Tanya Lando and Andrew Howard and Sri Gayatri Sundara Padmanabhan and Mayank Daswani and Andrea Banino and Michael Kilgore and Jonathan Heek and Ziwei Ji and Alvaro Caceres and Conglong Li and Nora Kassner and Alexey Vlaskin and Zeyu Liu and Alex Grills and Yanhan Hou and Roykrong Sukkerd and Gowoon Cheon and Nishita Shetty and Larisa Markeeva and Piotr Stanczyk and Tejas Iyer and Yuan Gong and Shawn Gao and Keerthana Gopalakrishnan and Tim Blyth and Malcolm Reynolds and Avishkar Bhoopchand and Misha Bilenko and Dero Gharibian and Vicky Zayats and Aleksandra Faust and Abhinav Singh and Min Ma and Hongyang Jiao and Sudheendra Vijayanarasimhan and Lora Aroyo and Vikas Yadav and Sarah Chakera and Ashwin Kakarla and Vilobh Meshram and Karol Gregor and Gabriela Botea and Evan Senter and Dawei Jia and Geza Kovacs and Neha Sharma and Sebastien Baur and Kai Kang and Yifan He and Lin Zhuo and Marija Kostelac and Itay Laish and Songyou Peng and Louis O'Bryan and Daniel Kasenberg and Girish Ramchandra Rao and Edouard Leurent and Biao Zhang and Sage Stevens and Ana Salazar and Ye Zhang and Ivan Lobov and Jake Walker and Allen Porter and Morgan Redshaw and Han Ke and Abhishek Rao and Alex Lee and Hoi Lam and Michael Moffitt and Jaeyoun Kim and Siyuan Qiao and Terry Koo and Robert Dadashi and Xinying Song and Mukund Sundararajan and Peng Xu and Chizu Kawamoto and Yan Zhong and Clara Barbu and Apoorv Reddy and Mauro Verzetti and Leon Li and George Papamakarios and Hanna Klimczak-Plucińska and Mary Cassin and Koray Kavukcuoglu and Rigel Swavely and Alain Vaucher and Jeffrey Zhao and Ross Hemsley and Michael Tschannen and Heming Ge and Gaurav Menghani and Yang Yu and Natalie Ha and Wei He and Xiao Wu and Maggie Song and Rachel Sterneck and Stefan Zinke and Dan A. Calian and Annie Marsden and Alejandro Cruzado Ruiz and Matteo Hessel and Almog Gueta and Benjamin Lee and Brian Farris and Manish Gupta and Yunjie Li and Mohammad Saleh and Vedant Misra and Kefan Xiao and Piermaria Mendolicchio and Gavin Buttimore and Varvara Krayvanova and Nigamaa Nayakanti and Matthew Wiethoff and Yash Pande and Azalia Mirhoseini and Ni Lao and Jasmine Liu and Yiqing Hua and Angie Chen and Yury Malkov and Dmitry Kalashnikov and Shubham Gupta and Kartik Audhkhasi and Yuexiang Zhai and Sudhindra Kopalle and Prateek Jain and Eran Ofek and Clemens Meyer and Khuslen Baatarsukh and Hana Strejček and Jun Qian and James Freedman and Ricardo Figueira and Michal Sokolik and Olivier Bachem and Raymond Lin and Dia Kharrat and Chris Hidey and Pingmei Xu and Dennis Duan and Yin Li and Muge Ersoy and Richard Everett and Kevin Cen and Rebeca Santamaria-Fernandez and Amir Taubenfeld and Ian Mackinnon and Linda Deng and Polina Zablotskaia and Shashank Viswanadha and Shivanker Goel and Damion Yates and Yunxiao Deng and Peter Choy and Mingqing Chen and Abhishek Sinha and Alex Mossin and Yiming Wang and Arthur Szlam and Susan Hao and Paul Kishan Rubenstein and Metin Toksoz-Exley and Miranda Aperghis and Yin Zhong and Junwhan Ahn and Michael Isard and Olivier Lacombe and Florian Luisier and Chrysovalantis Anastasiou and Yogesh Kalley and Utsav Prabhu and Emma Dunleavy and Shaan Bijwadia and Justin Mao-Jones and Kelly Chen and Rama Pasumarthi and Emily Wood and Adil Dostmohamed and Nate Hurley and Jiri Simsa and Alicia Parrish and Mantas Pajarskas and Matt Harvey and Ondrej Skopek and Yony Kochinski and Javier Rey and Verena Rieser and Denny Zhou and Sun Jae Lee and Trilok Acharya and Guowang Li and Joe Jiang and Xiaofan Zhang and Bryant Gipson and Ethan Mahintorabi and Marco Gelmi and Nima Khajehnouri and Angel Yeh and Kayi Lee and Loic Matthey and Leslie Baker and Trang Pham and Han Fu and Alex Pak and Prakhar Gupta and Cristina Vasconcelos and Adam Sadovsky and Brian Walker and Sissie Hsiao and Patrik Zochbauer and Andreea Marzoca and Noam Velan and Junhao Zeng and Gilles Baechler and Danny Driess and Divya Jain and Yanping Huang and Lizzie Tao and John Maggs and Nir Levine and Jon Schneider and Erika Gemzer and Samuel Petit and Shan Han and Zach Fisher and Dustin Zelle and Courtney Biles and Eugene Ie and Asya Fadeeva and Casper Liu and Juliana Vicente Franco and Adrian Collister and Hao Zhang and Renshen Wang and Ruizhe Zhao and Leandro Kieliger and Kurt Shuster and Rui Zhu and Boqing Gong and Lawrence Chan and Ruoxi Sun and Sujoy Basu and Roland Zimmermann and Jamie Hayes and Abhishek Bapna and Jasper Snoek and Weel Yang and Puranjay Datta and Jad Al Abdallah and Kevin Kilgour and Lu Li and SQ Mah and Yennie Jun and Morgane Rivière and Abhijit Karmarkar and Tammo Spalink and Tao Huang and Lucas Gonzalez and Duc-Hieu Tran and Averi Nowak and John Palowitch and Martin Chadwick and Ellie Talius and Harsh Mehta and Thibault Sellam and Philipp Fränken and Massimo Nicosia and Kyle He and Aditya Kini and David Amos and Sugato Basu and Harrison Jobe and Eleni Shaw and Qiantong Xu and Colin Evans and Daisuke Ikeda and Chaochao Yan and Larry Jin and Lun Wang and Sachin Yadav and Ilia Labzovsky and Ramesh Sampath and Ada Ma and Candice Schumann and Aditya Siddhant and Rohin Shah and John Youssef and Rishabh Agarwal and Natalie Dabney and Alessio Tonioni and Moran Ambar and Jing Li and Isabelle Guyon and Benny Li and David Soergel and Boya Fang and Georgi Karadzhov and Cristian Udrescu and Trieu Trinh and Vikas Raunak and Seb Noury and Dee Guo and Sonal Gupta and Mara Finkelstein and Denis Petek and Lihao Liang and Greg Billock and Pei Sun and David Wood and Yiwen Song and Xiaobin Yu and Tatiana Matejovicova and Regev Cohen and Kalyan Andra and David D'Ambrosio and Zhiwei Deng and Vincent Nallatamby and Ebrahim Songhori and Rumen Dangovski and Andrew Lampinen and Pankil Botadra and Adam Hillier and Jiawei Cao and Nagabhushan Baddi and Adhi Kuncoro and Toshihiro Yoshino and Ankit Bhagatwala and Marcáurelio Ranzato and Rylan Schaeffer and Tianlin Liu and Shuai Ye and Obaid Sarvana and John Nham and Chenkai Kuang and Isabel Gao and Jinoo Baek and Shubham Mittal and Ayzaan Wahid and Anita Gergely and Bin Ni and Josh Feldman and Carrie Muir and Pascal Lamblin and Wolfgang Macherey and Ethan Dyer and Logan Kilpatrick and Víctor Campos and Mukul Bhutani and Stanislav Fort and Yanif Ahmad and Aliaksei Severyn and Kleopatra Chatziprimou and Oleksandr Ferludin and Mason Dimarco and Aditya Kusupati and Joe Heyward and Dan Bahir and Kevin Villela and Katie Millican and Dror Marcus and Sanaz Bahargam and Caglar Unlu and Nicholas Roth and Zichuan Wei and Siddharth Gopal and Deepanway Ghoshal and Edward Lee and Sharon Lin and Jennie Lees and Dayeong Lee and Anahita Hosseini and Connie Fan and Seth Neel and Marcus Wu and Yasemin Altun and Honglong Cai and Enrique Piqueras and Josh Woodward and Alessandro Bissacco and Salem Haykal and Mahyar Bordbar and Prasha Sundaram and Sarah Hodkinson and Daniel Toyama and George Polovets and Austin Myers and Anu Sinha and Tomer Levinboim and Kashyap Krishnakumar and Rachita Chhaparia and Tatiana Sholokhova and Nitesh Bharadwaj Gundavarapu and Ganesh Jawahar and Haroon Qureshi and Jieru Hu and Nikola Momchev and Matthew Rahtz and Renjie Wu and Aishwarya P S and Kedar Dhamdhere and Meiqi Guo and Umang Gupta and Ali Eslami and Mariano Schain and Michiel Blokzijl and David Welling and Dave Orr and Levent Bolelli and Nicolas Perez-Nieves and Mikhail Sirotenko and Aman Prasad and Arjun Kar and Borja De Balle Pigem and Tayfun Terzi and Gellért Weisz and Dipankar Ghosh and Aditi Mavalankar and Dhruv Madeka and Kaspar Daugaard and Hartwig Adam and Viraj Shah and Dana Berman and Maggie Tran and Steven Baker and Ewa Andrejczuk and Grishma Chole and Ganna Raboshchuk and Mahdi Mirzazadeh and Thais Kagohara and Shimu Wu and Christian Schallhart and Bernett Orlando and Chen Wang and Alban Rrustemi and Hao Xiong and Hao Liu and Arpi Vezer and Nolan Ramsden and Shuo-yiin Chang and Sidharth Mudgal and Yan Li and Nino Vieillard and Yedid Hoshen and Farooq Ahmad and Ambrose Slone and Amy Hua and Natan Potikha and Mirko Rossini and Jon Stritar and Sushant Prakash and Zifeng Wang and Xuanyi Dong and Alireza Nazari and Efrat Nehoran and Kaan Tekelioglu and Yinxiao Li and Kartikeya Badola and Tom Funkhouser and Yuanzhen Li and Varun Yerram and Ramya Ganeshan and Daniel Formoso and Karol Langner and Tian Shi and Huijian Li and Yumeya Yamamori and Amayika Panda and Alaa Saade and Angelo Scorza Scarpati and Chris Breaux and CJ Carey and Zongwei Zhou and Cho-Jui Hsieh and Sophie Bridgers and Alena Butryna and Nishesh Gupta and Vaibhav Tulsyan and Sanghyun Woo and Evgenii Eltyshev and Will Grathwohl and Chanel Parks and Seth Benjamin and Rina Panigrahy and Shenil Dodhia and Daniel De Freitas and Chris Sauer and Will Song and Ferran Alet and Jackson Tolins and Cosmin Paduraru and Xingyi Zhou and Brian Albert and Zizhao Zhang and Lei Shu and Mudit Bansal and Sarah Nguyen and Amir Globerson and Owen Xiao and James Manyika and Tom Hennigan and Rong Rong and Josip Matak and Anton Bakalov and Ankur Sharma and Danila Sinopalnikov and Andrew Pierson and Stephen Roller and Geoff Brown and Mingcen Gao and Toshiyuki Fukuzawa and Amin Ghafouri and Kenny Vassigh and Iain Barr and Zhicheng Wang and Anna Korsun and Rajesh Jayaram and Lijie Ren and Tim Zaman and Samira Khan and Yana Lunts and Dan Deutsch and Dave Uthus and Nitzan Katz and Masha Samsikova and Amr Khalifa and Nikhil Sethi and Jiao Sun and Luming Tang and Uri Alon and Xianghong Luo and Dian Yu and Abhishek Nayyar and Bryce Petrini and Will Truong and Vincent Hellendoorn and Nikolai Chinaev and Chris Alberti and Wei Wang and Jingcao Hu and Vahab Mirrokni and Ananth Balashankar and Avia Aharon and Aahil Mehta and Ahmet Iscen and Joseph Kready and Lucas Manning and Anhad Mohananey and Yuankai Chen and Anshuman Tripathi and Allen Wu and Igor Petrovski and Dawsen Hwang and Martin Baeuml and Shreyas Chandrakaladharan and Yuan Liu and Rey Coaguila and Maxwell Chen and Sally Ma and Pouya Tafti and Susheel Tatineni and Terry Spitz and Jiayu Ye and Paul Vicol and Mihaela Rosca and Adrià Puigdomènech and Zohar Yahav and Sanjay Ghemawat and Hanzhao Lin and Phoebe Kirk and Zaid Nabulsi and Sergey Brin and Bernd Bohnet and Ken Caluwaerts and Aditya Srikanth Veerubhotla and Dan Zheng and Zihang Dai and Petre Petrov and Yichong Xu and Ramin Mehran and Zhuo Xu and Luisa Zintgraf and Jiho Choi and Spurthi Amba Hombaiah and Romal Thoppilan and Sashank Reddi and Lukasz Lew and Li Li and Kellie Webster and KP Sawhney and Lampros Lamprou and Siamak Shakeri and Mayank Lunayach and Jianmin Chen and Sumit Bagri and Alex Salcianu and Ying Chen and Yani Donchev and Charlotte Magister and Signe Nørly and Vitor Rodrigues and Tomas Izo and Hila Noga and Joe Zou and Thomas Köppe and Wenxuan Zhou and Kenton Lee and Xiangzhu Long and Danielle Eisenbud and Anthony Chen and Connor Schenck and Chi Ming To and Peilin Zhong and Emanuel Taropa and Minh Truong and Omer Levy and Danilo Martins and Zhiyuan Zhang and Christopher Semturs and Kelvin Zhang and Alex Yakubovich and Pol Moreno and Lara McConnaughey and Di Lu and Sam Redmond and Lotte Weerts and Yonatan Bitton and Tiziana Refice and Nicolas Lacasse and Arthur Conmy and Corentin Tallec and Julian Odell and Hannah Forbes-Pollard and Arkadiusz Socala and Jonathan Hoech and Pushmeet Kohli and Alanna Walton and Rui Wang and Mikita Sazanovich and Kexin Zhu and Andrei Kapishnikov and Rich Galt and Matthew Denton and Ben Murdoch and Caitlin Sikora and Kareem Mohamed and Wei Wei and Uri First and Tim McConnell and Luis C. Cobo and James Qin and Thi Avrahami and Daniel Balle and Yu Watanabe and Annie Louis and Adam Kraft and Setareh Ariafar and Yiming Gu and Eugénie Rives and Charles Yoon and Andrei Rusu and James Cobon-Kerr and Chris Hahn and Jiaming Luo and Yuvein and Zhu and Niharika Ahuja and Rodrigo Benenson and Raphaël Lopez Kaufman and Honglin Yu and Lloyd Hightower and Junlin Zhang and Darren Ni and Lisa Anne Hendricks and Gabby Wang and Gal Yona and Lalit Jain and Pablo Barrio and Surya Bhupatiraju and Siva Velusamy and Allan Dafoe and Sebastian Riedel and Tara Thomas and Zhe Yuan and Mathias Bellaiche and Sheena Panthaplackel and Klemen Kloboves and Sarthak Jauhari and Canfer Akbulut and Todor Davchev and Evgeny Gladchenko and David Madras and Aleksandr Chuklin and Tyrone Hill and Quan Yuan and Mukundan Madhavan and Luke Leonhard and Dylan Scandinaro and Qihang Chen and Ning Niu and Arthur Douillard and Bogdan Damoc and Yasumasa Onoe and Fabian Pedregosa and Fred Bertsch and Chas Leichner and Joseph Pagadora and Jonathan Malmaud and Sameera Ponda and Andy Twigg and Oleksii Duzhyi and Jingwei Shen and Miaosen Wang and Roopal Garg and Jing Chen and Utku Evci and Jonathan Lee and Leon Liu and Koji Kojima and Masa Yamaguchi and Arunkumar Rajendran and AJ Piergiovanni and Vinodh Kumar Rajendran and Marco Fornoni and Gabriel Ibagon and Harry Ragan and Sadh MNM Khan and John Blitzer and Andrew Bunner and Guan Sun and Takahiro Kosakai and Scott Lundberg and Ndidi Elue and Kelvin Guu and SK Park and Jane Park and Arunachalam Narayanaswamy and Chengda Wu and Jayaram Mudigonda and Trevor Cohn and Hairong Mu and Ravi Kumar and Laura Graesser and Yichi Zhang and Richard Killam and Vincent Zhuang and Mai Giménez and Wael Al Jishi and Ruy Ley-Wild and Alex Zhai and Kazuki Osawa and Diego Cedillo and Jialu Liu and Mayank Upadhyay and Marcin Sieniek and Roshan Sharma and Tom Paine and Anelia Angelova and Sravanti Addepalli and Carolina Parada and Kingshuk Majumder and Avery Lamp and Sanjiv Kumar and Xiang Deng and Artiom Myaskovsky and Tea Sabolić and Jeffrey Dudek and Sarah York and Félix de Chaumont Quitry and Jiazhong Nie and Dee Cattle and Alok Gunjan and Bilal Piot and Waleed Khawaja and Seojin Bang and Simon Wang and Siavash Khodadadeh and Raghavender R and Praynaa Rawlani and Richard Powell and Kevin Lee and Johannes Griesser and GS Oh and Cesar Magalhaes and Yujia Li and Simon Tokumine and Hadas Natalie Vogel and Dennis Hsu and Arturo BC and Disha Jindal and Matan Cohen and Zi Yang and Junwei Yuan and Dario de Cesare and Tony Bruguier and Jun Xu and Monica Roy and Alon Jacovi and Dan Belov and Rahul Arya and Phoenix Meadowlark and Shlomi Cohen-Ganor and Wenting Ye and Patrick Morris-Suzuki and Praseem Banzal and Gan Song and Pranavaraj Ponnuramu and Fred Zhang and George Scrivener and Salah Zaiem and Alif Raditya Rochman and Kehang Han and Badih Ghazi and Kate Lee and Shahar Drath and Daniel Suo and Antonious Girgis and Pradeep Shenoy and Duy Nguyen and Douglas Eck and Somit Gupta and Le Yan and Joao Carreira and Anmol Gulati and Ruoxin Sang and Daniil Mirylenka and Emma Cooney and Edward Chou and Mingyang Ling and Cindy Fan and Ben Coleman and Guilherme Tubone and Ravin Kumar and Jason Baldridge and Felix Hernandez-Campos and Angeliki Lazaridou and James Besley and Itay Yona and Neslihan Bulut and Quentin Wellens and AJ Pierigiovanni and Jasmine George and Richard Green and Pu Han and Connie Tao and Geoff Clark and Chong You and Abbas Abdolmaleki and Justin Fu and Tongzhou Chen and Ashwin Chaugule and Angad Chandorkar and Altaf Rahman and Will Thompson and Penporn Koanantakool and Mike Bernico and Jie Ren and Andrey Vlasov and Sergei Vassilvitskii and Maciej Kula and Yizhong Liang and Dahun Kim and Yangsibo Huang and Chengxi Ye and Dmitry Lepikhin and Wesley Helmholz},
      year={2025},
      eprint={2507.06261},
      archivePrefix={arXiv},
      primaryClass={cs.CL},
      url={https://arxiv.org/abs/2507.06261}, 
}

@misc{deepseekcoderv2,
      title={DeepSeek-Coder-V2: Breaking the Barrier of Closed-Source Models in Code Intelligence}, 
      author={DeepSeek-AI and Qihao Zhu and Daya Guo and Zhihong Shao and Dejian Yang and Peiyi Wang and Runxin Xu and Y. Wu and Yukun Li and Huazuo Gao and Shirong Ma and Wangding Zeng and Xiao Bi and Zihui Gu and Hanwei Xu and Damai Dai and Kai Dong and Liyue Zhang and Yishi Piao and Zhibin Gou and Zhenda Xie and Zhewen Hao and Bingxuan Wang and Junxiao Song and Deli Chen and Xin Xie and Kang Guan and Yuxiang You and Aixin Liu and Qiushi Du and Wenjun Gao and Xuan Lu and Qinyu Chen and Yaohui Wang and Chengqi Deng and Jiashi Li and Chenggang Zhao and Chong Ruan and Fuli Luo and Wenfeng Liang},
      year={2024},
      eprint={2406.11931},
      archivePrefix={arXiv},
      primaryClass={cs.SE},
      url={https://arxiv.org/abs/2406.11931}, 
}

@inproceedings{mamedov-etal-2025-gigachat,
    title = "{G}iga{C}hat Family: Efficient {R}ussian Language Modeling Through Mixture of Experts Architecture",
    author = "Mamedov, Valentin  and
      Kosarev, Evgenii  and
      Leleytner, Gregory  and
      Shchuckin, Ilya  and
      Berezovskiy, Valeriy  and
      Smirnov, Daniil  and
      Kozlov, Dmitry  and
      Averkiev, Sergei  and
      Ivan, Lukyanenko  and
      Proshunin, Aleksandr  and
      Israfilova, Ainur  and
      Baskov, Ivan  and
      Chervyakov, Artem  and
      Shakirov, Emil  and
      Kolesov, Mikhail  and
      Khomich, Daria  and
      Latortseva, Daria  and
      Porkhun, Sergei  and
      Fedorov, Yury  and
      Kutuzov, Oleg  and
      Kudriavtseva, Polina  and
      Soldatova, Sofiia  and
      Egor, Kolodin  and
      Pyatkin, Stanislav  and
      Menshykh, Dzmitry  and
      IUrevich, Grafov Sergei  and
      Damirov, Eldar  and
      Karlov, Vladimir  and
      Gaitukiev, Ruslan  and
      Shatenov, Arkadiy  and
      Fenogenova, Alena  and
      Savushkin, Nikita  and
      Minkin, Fedor",
    editor = "Mishra, Pushkar  and
      Muresan, Smaranda  and
      Yu, Tao",
    booktitle = "Proceedings of the 63rd Annual Meeting of the Association for Computational Linguistics (Volume 3: System Demonstrations)",
    month = jul,
    year = "2025",
    address = "Vienna, Austria",
    publisher = "Association for Computational Linguistics",
    url = "https://aclanthology.org/2025.acl-demo.10/",
    doi = "10.18653/v1/2025.acl-demo.10",
    pages = "93--106",
    ISBN = "979-8-89176-253-4"
}

@article{Tikhomirov_Chernyshov_2024, title={Facilitating Large Language Model Russian Adaptation with Learned Embedding Propagation}, volume={10}, url={https://jle.hse.ru/article/view/22224}, DOI={10.17323/jle.2024.22224}, abstractNote={&lt;p&gt;&lt;strong&gt;Background: &lt;/strong&gt;Recent advancements in large language model (LLM) technologies have introduced powerful open-source instruction-tuned LLMs that match the text generation quality of leading models like GPT-4. Despite accelerating LLM adoption in sensitive-information environments, the lack of disclosed training data hinders replication and makes these achievements exclusive to specific models.&lt;/p&gt; &lt;p&gt;&lt;strong&gt;Purpose: &lt;/strong&gt;Given the multilingual nature of the latest iteration of open-source LLMs, the benefits of training language-specific LLMs diminish, leaving computational efficiency as the sole guaranteed advantage of this computationally-expensive procedure. This work aims to address the language-adaptation limitations posed by restricted access to high-quality instruction-tuning data, offering a more cost-effective pipeline.&lt;/p&gt; &lt;p&gt;&lt;strong&gt;Method: &lt;/strong&gt;To tackle language-adaptation challenges, we introduce Learned Embedding Propagation (LEP), a novel method with lower training data requirements and minimal disruption of existing LLM knowledge. LEP employs an innovative embedding propagation technique, bypassing the need for instruction-tuning and directly integrating new language knowledge into any instruct-tuned LLM variant. Additionally, we developed Darumeru, a new benchmark for evaluating text generation robustness during training, specifically tailored for Russian adaptation.&lt;/p&gt; &lt;p&gt;&lt;strong&gt;Results: &lt;/strong&gt;We applied the LEP method to adapt LLaMa-3-8B and Mistral-7B for Russian, testing four different vocabulary adaptation scenarios. Evaluation demonstrates that LEP achieves competitive performance levels, comparable to OpenChat 3.5 and LLaMa-3-8B-Instruct. Further improvements were observed through self-calibration and additional instruction-tuning steps, enhancing task-solving capabilities beyond the original models.&lt;/p&gt; &lt;p&gt;&lt;strong&gt;Conclusion: &lt;/strong&gt;LEP offers a viable and efficient alternative to traditional language-specific instruction-tuning, significantly reducing the costs associated with language adaptation while maintaining or surpassing the performance benchmarks set by contemporary LLMs.&lt;/p&#38;gt;}, number={4}, journal={Journal of Language and Education}, author={Tikhomirov, Mikhail and Chernyshov, Daniil}, year={2024}, month={Dec.}, pages={130-145} }

@misc{yang2025qwen3technicalreport,
      title={Qwen3 Technical Report}, 
      author={An Yang and Anfeng Li and Baosong Yang and Beichen Zhang and Binyuan Hui and Bo Zheng and Bowen Yu and Chang Gao and Chengen Huang and Chenxu Lv and Chujie Zheng and Dayiheng Liu and Fan Zhou and Fei Huang and Feng Hu and Hao Ge and Haoran Wei and Huan Lin and Jialong Tang and Jian Yang and Jianhong Tu and Jianwei Zhang and Jianxin Yang and Jiaxi Yang and Jing Zhou and Jingren Zhou and Junyang Lin and Kai Dang and Keqin Bao and Kexin Yang and Le Yu and Lianghao Deng and Mei Li and Mingfeng Xue and Mingze Li and Pei Zhang and Peng Wang and Qin Zhu and Rui Men and Ruize Gao and Shixuan Liu and Shuang Luo and Tianhao Li and Tianyi Tang and Wenbiao Yin and Xingzhang Ren and Xinyu Wang and Xinyu Zhang and Xuancheng Ren and Yang Fan and Yang Su and Yichang Zhang and Yinger Zhang and Yu Wan and Yuqiong Liu and Zekun Wang and Zeyu Cui and Zhenru Zhang and Zhipeng Zhou and Zihan Qiu},
      year={2025},
      eprint={2505.09388},
      archivePrefix={arXiv},
      primaryClass={cs.CL},
      url={https://arxiv.org/abs/2505.09388}, 
}

@misc{kimiteam2025kimik2openagentic,
      title={Kimi K2: Open Agentic Intelligence}, 
      author={Kimi Team and Yifan Bai and Yiping Bao and Guanduo Chen and Jiahao Chen and Ningxin Chen and Ruijue Chen and Yanru Chen and Yuankun Chen and Yutian Chen and Zhuofu Chen and Jialei Cui and Hao Ding and Mengnan Dong and Angang Du and Chenzhuang Du and Dikang Du and Yulun Du and Yu Fan and Yichen Feng and Kelin Fu and Bofei Gao and Hongcheng Gao and Peizhong Gao and Tong Gao and Xinran Gu and Longyu Guan and Haiqing Guo and Jianhang Guo and Hao Hu and Xiaoru Hao and Tianhong He and Weiran He and Wenyang He and Chao Hong and Yangyang Hu and Zhenxing Hu and Weixiao Huang and Zhiqi Huang and Zihao Huang and Tao Jiang and Zhejun Jiang and Xinyi Jin and Yongsheng Kang and Guokun Lai and Cheng Li and Fang Li and Haoyang Li and Ming Li and Wentao Li and Yanhao Li and Yiwei Li and Zhaowei Li and Zheming Li and Hongzhan Lin and Xiaohan Lin and Zongyu Lin and Chengyin Liu and Chenyu Liu and Hongzhang Liu and Jingyuan Liu and Junqi Liu and Liang Liu and Shaowei Liu and T. Y. Liu and Tianwei Liu and Weizhou Liu and Yangyang Liu and Yibo Liu and Yiping Liu and Yue Liu and Zhengying Liu and Enzhe Lu and Lijun Lu and Shengling Ma and Xinyu Ma and Yingwei Ma and Shaoguang Mao and Jie Mei and Xin Men and Yibo Miao and Siyuan Pan and Yebo Peng and Ruoyu Qin and Bowen Qu and Zeyu Shang and Lidong Shi and Shengyuan Shi and Feifan Song and Jianlin Su and Zhengyuan Su and Xinjie Sun and Flood Sung and Heyi Tang and Jiawen Tao and Qifeng Teng and Chensi Wang and Dinglu Wang and Feng Wang and Haiming Wang and Jianzhou Wang and Jiaxing Wang and Jinhong Wang and Shengjie Wang and Shuyi Wang and Yao Wang and Yejie Wang and Yiqin Wang and Yuxin Wang and Yuzhi Wang and Zhaoji Wang and Zhengtao Wang and Zhexu Wang and Chu Wei and Qianqian Wei and Wenhao Wu and Xingzhe Wu and Yuxin Wu and Chenjun Xiao and Xiaotong Xie and Weimin Xiong and Boyu Xu and Jing Xu and Jinjing Xu and L. H. Xu and Lin Xu and Suting Xu and Weixin Xu and Xinran Xu and Yangchuan Xu and Ziyao Xu and Junjie Yan and Yuzi Yan and Xiaofei Yang and Ying Yang and Zhen Yang and Zhilin Yang and Zonghan Yang and Haotian Yao and Xingcheng Yao and Wenjie Ye and Zhuorui Ye and Bohong Yin and Longhui Yu and Enming Yuan and Hongbang Yuan and Mengjie Yuan and Haobing Zhan and Dehao Zhang and Hao Zhang and Wanlu Zhang and Xiaobin Zhang and Yangkun Zhang and Yizhi Zhang and Yongting Zhang and Yu Zhang and Yutao Zhang and Yutong Zhang and Zheng Zhang and Haotian Zhao and Yikai Zhao and Huabin Zheng and Shaojie Zheng and Jianren Zhou and Xinyu Zhou and Zaida Zhou and Zhen Zhu and Weiyu Zhuang and Xinxing Zu},
      year={2025},
      eprint={2507.20534},
      archivePrefix={arXiv},
      primaryClass={cs.LG},
      url={https://arxiv.org/abs/2507.20534}, 
}

@misc{hui2024qwen25codertechnicalreport,
      title={Qwen2.5-Coder Technical Report}, 
      author={Binyuan Hui and Jian Yang and Zeyu Cui and Jiaxi Yang and Dayiheng Liu and Lei Zhang and Tianyu Liu and Jiajun Zhang and Bowen Yu and Keming Lu and Kai Dang and Yang Fan and Yichang Zhang and An Yang and Rui Men and Fei Huang and Bo Zheng and Yibo Miao and Shanghaoran Quan and Yunlong Feng and Xingzhang Ren and Xuancheng Ren and Jingren Zhou and Junyang Lin},
      year={2024},
      eprint={2409.12186},
      archivePrefix={arXiv},
      primaryClass={cs.CL},
      url={https://arxiv.org/abs/2409.12186}, 
}

@misc{seed2025seedcoderletcodemodel,
      title={Seed-Coder: Let the Code Model Curate Data for Itself}, 
      author={ByteDance Seed and Yuyu Zhang and Jing Su and Yifan Sun and Chenguang Xi and Xia Xiao and Shen Zheng and Anxiang Zhang and Kaibo Liu and Daoguang Zan and Tao Sun and Jinhua Zhu and Shulin Xin and Dong Huang and Yetao Bai and Lixin Dong and Chao Li and Jianchong Chen and Hanzhi Zhou and Yifan Huang and Guanghan Ning and Xierui Song and Jiaze Chen and Siyao Liu and Kai Shen and Liang Xiang and Yonghui Wu},
      year={2025},
      eprint={2506.03524},
      archivePrefix={arXiv},
      primaryClass={cs.CL},
      url={https://arxiv.org/abs/2506.03524}, 
}

@misc{deepseekai2025deepseekv3technicalreport,
      title={DeepSeek-V3 Technical Report}, 
      author={DeepSeek-AI and Aixin Liu and Bei Feng and Bing Xue and Bingxuan Wang and Bochao Wu and Chengda Lu and Chenggang Zhao and Chengqi Deng and Chenyu Zhang and Chong Ruan and Damai Dai and Daya Guo and Dejian Yang and Deli Chen and Dongjie Ji and Erhang Li and Fangyun Lin and Fucong Dai and Fuli Luo and Guangbo Hao and Guanting Chen and Guowei Li and H. Zhang and Han Bao and Hanwei Xu and Haocheng Wang and Haowei Zhang and Honghui Ding and Huajian Xin and Huazuo Gao and Hui Li and Hui Qu and J. L. Cai and Jian Liang and Jianzhong Guo and Jiaqi Ni and Jiashi Li and Jiawei Wang and Jin Chen and Jingchang Chen and Jingyang Yuan and Junjie Qiu and Junlong Li and Junxiao Song and Kai Dong and Kai Hu and Kaige Gao and Kang Guan and Kexin Huang and Kuai Yu and Lean Wang and Lecong Zhang and Lei Xu and Leyi Xia and Liang Zhao and Litong Wang and Liyue Zhang and Meng Li and Miaojun Wang and Mingchuan Zhang and Minghua Zhang and Minghui Tang and Mingming Li and Ning Tian and Panpan Huang and Peiyi Wang and Peng Zhang and Qiancheng Wang and Qihao Zhu and Qinyu Chen and Qiushi Du and R. J. Chen and R. L. Jin and Ruiqi Ge and Ruisong Zhang and Ruizhe Pan and Runji Wang and Runxin Xu and Ruoyu Zhang and Ruyi Chen and S. S. Li and Shanghao Lu and Shangyan Zhou and Shanhuang Chen and Shaoqing Wu and Shengfeng Ye and Shengfeng Ye and Shirong Ma and Shiyu Wang and Shuang Zhou and Shuiping Yu and Shunfeng Zhou and Shuting Pan and T. Wang and Tao Yun and Tian Pei and Tianyu Sun and W. L. Xiao and Wangding Zeng and Wanjia Zhao and Wei An and Wen Liu and Wenfeng Liang and Wenjun Gao and Wenqin Yu and Wentao Zhang and X. Q. Li and Xiangyue Jin and Xianzu Wang and Xiao Bi and Xiaodong Liu and Xiaohan Wang and Xiaojin Shen and Xiaokang Chen and Xiaokang Zhang and Xiaosha Chen and Xiaotao Nie and Xiaowen Sun and Xiaoxiang Wang and Xin Cheng and Xin Liu and Xin Xie and Xingchao Liu and Xingkai Yu and Xinnan Song and Xinxia Shan and Xinyi Zhou and Xinyu Yang and Xinyuan Li and Xuecheng Su and Xuheng Lin and Y. K. Li and Y. Q. Wang and Y. X. Wei and Y. X. Zhu and Yang Zhang and Yanhong Xu and Yanhong Xu and Yanping Huang and Yao Li and Yao Zhao and Yaofeng Sun and Yaohui Li and Yaohui Wang and Yi Yu and Yi Zheng and Yichao Zhang and Yifan Shi and Yiliang Xiong and Ying He and Ying Tang and Yishi Piao and Yisong Wang and Yixuan Tan and Yiyang Ma and Yiyuan Liu and Yongqiang Guo and Yu Wu and Yuan Ou and Yuchen Zhu and Yuduan Wang and Yue Gong and Yuheng Zou and Yujia He and Yukun Zha and Yunfan Xiong and Yunxian Ma and Yuting Yan and Yuxiang Luo and Yuxiang You and Yuxuan Liu and Yuyang Zhou and Z. F. Wu and Z. Z. Ren and Zehui Ren and Zhangli Sha and Zhe Fu and Zhean Xu and Zhen Huang and Zhen Zhang and Zhenda Xie and Zhengyan Zhang and Zhewen Hao and Zhibin Gou and Zhicheng Ma and Zhigang Yan and Zhihong Shao and Zhipeng Xu and Zhiyu Wu and Zhongyu Zhang and Zhuoshu Li and Zihui Gu and Zijia Zhu and Zijun Liu and Zilin Li and Ziwei Xie and Ziyang Song and Ziyi Gao and Zizheng Pan},
      year={2025},
      eprint={2412.19437},
      archivePrefix={arXiv},
      primaryClass={cs.CL},
      url={https://arxiv.org/abs/2412.19437}, 
}

@misc{ai2025yiopenfoundationmodels,
      title={Yi: Open Foundation Models by 01.AI}, 
      author={01. AI and : and Alex Young and Bei Chen and Chao Li and Chengen Huang and Ge Zhang and Guanwei Zhang and Guoyin Wang and Heng Li and Jiangcheng Zhu and Jianqun Chen and Jing Chang and Kaidong Yu and Peng Liu and Qiang Liu and Shawn Yue and Senbin Yang and Shiming Yang and Wen Xie and Wenhao Huang and Xiaohui Hu and Xiaoyi Ren and Xinyao Niu and Pengcheng Nie and Yanpeng Li and Yuchi Xu and Yudong Liu and Yue Wang and Yuxuan Cai and Zhenyu Gu and Zhiyuan Liu and Zonghong Dai},
      year={2025},
      eprint={2403.04652},
      archivePrefix={arXiv},
      primaryClass={cs.CL},
      url={https://arxiv.org/abs/2403.04652}, 
}

@misc{jiang2024mixtralexperts,
      title={Mixtral of Experts}, 
      author={Albert Q. Jiang and Alexandre Sablayrolles and Antoine Roux and Arthur Mensch and Blanche Savary and Chris Bamford and Devendra Singh Chaplot and Diego de las Casas and Emma Bou Hanna and Florian Bressand and Gianna Lengyel and Guillaume Bour and Guillaume Lample and Lélio Renard Lavaud and Lucile Saulnier and Marie-Anne Lachaux and Pierre Stock and Sandeep Subramanian and Sophia Yang and Szymon Antoniak and Teven Le Scao and Théophile Gervet and Thibaut Lavril and Thomas Wang and Timothée Lacroix and William El Sayed},
      year={2024},
      eprint={2401.04088},
      archivePrefix={arXiv},
      primaryClass={cs.LG},
      url={https://arxiv.org/abs/2401.04088}, 
}

@article{roziere2023code,
  title={Code llama: Open foundation models for code},
  author={Roziere, Baptiste and Gehring, Jonas and Gloeckle, Fabian and Sootla, Sten and Gat, Itai and Tan, Xiaoqing Ellen and Adi, Yossi and Liu, Jingyu and Sauvestre, Romain and Remez, Tal and others},
  journal={arXiv preprint arXiv:2308.12950},
  year={2023}
}
